\documentclass[preprint2]{proto}
\usepackage{times}
\usepackage{amsmath}
\usepackage{xcolor}
\usepackage{ulem}
\usepackage{hyperref}

\usepackage[top=17truemm,bottom=17truemm,left=20truemm,right=15truemm]{geometry}

\usepackage[switch]{lineno}         

\newcommand\amrev[1]{{{#1}}}
\newcommand\am[1]{{{#1}}}
\newcommand\yt[1]{{{#1}}}

\newcommand\bc[1]{{{#1}}}

\begin{document}

\title{\textbf{\LARGE The role of magnetic fields in the formation of protostars, disks, and outflows }}

\author {\textbf{\large Yusuke Tsukamoto }}
\affil{\small Kaghoshima University,Department of Physics and Astronomy
Graduate School of Science and Engineering, Korimoto, Kagoshima, Japan }
\author {\textbf{\large Anaëlle Maury}}
\affil{\small CEA/DRF/IRFU Astrophysics department, Université Paris-Saclay, UMR AIM, F-91191 Gif-sur-Yvette, France}
\author {\textbf{\large Beno\^it Commer\c con}} 
\affil{\small Univ Lyon, Ens de Lyon, Univ Lyon1, CNRS, Centre de Recherche Astrophysique de Lyon UMR5574, F-69007, Lyon, France}
\author {\textbf{\large Felipe O. Alves}}
\affil{\small Max-Planck-Institut f\"ur extraterrestrische Physik, Gie{\ss}enbachstra{\ss}e 1, Garching, 85748, Germany}
\author {\textbf{\large Erin G. Cox}}
\affil{\small Center for Interdisciplinary Exploration and Research in Astrophysics (CIERA), 1800 Sherman Avenue, Evanston, IL 60201, USA}
\author {\textbf{\large Nami Sakai}}
\affil{\small RIKEN Cluster for Pioneering Research, 2-1 Hirosawa, Wako, Saitama, Japan}
\author {\textbf{\large Tom Ray}}
\affil{\small Astronomy and Astrophysics Section, School of Cosmic Physics, Dublin Institute for Advanced Studies, Ireland}
\author {\textbf{\large Bo Zhao}}
\affil{\small Max-Planck-Institut für extraterrestrische Physik,
Gie{\ss}enbachstra{\ss}e 1, Garching, 85748, Germany}
\author {\textbf{\large Masahiro N. Machida}}
\affil{\small Kyushu University, Department of Earth and Planetary Sciences, Motoka, Fukuoka, Japan}

\begin{abstract}
\baselineskip = 11pt
\leftskip = 0.65in 
\rightskip = 0.65in
\parindent=1pc
{\small Abstract}
We present our current understanding of the formation and early evolution of protostars, protoplanetary disks, and the driving of outflows as dictated by the interplay of magnetic fields and partially ionized gas in molecular cloud cores.
In recent years, the field has witnessed enormous development through sub-millimeter observations which in turn have constrained models of protostar formation. 
As a result of these observations 
the state-of-the-art theoretical understanding of the formation and evolution of young stellar objects is described.
In particular, we emphasize the importance of the coupling, decoupling, and re-coupling between weakly ionized gas and the magnetic field on appropriate scales. This highlights the complex and intimate relationship between gravitational collapse and magnetic fields in young protostars.\\

\end{abstract}  

\section{Introduction}

\am{The evolution of solar-type protostars is classified in stages following an empirical sequence. It starts from the most embedded objects, Class 0 protostars, where the envelope dominates the mass: this main accretion stage is believed to last less than 0.1\,Myr \citep[e.g.][]{2009ApJS..181..321E,2011A&A...535A..77M}. Once the envelope mass and the mass accumulated in the central embryo are roughly equal, protostars enter the Class I phase which sees most of the second half of the final stellar mass accreted, over a fraction of a Myr. Finally, the young stellar object is no longer enbedded and becomes visible during the T-Tauri phase that precedes its arrival onto the Main Sequence. In this review, we focus on the physics at work during the early stages of star formation, in embedded protostars.}

Recent years have witnessed the major development of realistic magneto-hydrodynamics simulations modelling the physics of protostellar formation, as well as a cascade of detailed, sensitive observations of large samples of protostellar objects. These advances have transformed our understanding and highlighted the paramount role of the magnetic field during the early evolution of young stellar objects.
Indeed, radio interferometers sensitive to the cold and dense material typical of embedded protostars, e.g., the Atacama Large Millimeter Array (ALMA), Northern Extended Millimeter Aray (NOEMA), and the Very Large Array (VLA), have significantly extended their capabilities since \textit{Protostars and Planets VI} (henceforth PP\,VI). 
Exquisite maps probing the gas, dust, and magnetic field properties on the very small scales of protostellar cores, where material is accreted into a stellar embryo, stored in a disk, and partially ejected via outflows and jets, have placed unprecedented constraints on theoretical models. 

In parallel, state-of-the art computations now provide a comprehensive theoretical landscape thanks to recent progress in 3-D magneto-hydrodynamic simulations which are able to follow the time evolution of the physics from the large spatial scales typical of dense cores ($\sim 10^4$ au) to the protostar-disk system ($\sim 10^2$ au), over long timescales from the onset of prestellar collapse to the end of the Class I phase ($\sim 10^5$ years after protostar formation). They revealed the essential physics in the complex processes involved in the formation of protostar, disk, and outflow; i.e., the coupling, decoupling, and re-coupling between weakly ionized gas and magnetic field on appropriate scales.
\yt{in these processes, microscopic ion chemistry, especially adsorption of charged particles by dust, plays an essential role.}

In this chapter, we aim to present 
a synthesis of the observational and theoretical progresses made in the past decade to enrich our understanding of the formation of protostars, their protostellar disks, outflows, and jets.


%

This chapter is organized as follows: Section \ref{sec:observations} presents the state-of-the-art observational characterization of magnetic fields, gas and dust properties in protostellar environments. Section \ref{sec:theory} describes our current understanding of the formation and early evolution of protostars, protoplanetary disks, outflows, and jets based on recent theoretical studies. We present theories that are in the main successful in reproducing current observations, but also highlight some remaining discrepancies and key challenges for both observations and modelling to tackle in the near future. In Section \ref{sec:dust}, we describe a promising avenue for future research regarding the evolution of dust in magnetized protostellar environments, and its relevance to understanding 
not only the formation of planetesimals but also the pristine properties of the disks and stars that host them.

\section{\textbf{Observations of embedded protostars: envelopes, young disks, outflows, and magnetic fields}}
\label{sec:observations}
\subsection{Observing magnetic fields: techniques}\label{magnetic_measuring_techniques}

Since {\it Protostar and Planets VI} (see chapter by \citealt{2014prpl.conf..101L}), our view of interstellar magnetic fields as a key ingredient in threading the interstellar medium from clouds, to filaments and their embedded cores has been firmly established by multi-wavelength analysis of polarized light. Several observational techniques have been developed or \amrev{improved}. We summarise them below and the constraints they allow us to put on the strength and topology of magnetic fields in star-forming structures.

The linear polarization of dust thermal emission and absorption is widely used to trace the magnetic field topology from the plane-of-the-sky component $B_\mathrm{pos}$ of the magnetic vector {\bf{B}}, in dense ISM structures. The polarized dust thermal emission largely maps the field topology at mm/sub-mm wavelengths: assuming dust grains populating the star-forming structures are elongated, they are expected to produce polarized dust emission if they are aligned along B-field lines in an anisotropic radiation field \citep[radiative torques alignment,][]{2007MNRAS.378..910L}. Optical and near-IR observations of the dichroic polarization of background stars are also used to probe magnetic fields along highly extincted lines-of-sight, such as those typical of prestellar cores ($A_\mathrm{v} >10$).
The magnetic field strength can be estimated from these polarimetric observations, by performing a statistical analysis of the dispersion of polarization angles compared to the velocity dispersion of the gas, such as in the Davis-Chandrasekhar-Fermi method \citep[DCF,][]{1951PhRv...81..890D,1953ApJ...118..113C}. Although this method relies heavily on the assumed physical properties of the medium and how the latter couples to the magnetic field \citep{2008ApJ...679..537F,2021ApJ...919...79L,2022ApJ...925...30L}, it has been widely used in studies from clouds to cores scales.

The only direct technique to estimate the strength of the magnetic field in star-forming structures is through utilizing the Zeeman effect via spectro-polarimetry. Observations of molecular species with significant magnetic susceptibility, i.e., those with an unpaired electron, can be used to trace the magnitude of the line-of-sight component $B_\mathrm{los}$ of the magnetic vector {\bf{B}} in star-forming clouds, dense cores, and disks \citep{2012ARA&A..50...29C,2019FrASS...6...66C}. 
Spectral-line polarization can also arise from the so-called Goldreich-Kylafis (GK) effect \citep{1982ApJ...253..606G}, because individual molecular rotational levels, in the presence of a magnetic field, split into sub-levels \amrev{and the resulting line emission can be polarized either parallel or perpendicular to the magnetic field depending on the sign of $\rm{n}_u - \rm{n}_{\pm}$, where $\rm{n}_u$ and $\rm{n}_{\pm}$ are the populations of an upper energy level and its magnetic sub-levels, respectively}. It is worth pointing out however that the GK effect requires optical depths $\tau \approx 1$ to be observed, i.e.\ the radiation field has to be anisotropic. It is not usable as a means of \amrev{measuring the} magnetic field if the optical depth is large. 

When magnetic field strengths can be estimated, they are typically used to compute two physical quantities that tell about the importance of magnetic field against other key ingredients, turbulence and gravity. 
The Alfv{\'e}nic Mach number captures the relative importance of magnetic and turbulent energies: it is the velocity dispersion in the gas flow divided by the Alfv{\'e}n waves velocity, which is directly computed from the B-field strength.
In the densest star-forming structures, observations are used to estimate the ratio of gravitational to magnetic energies, a.k.a.\ the mass-to-flux ratio $\mu$ expressed in units of the critical value when the magnetic energy equals the gravitational energy \citep{1978PASJ...30..671N}. Gravitational collapse to form a stellar system requires $\mu>1$, otherwise the magnetic forces will dominate over gravity and stabilize the structure. In the latter case evolution out of equilibrium can only happen over long timescales by diffusive processes that allow to remove the magnetic flux from the central region of the core, for example by ambipolar diffusion \citep[see, e.g.][]{1983ApJ...273..202S}.

While many interesting constraints have been obtained regarding the magnetic field in star-forming clouds \citep[see for example the reviews by][]{2019FrASS...6...15P,2019FrASS...6....5H}, in the following we focus on observations that probe the magnetic properties of dense cores, protostars, their disks and outflows.

\subsection{Magnetic fields in prestellar cores and protostars}

In prestellar cores, both near-IR starlight polarization and sub-mm polarized dust emission have revealed a large fraction of cores with highly organized magnetic field topologies, at the typical $\sim 0.1$\,pc core scales \citep[see e.g.][and references therein]{2006MNRAS.369.1445K,2015AJ....149...31J, 2018ApJ...857..100K}. In the few cores where a magnetic field strength can be estimated, values are typically between 10 and 100 $\mu$G. A comparison of gravitational and magnetic energies then suggests a dynamical state in which the mass is close to the magnetic critical mass, and this could significantly delay gravitational collapse \citep{2013ApJ...769L..15S, 2006Sci...313..812G, 2017ApJ...848..110K, 2020PASJ...72....8K, 2020ApJ...891...55K, 2021arXiv210402597M}. 
The orientations of B-field lines mapped with optical/near-IR polarimetry are sometimes smoothly connected to the ones recovered with sub-mm single-dish observations on core scales \citep{2016A&A...596A..93S,2019ApJ...883....9S}, but are sometimes significantly different \citep{2014A&A...569L...1A,2016ApJ...833..176C}. Detailed analyses using statistics from large single-dish surveys, such as the JCMT BISTRO survey \citep{2017ApJ...842...66W,2019ApJ...877...43L}, and starlight polarimetry, may give more insights into the role and structure of magnetic fields in prestellar cores.

Thanks to the Zeeman effect, the OH molecule is used to probe the strength of magnetic fields in molecular clouds \citep{2008ApJ...680..457T}. However, OH cannot be used in the same way to investigate magnetic field strength in cores, due to their high gas densities. Here other tracers, such as CN, must be employed. 
Only the most massive protostellar cores have detections of the Zeeman effect in CN single-dish surveys: the values found for the mean total magnetic field, $B_\mathrm{tot}$, are typically around a few hundred $\mu$G \citep{2008A&A...487..247F}. They seem in broad agreement with statistical estimates derived from polarized dust emission in the mm, which suggest values up to 1 mG towards massive cores. Observed mass-to-magnetic flux ratios, derived from Zeeman measurements are found to be super-critical with typical values $\mu \sim 2-3$.
The average Alfv{\'e}nic Mach number in star-forming gas at densities $\sim 10^{3}-10^{6}$ cm$^{-3}$ inferred from these observations is then found to be $M_\mathrm{A} \sim 1.5$ \citep{2008A&A...487..247F,2009Sci...324.1408G}. This suggests that from clouds to cores scales, the turbulent and magnetic energies may be in approximate equipartition.
The magnetic field strengths observed are systematically higher in the CN cores than in the lower density clouds (typical magnetic field strengths of $\sim$ 10 to 30 $\mu$G for densities of a few $10^{3}$ cm$^{-3}$, and on $\sim$0.1\,pc scales, see e.g., Figure 1 in \citealt{2010ApJ...725..466C}). 
This suggests intrinsically stronger magnetic fields in denser gas, but there could also be a bias through the specific conditions necessary to form the most massive cores. 
Note that it is still unclear however whether the switch in Zeeman molecular tracer is responsible for the steep increase in the observed $B_\mathrm{tot}$ between clouds and protostellar cores.
Finally, the total magnetic pressure of a core with a complex magnetic field could be higher than what is inferred from the Zeeman effect, due to the cancelling effect of field vectors with opposite directions along the line of sight. With all these caveats in mind, nevertheless these measurements suggest magnetic fields are relatively strong in most massive star-forming cores. 

\begin{figure*}[ht]
\begin{center}
\includegraphics[trim=0.2cm 2.5cm 0.2cm 2cm, clip, width=140mm]{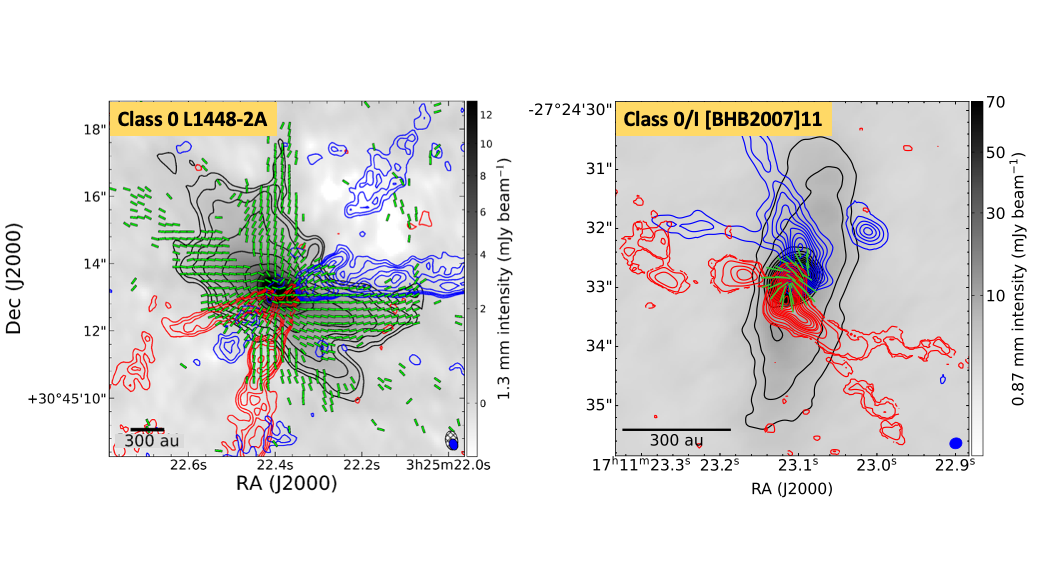}
\end{center}
\vspace{-8truemm}
\caption{Illustrative examples of protostellar magnetic fields mapped with millimeter/sub-millimeter polarized dust continuum emission on similar envelope/disk scales in the L1448-2A Class 0 protostar \citep[left panel, from][]{2019ApJ...879...25K} and the [BHB2007]11 Class 0/I protostar \citep[right panel, data published in][]{2017A&A...603L...3A,2018A&A...616A..56A}. Both panels show the ALMA dust continuum emission at sub-millimeter wavelength as a background image, and contours of integrated CO ($2\rightarrow 1$) emission tracing the outflow at blue/red velocities. The B-field lines, obtained from the polarized dust emission, are shown towards both objects as green line segments.
The synthesized beam of the observations are shown in the lower right corners.} 
\label{fig:Bfields}
\end{figure*}

Single-dish observations of the polarized dust emission have allowed us to map the magnetic field line spatial distribution in several low-mass protostellar cores, particularly in recent times as a result of the extensive JCMT BISTRO survey \citep[see, e.g.][]{2021ApJ...912L..27E}. In the inner envelope, on typical scales $\sim 100-5000$ au, polarized dust emission can be probed using (sub-) millimeter interferometry and has been detected at the few percent level \citep[see e.g.\,][]{2018A&A...616A.139G, 2018ApJ...855...92C}. The wealth of recent interferometric observations of magnetic field lines have suggested that the early detections of the ``hourglass" pattern \citep[e.g.\,][]{2006Sci...313..812G,2009ApJ...707..921R} may actually be a common feature in low-mass cores \citep{2018MNRAS.477.2760M,2019ApJ...879...25K,2019ApJ...871..243Y, 2020PASJ...72....8K}. An example can be found in the left panel of Figure \ref{fig:Bfields} showing the B-field topology reconstructed from ALMA observations of the millimeter polarized dust emission in a Class 0 protostar. Multi-scale observations and comparison of the magnetic field topology with the gas kinematics in the inner envelope suggest this hourglass shape may be a natural result from an initially quasi-poloidal configuration being affected by the collapse of the inner envelope: magnetic field lines efficiently coupled to the infalling gas are pulled towards the core center as the collapse proceeds. 
However, this morphology is not seen as ubiquitously as one might expect from a simple strongly magnetized scenario for low-mass star formation \citep{2019FrASS...6....3H} and this may stem from several causes. First, identifying hourglass fields is not simple due to uncertainties caused by projection effects and limited instrument sensitivity; thus a robust survey probing the magnetic field in star-forming cores with high spatial dynamic range still needs to be carried out. 
Second, the simple view that protostellar accretion proceeds largely in a purely symmetric fashion, with gas flowing down to the central embryo along the equatorial plane and creating a distinguishable hourglass pattern, has received much criticism, and contradicts some recent observations (see the PPVII review chapter by Pineda et al., and references in $\S$\ref{ssec:obs_disks} of the present chapter). 
The early development of non-isotropic infall and differences in gas ionization levels affecting the coupling at different locations in the protostellar cores may lead to the complex field geometries observed. Finally, local irradiation conditions may enhance polarized dust emission towards specific locations, making it difficult to assess the global 3-D B-field topology. Observations of the magnetic field on core scales have suggested in a few cases that the magnetic field lines could correlate with density and/or kinematic structures such as (gravo-)turbulence \citep[e.\,g., ][]{2017ApJ...842L...9H}, accretion streamers \citep[e.\,g.,][]{2018A&A...616A..56A, 2019ApJ...885..106L}, irradiated layers \citep{2020A&A...644A..11L} and protostellar outflows \citep[e.\,g.][]{2017ApJ...847...92H}.

Protostars are actively accreting envelope material onto the central stellar embryo: as such they are expected to be primarily super-critical ($\mu>1$). Unfortunately magnetic field strengths in solar-type cores have not been widely explored using the Zeeman effect, because it is a subtle effect in relatively weak spectral lines. Only recent sensitive observations have allowed us to detect large numbers of independent B-field orientations within single low-mass cores, and use the DCF method to estimate B-field strengths. Measured values suffer from large uncertainties due to the gravitational energies associated with these collapsing objects, but typical field strengths are around a few tens of $\mu$G \citep{2019FrASS...6....3H}. Finally, since the polarized flux is a cumulative quantity that is prone to cancellation if the polarization angle is highly disorganized along the line of sight, observations of a few percent of polarization suggests the magnetic field may be strong enough to remain at least partly organized inside star-forming cores \citep{2020A&A...644A..11L}.

\subsection{Magnetic fields observed in protostellar disks and outflows}

Polarization observations at high angular resolution should be able to differentiate the properties (i.e., polarization fraction and position angle) between components from the disk and the envelope \citep[e.g.\,][]{2018ApJ...855...92C,2019ApJS..245....2S}. 
However, although a small number of embedded disks show a few detectable polarization measurements, most observations at these small scales are still limited by sensitivity \citep[e.\,g., Oph-emb-1,][]{2019ApJS..245....2S}. Moreover, in relevant surveys the disk emission is often optically thick and B-field aligned grains are not the only plausible origin for the polarized emission, complicating the interpretation. 
Indeed, the scattering opacity of large grains (a$_{\mathrm{max}} > 100~\mu$m) is predicted to prevail over their absorption opacity \citep{2015ApJ...809...78K}, in most conditions typical of disks. This results in polarized emission due to self-scattering of the dust grains, up to a few percent at millimeter wavelengths in the most extreme cases.
In edge-on disks such as HH\,212 or L1527, the impact of large optical depth is clear on the polarization maps, where polarization angles are predominantly parallel to the disk minor axis \citep{2015ApJ...798L...2S,2018ApJ...861...91H,2021ApJ...910...75L}.
Hence, only very few measurements of dust polarization from disks in young protostars can be interpreted as magnetically aligned grains rather than by the polarized emission from dust self-scattering (this is usually probed by large polarization fractions and the combined analysis of the polarization properties at varying wavelengths). For example, ALMA observations of the Class I protostar TMC-1A shows polarization components from the disk poles that can be interpreted as arising from toroidal magnetic fields associated the disk or the outflow or magnetic accretion flows within the disk \citep{2021arXiv210710646A}. 
Another example is the disk around the Class 0/I [BHB2007] 11 protostar, the magnetic field of which is  modelled as a combination of poloidal and toroidal components produced by disk rotation and infalling material from the envelope, respectively \citep[see Fig. \ref{fig:Bfields} and][]{2018A&A...616A..56A}. However, if the dust population in this source is dominated by mm-size grains, the polarization observed at mm-wavelengths (i.e., in the Mie regime) is expected to be negative, which in practical terms means a 90-degree flip in the polarization direction \citep{2020A&A...634L..15G}. In this scenario, the disk polarization is parallel to the magnetic fields, potentially following the accretion streamers seen within the circumbinary disk \citep{2019Sci...366...90A}. 
A poloidal hour-glass magnetic field was also interpreted as causing the polarized dust emission around the  disk of VLA 1623A, at scales of 200 au \citep{2018ApJ...859..165S}. The non-detection of a toroidal component in the dust polarization signal in this protostar could be due to the predominance of scattering polarization from large grains at disk scales, non-ideal MHD effects decoupling the magnetic field from the rotating gas in the disk at small scales, or a combination of an intrinsically weak magnetic field and a weakly ionized medium on disk scales \citep{2018A&A...615A...5V,2020A&A...635A..67H}.

Spectro-polarimetry may become a powerful tool to measure the magnetic strength in embedded disks. The high abundance of the Zeeman-sensitive CN molecule in their midplane  \citep{2012A&A...537A..60C}, for example, make this molecule a potential tracer of magnetic fields in disks as well \citep{2017A&A...607A.104B}.  
Several attempts to carry out Zeeman measurements in mm lines with ALMA have provided stringent upper limits of a few mG for the line-of-sight magnetic field intensity in a couple of evolved disks \citep{2019A&A...624L...7V,2021ApJ...908..141H}, while polarized emission from spectral lines has been tentatively linked to the magnetic field morphology in the TW-Hya disk \citep{2021ApJ...922..139T}. In the future, it may be possible to extend this technique to embedded disks, and hopefully lift the current challenge to measuring their B-fields. 
Maser emission has also proved to be a powerful tool to study magnetic fields in the densest portions of star-forming regions \citep{2018NatAs...2..145L}. As masers are often very bright, polarimetric observations can detect the line-of-sight component of the magnetic field, reliably determining its strength (and direction, if VLBI techniques are employed) in high-mass \citep[e.g., using 6.7 GHz methanol masers, ][]{2013A&A...556A..73S} and low-mass  \citep[e.g., using 22 GHz water masers,][]{2012A&A...542A..14A} disks. Despite these observations being sensitive to very specific conditions such as shocks, where magnetic fields may be amplified to several mG, comparison to models may allow for a finer characterization of maser emission over a wide range of pumping and excitation conditions, using sources such as SiO and H$_2$O masers, in the near future \citep{2019A&A...628A..14L}.

\subsection{The physical and chemical conditions in inner envelopes: observational characterization}
\label{sec:kin}

The rotating/infalling envelopes of embedded protostars are the cradles where the most pristine disks form and evolve, at the same time as the accretion of envelope material onto the protostellar embryo. Both the gas kinematics and the chemical processes at work on disk-forming scales can directly affect the disk properties. We refer the reader to the dedicated PPVII chapters devoted to the gas kinematics in star-forming structures (Pineda et al.) and the chemical conditions in protostellar environments (Ceccarelli et al.) for detailed reviews. We briefly describe here the gas properties in the inner envelopes which have important consequences for how efficiently magnetic field couple to the protostellar gas, the transport and redistribution of angular momentum, and ultimately the properties of disk-forming material.

The recent development of observational capabilities (increased bandwidth of the Sub-millimeter Array (SMA), the upgrade of the NOEMA interferometer, and development of the ALMA observatory) have allowed us to observe a wide range of molecular lines as probes, not only to study the pristine chemistry in star-forming cores, but also to selectively study the physical processes at work in protostellar interiors, down to scales where disks form \citep{2020ARA&A..58..727J,2021PhR...893....1O}.
The warm inner envelope, apart from showing emission from complex organic molecules (COMs), also presents compact emission from small molecules like H$\rm{_{2}}$S, SO, OCS and H$\rm{^{13}}$CN, most likely related to ice sublimation and high-temperature chemistry \citep{2021arXiv210703696T}. 

\begin{figure*}[ht]
\begin{center}
\includegraphics[width=130mm]{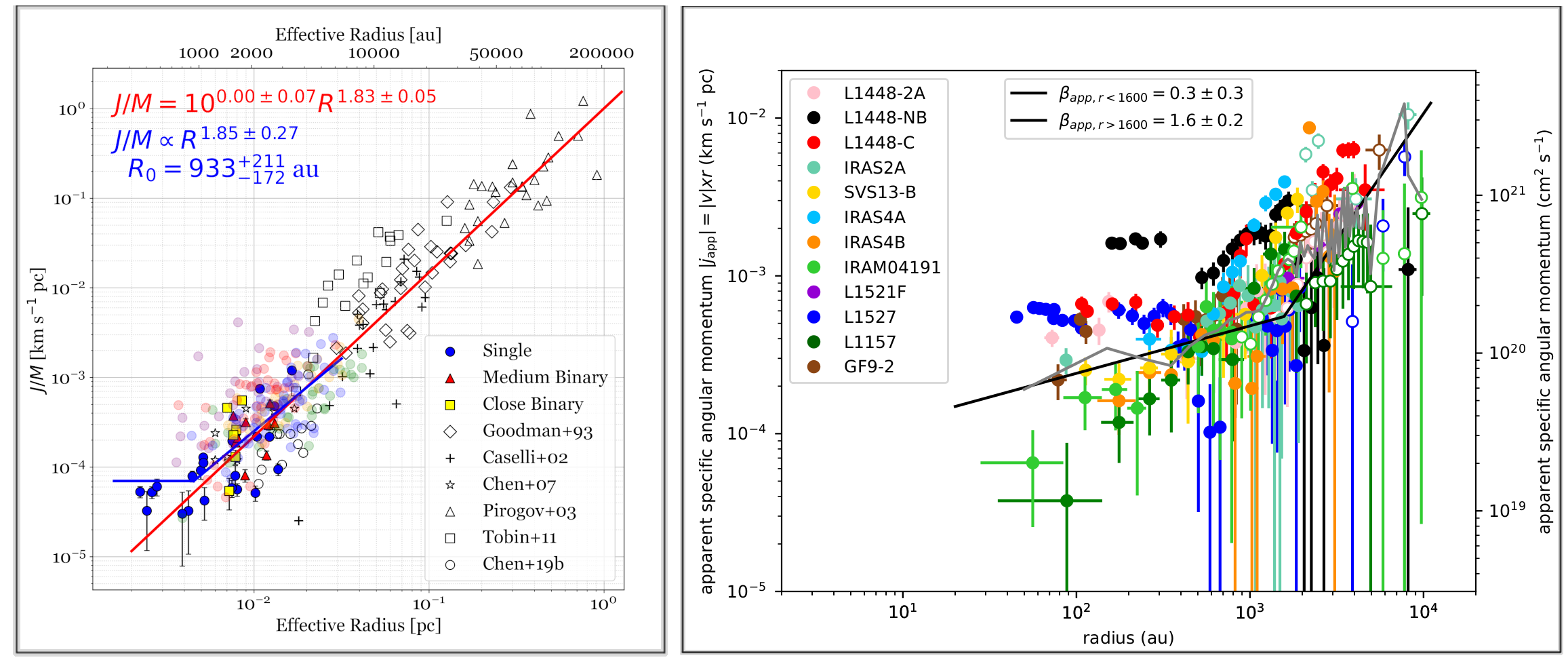}
\end{center}
\vspace{-8truemm}
\caption{Specific angular momentum of the gas measured at clouds and core scales (left panel, from \citealt{2021arXiv211209848H}) and inside protostellar envelopes (right panel, from \citealt{2020A&A...637A..92G}).} \label{fig:J-obs}
\end{figure*}

Molecular line observations are also used to characterize the angular momentum contained in the star-forming gas, ultimately responsible for the formation of the protostellar disks. 
Recent observations of Class 0 protostellar envelopes at large scales ($>$1000\,au) suggest the specific angular momentum scales with envelope radii, following a power-law relation $j \propto r^{1.8}$ between 1000 and 10000 au, with values ranging from $10^{-5}$ to a few $10^{-3}$ $\mathrm{km}\,{{\rm{s}}}^{-1}\,\mathrm{pc}$ at 1000 au \citep{2015ApJ...799..193Y,2019ApJ...882..103P,2021arXiv211209848H}.
In the CALYPSO survey, molecular line observations at sub-arcsecond resolution have allowed characterization of protostellar gas motions down to $\sim 50$ au scales. \citet{2020A&A...637A..92G} finds a break in the angular momentum evolution with radius, from a steep profile $j \propto r^{1.6}$ at radii $r>1600$ au, to a quasi-flat profile at radii 50--1600 au. These profiles are shown in the right panel of Figure \ref{fig:J-obs}, together with the SMA MASSES measurements from \citet{2021arXiv211209848H} in the left panel. 

Envelope mass infall rates are estimated by modelling molecular lines showing inverse P-Cygni profiles \citep{2012A&A...544L...7P,2013A&A...558A.126M,2015ApJ...814...22E} but these are rare due to he intrinsic complexity of the velocity field of the gas in the inner envelopes. Typical values measured on scales of a few hundred au suggest a gas mass accretion rate $\sim 10^{-5}$\,$\rm{M}_\odot$\ per year in Class 0 protostars. In the Class I protostars of the ALMA Ophiuchus survey, typical mass accretion rates are a few $10^{-7}$ $\rm{M}_\odot$\ per year \citep{2019A&A...626A..71A}. The rather high rates found in Class 0 protostars at first sight seem inconsistent with their observed protostellar luminosities because one would expect typical accretion luminosities of a few tens of $\rm{L}_\odot$\ while the median luminosity is observed to be around a few $\rm{L}_\odot$\ (with large variations from object to object). \am{This conundrum, known as the luminosity problem \citep{2012ApJ...747...52D,2014prpl.conf..195D}, may be solved by} by invoking episodic accretion, an idea that is also supported by observations of molecular species that are found at radii larger than expected based on their gas temperature. Assuming the chemical timescale is considerably longer than the cooling timescale of the gas, sudden temperature changes due to short but vigorous accretion bursts would explain why CO is observed where the current envelope temperature predicts it should be frozen onto dust grains ($<30$K) for example \citep{2016A&A...591A...3A,2016Natur.535..258C, 2017A&A...602A.120F}. A recent ALMA survey of N$\rm{_{2}}$H$\rm{^{+}}$ and HCO$\rm{^{+}}$ toward 39 Class\,0 and I protostars in Perseus suggest almost all sources in the sample show evidence for post-burst signatures in N$\rm{_{2}}$H$\rm{^{+}}$ while the frequency of the bursts may decrease with protostellar evolution, from a burst every 2400 yr in the Class 0 stage to 8000 yr in the Class I stage \citep{2019ApJ...884..149H}.
Moreover, observations show that the velocity field of the gas at radii $r<5000$ au is not organized as expected from the collapse of axisymmetric rotating cores, with many envelopes exhibiting reversal of their velocity fields or multiple velocity components \citep{2020A&A...637A..92G,2017ApJ...849...89M}, and well-developed asymmetric features which may trace preferred pathways that funnel accretion, such as streamers  \citep{2020NatAs...4.1158P} or supersonic infall along outflow cavity walls \citep{2021arXiv210701986C}.

The study of the molecular line emission at the disk surface and at the interface between the rotationally-supported disk and the infalling-rotating envelope may reveal the physical conditions of the accreted gas, at scales where the kinetic energy of the infalling gas is converted into thermal energy and rotational motions.
The gas from the infalling envelope landing onto the disk can create shocks \am{for example by friction between the rotating material and the infalling material}, raising the temperatures of the gas and dust at the disk-envelope interface to values much higher than from heating by stellar photons alone. These accretion shocks may therefore be crucial in setting the pristine chemical composition of young disks\citep{2014Natur.507...78S}.
\am{Observational signatures of} accretion shocks have been \am{reported, on scales between the centrifugal barrier (where most of the gas kinetic energy contained in infalling motion is converted to rotational motion, producing an azimuthal velocity,  $v_{\theta}$, larger than the Keplerian velocity) and the centrifugal radii (inside which the circular gas motions are nearly Keplerian)}, in a handful of young protostars.
For instance, SO and other sulfur-containing species were found to be enhanced at a radius $\sim 100$\,au around the L1527 disk as a result of sublimation of grain mantles due to a weak accretion shock \citep{2014ApJ...791L..38S,2017ApJ...839...47M}. Gas temperatures are also found to be enhanced in the shocked region from $\sim$30 K \am{pre-shock} to $>$60 K \am{inside the shocked region} \citep{2017MNRAS.467L..76S}. In IRAS16293-2422A, for comparison, it is enhanced from 80-120 K to 130-160 K  \citep{2016ApJ...824...88O}. 
Equivalently, warm SO$\rm{_{2}}$ emission, possibly related to an accretion shock, was observed toward the B335 Class 0 protostar (Bjerkeli et al. 2019) and a few Class I sources \citep{2019ApJ...881..112O,2019A&A...626A..71A}. 
The observed abundances of SO can be reproduced either with low-density shocks ($<10^{6}$ cm$^{-3}$) in a weak UV field, if the SO ice is not thermally sublimated, or high density shocks in strong UV fields \citep{2021arXiv210709750V}. This second hypothesis seem to be supported by tentative evidence of a correlation between the amount of warm SO$\rm{_{2}}$ and the bolometric luminosity, as well as observed SO$\rm{_{2}}$/SO column density ratios in \am{the Class I source} Elias 29 \citep{2019A&A...626A..71A}. 
However, high-angular resolution observations ($\sim 30$ au) of the Class I source TMC1-A show narrow SO emission lines coming from a ring-shaped morphology which may be linked to the warm inner envelope \citep{2021A&A...646A..72H} and SO is also found to trace molecular outflows and jets in some objects \citep{2016A&A...593L...4P, 2017A&A...607L...6T, 2020A&ARv..28....1L, 2021A&A...648A..45P}: emission from warm SO and SO$\rm{_{2}}$ is thus not an unambiguous tracer of accretion shocks. Observations of mid-infrared shock tracers such as H$\rm{_{2}}$O, high-J CO, [S I], and [O I] with the James Webb Space Telescope may allow a finer characterization of shock conditions and put constraints on both the chemical composition of the gas incorporated into the protostellar disks and the magnetic field coupling at the disk/envelope interface.

\subsection{Protostellar disks: observed properties}
\label{ssec:obs_disks}

In the standard paradigm of star and planet formation, circumstellar disks have long since been described as a natural, almost unavoidable, outcome of the envelope's angular momentum conservation during its collapse, ultimately leading to the formation of a young star \citep[see e.g. the reviews by][]{1991ASIC..340....1B,2010RPPh...73a4901L}. As such, they were a sound and convenient solution to the so-called ``angular momentum problem" in star formation \citep[see e.g.,][for early reviews]{1965QJRAS...6..161M,1971MNRAS.151..177P}, as buffer structures preventing the transfer of all the envelope's angular momentum to the forming star. In the traditional analytical description of the ``inside-out" collapse of a singular isothermal sphere (SIS) in solid-body rotation \citep{1984ApJ...286..529T}, the centrifugal radius grows very quickly \am{with time} with $t^3$ as a result of incoming material with increasingly larger specific angular momentum, producing disk radii of a few hundreds of au in a few thousands years \citep{1994ApJ...431..341S}. In a magnetized scenario, however, disk growth is mitigated by magnetic braking and \am{diffusive processes, such as ambipolar diffusion, regulating magnetic flux evolution}, and the centrifugal radius grows more slowly, as $t$ \citep{1997ApJ...485..240B,1998ApJ...509..229B}. Although different initial conditions may better reproduce observations of prestellar and protostellar cores \citep[such as, e.g., Bonnort-Ebert spheres, see for example][]{2004MNRAS.355..248B, 2015MNRAS.446.3731K}, and affect the details of disk formation during the main accretion phase, it is a common feature of all protostellar formation models that the introduction of magnetic fields will produce somewhat smaller rotationally-supported disks. While models and their main features are described in the following section \ref{sec:theory}, we outline here how observations of the youngest embedded protostars have been used to discriminate between them, and ultimately shed light on the role of key physical ingredients in disk formation, and in determining pristine properties of planet-forming disks.


\subsubsection{Disk radii}
\label{ssec:obs_radii}

\begin{figure}[ht]
\begin{center}
\includegraphics[width=\linewidth]{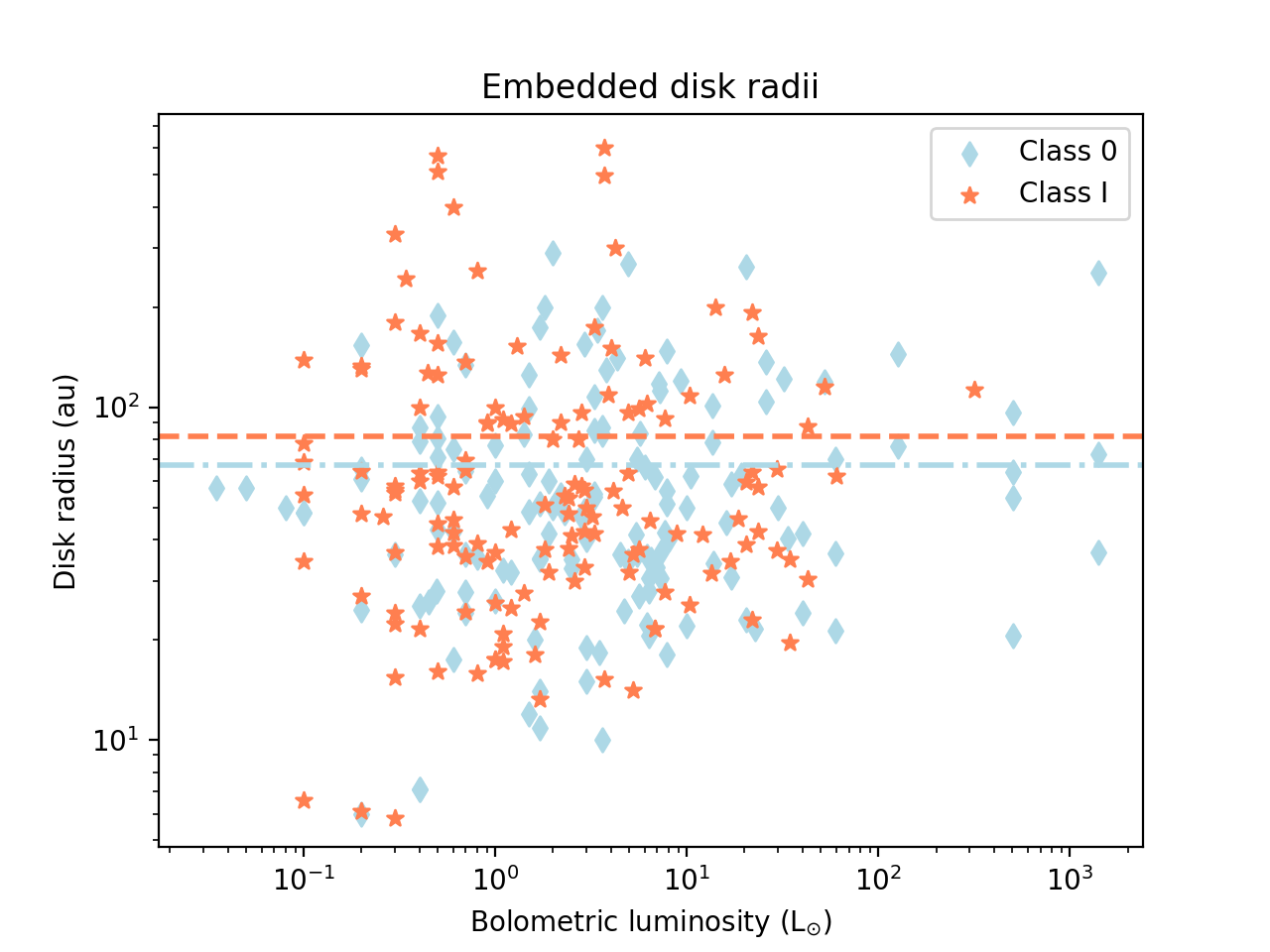}
\end{center}
\caption{Compilation of Class 0 (in blue) and Class I (in orange) protostellar disk radii, from the literature, observed from the dust continuum emission at (sub-)millimeter wavelengths ($\lambda<2.7$mm). From \citet{2014ApJ...795..152O, 2018ApJ...863...94L, 2015ApJ...812..129Y, 2017ApJ...834..178Y, 2015ApJ...812...27A, 2014ApJ...796..131O, 2013A&A...559A..82B, 2014A&A...562A..77H, 2014A&A...567A..32M,2020ApJ...890..130T, 2017ApJ...851...45S, 2019A&A...621A..76M}. The two horizontal lines show the median disk radii of 61\,au and 81\,au for each Class (0 in blue, I in orange) respectively, from these distributions. We stress that upper-limit radii from (yet?) unresolved disks are not included in those statistics so these median values should certainly be interpreted as upper-bounds.}
\label{fig:Rdisks}
\end{figure}

Disentangling the disk from the envelope contribution on scales where protostellar disks are expected to form has long been an observational challenge \citep{2000ApJ...529..477L,2009A&A...507..861J}.
Since PP\,VI, many sensitive high-resolution interferometer surveys of protostellar populations have been carried out. Observations of the dust continuum emission at small envelope radii, and  comparison with protostellar models of dust emission, has allowed us to determine which protostars likely host a disk, and to estimate dusty disk radii down to the observational limit \am{(typically a few tens of au).} 
We review here of some of the largest recent surveys, and their conclusions. Figure \ref{fig:Rdisks} presents all dust disk sizes reported in the literature at short wavelengths $\lambda<2.7$mm, associated with embedded protostars.

The CALYPSO survey \citep{2010A&A...512A..40M, 2019A&A...621A..76M} used the Plateau de Bure Interferometer, at 1.3\,mm and 2.7\,mm, to characterize disk properties of 26 Class 0 and Class I protostars. Modeling the visibility profiles of the millimeter dust continuum emission with a combination of envelope and disk contributions, they found an average disk size of $<$ 50 au $\pm$ 10 au in the Class 0 objects, and 115 $\pm$ 15 au in the Class I objects. While this survey had the advantage of samples from several different star forming regions, it is likely biased towards the more luminous sources that were easily observed in the pre-ALMA era. 

\am{The VANDAM survey produced} VLA observations of the 8\,mm dust continuum emission which were used to measure the radii of all observed Class 0 and Class I disks in Perseus using a power-law intensity model. \citet{2018ApJ...866..161S} find 14/43 ($33\%$) Class 0 and 4/37 ($11\%$) Class I sources have resolved disks with r $>$ 12 au, and 62/80 ($78\%$) of Class 0 and I protostars are not associated with resolved disk emission within r $>$ 8 au. \citet{2017ApJ...851...83C} observed a sample consisting largely of Class II YSOs, Flat and Class I objects in Ophiuchus with ALMA in the 870 $\mu$m dust continuum emission and found a median radius of 12.6 au for the Class I/Flat Sources by fitting a 2D Gaussian to the images. A complementary study of Ophiuchus done by \citet{2021ApJ...913..149E} found a median radius of 23.5 au for Class I protostars. The VANDAM survey, \am{also} \citep{2020ApJ...890..130T}, looked at Orion protostars at 870 $\mu$m using ALMA: from the Gaussian fitting of the dust continuum emission, they find the median dust disk radii for the 69 Orion Class 0 protostars  $\sim 48$ au and $\sim 38.0$ au for the 110 Class I sources. \am{While these results confirm Class 0 disks are typically compact, they reveal median dust disk sizes smaller than the median radii computed by using all the literature available (see Figure \ref{fig:Rdisks})}. This difference may stem from the statistical completeness of the VANDAM survey in one single region, while other Class I disk sizes estimates such as the ones reported in \citet{2019A&A...621A..76M} may be biased with mostly bright Class I protostars in the sample. 
Note however that Orion is a unique cloud due to its very active environment as well as the number of massive stars forming therein: this could be the cause of the peculiar decrease in disk sizes from Class 0 to Class I protostars, which do not seem to be routinely observed in other star-forming regions.

We stress that characterising disk size depends markedly on the choice of observations performed, and the analysis carried out. For example in some cases, disk radii are estimated from the additional thermal component which cannot be accounted by small-radii envelope emission. \am{Alternatively, some} studies assume the envelope emission is completely filtered out on sub-arcsecond scales and attribute the entire emission obtained from long-baseline maps as exclusively due to disk emission. 
While simple Gaussian models measure the spatial extent of the dust thermal emission, they ignore the potential contribution from envelope emission, which may differ, statistically, from Class 0 to Class I with the envelope dispersal/accretion. This method may overlook complex structures (streamers, outflow cavities, etc.) which are often reported when carefully examining the spatial distribution in sensitive dust emission maps. 
Finally, the choice of wavelength where to observe the thermal emission of the dust may also influence the dusty disk sizes observed. Using long wavelengths may be more sensitive to the spatial distribution of large grains, especially at more advanced protostellar stages when dust settling may have had time to operate. \am{This was noted by \citet{2020ApJ...890..130T} who finds systematically more extended 0.8mm emission than 8mm emission in their sample of Orion disks.}
The use of a shorter wavelength may bias the sizes of \am{more evolved} sources because optically-thick emission is better modeled by flatter Gaussian sources with larger FWHMs. 

These aforementioned studies rely mainly on the analysis of the thermal dust emission, but protostellar disks are gaseous structures, which sizes should be evaluated also using the gas emission.
This seems even more important as recent observations towards Class II YSOs have revealed a large population of compact dusty disks but suggest as well that the gaseous disks may be on average a factor of two larger than the dust extent \citep[see, e.g.,][, as well as the PPVII review chapter by Miotello et al.]{2016ApJ...827..142B,2021A&A...649A..19S,2016ApJ...828...46A,2018ApJ...859...21A}. 
This \am{could suggest the existence of a potential bias in the \am{youngest} protostellar disk size distribution, which are mostly obtained from dust emission. The dusty disk sizes could be different from} gaseous disk sizes, although at the high densities typical of Class 0 inner envelopes, no significant gas/dust decoupling can be expected if the grain sizes remain relatively small \citep{2020A&A...641A.112L}. While the small sample of protostellar disks characterized in dust and molecular line emission appears to show gaseous radii similar to dust radii, only a larger sample can confirm this.

Embedded disks are expected to be largely rotationally supported and as such their gaseous sizes could be measured with the detection of (nearly) Keplerian gas motions around the central stellar embryo. 
A few studies have been able to detect Keplerian rotation from protostellar dusty disks, finding gaseous radii close to the dust radii in Class 0 protostars \citep{2014ApJ...796..131O, 2017ApJ...834..178Y,2019A&A...631A..64B,2020A&A...635A..15M} and Class I protostars \citep{2014A&A...562A..77H, 2014ApJ...796...70C, 2014ApJ...793....1Y, 2015ApJ...812...27A}. Observations may also be able to trace the location of the centrifugal radius and centrifugal barrier of the gas, thanks to shock tracers, as detailed in the previous section \ref{sec:kin} \citep{2014ApJ...791L..38S,2017A&A...603L...3A, 2017SciA....3E2935L,2019ApJ...873L..21I}. 

The small numbers are due to the complexity of the velocity field in embedded protostars on the smallest scales. Deciphering the kinematic signature of disks in embedded protostars is hampered by the intrinsic complexity of the protostellar environment which naturally produces entangled contributions from e.g., envelope infall, rotation, and outflowing gas, on overlapping scales and mixed along the line of sights.
For example in the HH\,212 Class 0/I protostar the sequential analysis of ALMA observations providing increased spatial resolution and sensitivity have led to a revision downwards of the disk radius from $\sim$100 au \citep{2014A&A...568L...5C,2014ApJ...786..114L} to a smaller $<44$ au radius \citep{2017SciA....3E2935L}, thanks to the characterization of complex kinematic structures contributing at intermediate radii 50-200 au \citep{2015A&A...581A..85P}.  Similarly in L483, where \citet{2017ApJ...837..174O} argued that the centrifugal barrier in this source is at radii of 30–200 au while \citet{2019A&A...629A..29J} suggest molecular lines trace infalling gas down to radii of 10–15 au, rather than Keplerian rotation. 
 
Moreover, the kinematic signature of embedded and heavily accreting disk structures may present significant deviations from purely axisymetric Keplerian rotation curves (e.g. because of the generation of relatively massive tidal streams and spiral arms, \citealt{2014ApJ...796....1T, 2016Sci...353.1519P, 2016Natur.538..483T,2017ApJ...837...86T,2019Sci...366...90A}), making it difficult to characterize young embedded disks as easily as their older counterparts. An example is found in the L1527 Class 0/I protostar where the $\sim 60$ au disk, although one of the first resolved kinematically and seen in a favorable edge-on configuration, was recently proposed to be warped because of either anisotropic accretion of gas with different rotational axes, or misalignment of the rotation axis of the disk with the magnetic field direction \citep{2019Natur.565..206S}.

Confirming the rotationally-supported nature of candidate protostellar disks observed in the dust continuum emission is also observationally expensive as it requires sensitive observations of faint molecular emission, from species which remain optically thin at high densities. Future observational studies should not only rely on observations of multiple molecular species, but also model them jointly to characterize the gaseous disks in the embedded phases.

\subsubsection{Disk masses}
\label{ssec:obs_masses}

Several observational studies have measured disk masses in embedded protostars, using the emission due to dust which is then corrected to a gas mass using an assumed gas-to-dust ratio. 
In Orion, \cite{2020ApJ...890..130T} use the 870 $\mu$m ALMA dust continuum fluxes to measure the disk masses, finding a median dust mass $52.5 {\rm\,M_\oplus}$ for the 69 Class 0 single protostars, and $15.2 {\rm\,M_\oplus}$ for the 110 Class I protostars.  
\am{The choice of wavelength(s) for the dust observations is crucial not only for estimating disk sizes but also their masses.  For example} in Perseus, \citet{2020A&A...640A..19T} measure median dust masses of the embedded disks $\sim 158 {\rm\,M_\oplus}$ for Class 0 and $52 {\rm\,M_\oplus}$ for Class I from the VLA dust continuum emission at centimeter wavelengths. \am{In comparison}, the lower limits on the median dust disk masses extrapolated from sub-millimeter dust emission in ALMA bands are in better agreement with the masses found in Orion, $47 {\rm\,M_\oplus}$ and $12 {\rm\,M_\oplus}$ for 38 Class 0 and 39 Class I respectively in Perseus.

These disk dust masses are however subject to several caveats due to the possible presence of optically thick dust emission and what is assumed as regards the dust opacity and temperatures in these yet largely unconstrained environments.
First, most studies are based on a single sub-millimeter flux measurement, the mass estimates assume optically thin dust emission: if the dust emission is not optically thin (as shown in several cases, see \citealt{2020ApJ...889..172K} and discussions therein), then the dust masses will be lower limits. We stress that this could cause a bias leading to an underestimate of the masses of the most evolved protostellar disks, which could be optically-thick as collapse proceeds \citep{2018ApJ...868...39G,2019ApJ...877L..18Z}.
Regarding mass estimates based on dust emission at longer (cm) wavelengths, the contribution of large dust grains to the observed emission could be enormously variable from object to object, and tracing mostly the compact emission from the inner disk. Obtaining a dust mass from cm dust emission is thus also highly uncertain. 
(Sub-)millimeter dust emission is also a uncertain probe of dust masses, unfortunately, as it suffers from uncertainties on dust emissivity properties which have been found to be very different from those of ISM dust, see section \ref{sec:dust}. Finally, the considerable unknowns regarding the typical gas and dust temperature of these young disks, actively accreting and bombarded by envelope material, also impair the dust continuum observations in providing accurate disk masses. That said a few clues have been recently come to light suggesting these disks are  warmer than their older counterparts, although these conclusions still suffer from poor statistics \citep{2020ApJ...901..166V}. 

Demographics of older, Class II, disks appear to imply that most do not contain enough material to form the known census of exoplanets \citep{2016ApJ...828...46A, 2018A&A...618L...3M}. Both dust and gas mass estimates (usually from CO emission) suffer from large uncertainties and their reliability is still widely debated by the community \citep[e.g., ][]{1996ApJ...467..684A,2016ApJ...833..105Y, 2001ApJ...561.1074T,2018ApJ...868L..37D,2019ApJ...878..116P,2013Natur.493..644B,2013ApJ...776L..38F, 2016A&A...592A..83K, 2016ApJ...831..167M}. Details are beyond the scope of this review as we focus on the youngest disks. Nevertheless, the resultant uncertainties highlight the importance of determining disk masses \am{from the earliest phases}, as such masses determine how early planets might form.
Recent observations largely suggest the dust masses of young Class 0 and I disks are larger, by at least a factor of a few, compared to more evolved Class II disks. In Ophiuchus, for example, the mean Class I disk dust mass is found to be significantly lower than in Perseus and Orion, $\sim 2.8-3.8 {\rm\,M_\oplus}$ \citep{2019ApJ...875L...9W, 2021ApJ...913..149E}. Nevertheless it is still about 5 times greater than the mean Class II disk mass in the same region, although the dispersion in each class is so high that there is a large overlap between the two distributions. 
\am{Interestingly, some annular structures in the dust continuum emission of Class I protostars have recently been observed \citep{2018NatAs...2..646H,2020Natur.586..228S,2020ApJ...902..141S}. While their origin is still debated, it is possible they are carved by the early planetary seeds. Observations in the coming years of the gas kinematics in young disks should help constrain whether this is the case.}
Moreover, these preliminary findings, regarding both dust masses and their spatial distributions, still need to be validated through the use of large, high spatial resolution, multi-wavelength samples that allow one to remove current degeneracies. If confirmed, they would make embedded protostars the likely cradles of planetary formation.

\subsection{Outflows and jets in embedded protostars}
\label{SR_MHD_Outflows}
Jets from young stars have been observed from the X-ray regime right down in frequency to the radio band. Historically the emphasis has been on studying them at optical and near-infrared wavelengths but with the advent of high spatial and spectral resolution millimeter arrays such as ALMA, increasingly the focus has shifted to molecular emission lines and more embedded sources. Whether we are looking at the ionized, neutral atomic or molecular component, the presence of line emission ensures we can determine fundamental outflow parameters such as velocity, density, temperature, etc. Of course these various species need not be co-located, nor are they expected to be, and so no single set of waveband limited derived parameters can characterise an outflow. Outflow emission arises primarily from the cooling zones of shocks, shocks that are generated by higher velocity gas catching up with slower material ahead of it. This means the strength of the shock is determined by velocity differences rather than the actual outflow velocity with respect to the source. 

There is insufficient space here to provide an in-depth review of the properties of outflows generated by the wide variety of YSOs, or even proto-brown dwarfs, from which they are observed, the reader instead is referred to \citet{2014prpl.conf..451F}, \citet{2016ARA&A..54..491B} and \citet{2021NewAR..9301615R}. Here we will concentrate on a few salient properties of those from the most embedded (Class 0/I) sources and describe in general terms how observations can give us clues to their launching mechanism. Note that as Class 0/I sources are highly embedded, observations have largely been confined to millimeter wavelengths, i.e.\ their molecular emission, although not exclusively.

\begin{figure}[ht]
\begin{center}
\includegraphics[width=80mm]{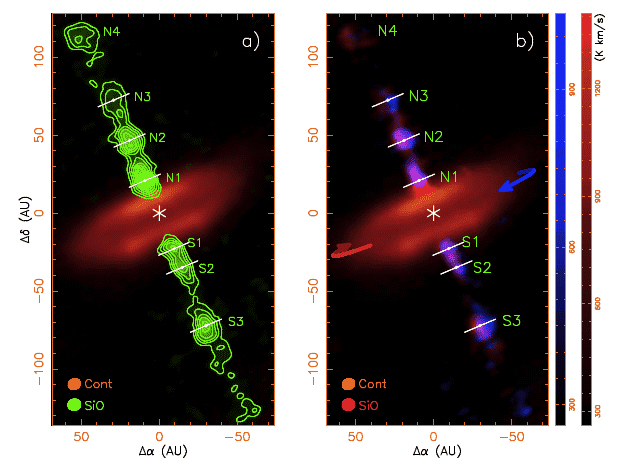}
\end{center}
\caption{ALMA SiO and continuum observations of the rotating outflow from the Class 0 protostar HH\,212. While the outflow itself extends to much larger scales, this shows the region within $\approx$ 120 au of the
central source, at a  resolution of $\approx$ 8\,au on top of the continuum map of the
disk. The maps show the intensity (in unit of K\,km s$^{\rm -1}$) integrated over the outflow velocity range. (a) A chain of SiO knots trace the primary jet emanating from the disk testifying to the episodic nature of the outflow. (b) Blueshifted and redshifted SiO emission of the jet plotted with the
continuum (disk) emission. The direction of rotation of the disk is shown with blue and red arrows and this is the same as the direction of rotation of the red and blueshifted jets. From \cite{2017NatAs...1E.152L} reproduced with permission \textcircled{c} Springer Nature.} \label{HH212}
\end{figure}

Of course molecular outflows from young stars have been known for several decades \citep[e.g.,][]{2007prpl.conf..245A} but the realisation that such outflows can appear (e.g.\ in SiO, SO or even CO) just as jet-like as their ionized/neutral atomic counterparts was slower in coming \citep[e.g.,][]{2015Natur.527...70P, 2017NatAs...1E.152L, 2021ApJ...909...11J}. Studies using mm interferometers of Class 0/I sources show their outflows can be traced right back to the protostar with in some cases very narrow opening angles \citep{2019A&A...631A..64B} and velocities comparable to the young star's escape velocity \citep{2010ApJ...717...58H}. All of this suggests such outflows, or at least their most highly collimated component i.e. the jet,  arise from a region no greater than $\approx$1\,au centred on the YSO. A classic example, and something of a poster child is HH\,212 \citep[e.g.,][and see Fig.\  \ref{HH212}]{2020A&ARv..28....1L}. While extended emission in H$_{\rm 2}$ is observed on pc scales, SiO can be traced back to at least 10\,au from the source. Certainly the presence of a disk is a necessary, although not necessarily sufficient, condition for jet launching. 

The other ingredient needed, as in almost all astrophysical models for the generation of jets is a magnetic field \citep[e.g.,][]{2011AIPC.1358..329L}. In \ref{Obs_v_Pred}, we will examine the evidence for magnetic fields in outflows but here we mention some plausibility arguments as to why we think they are present and why they are thought to be dynamically important \am{as they transfer angular momentum from infalling/rotating motions to outflowing motions}, at least close to a YSO.  While there is very little data on \am{magnetic fields threading the base of the jets and protostellar disk surfaces where disk winds are launched in} embedded sources, we know that the surface magnetic fields of Class II/III young stars have strengths of several kG \citep[e.g.,][]{2020MNRAS.491.5660D} as derived from Zeeman splitting measurements. Data on field strengths in the innermost region of protostellar disks (i.e.\,$\lessapprox$\,a few stellar radii) is sparse but observations obtained from spectro-polarimetry in the disk of FU~Ori \citep{2005Natur.438..466D} lead to similar values as derived for the parent star. In addition studies of paleomagnetism in chondrules from the early period in the formation of the Solar System \citep{2021SciA....7.5967W} suggest field strengths of $\approx$\,1\,G at 1\,au. Such values, in combination with the expected density, temperature and velocity of a stellar/disk wind, are clearly capable of collimating a wind into a jet, at least very 
close to the star.  
The most efficient means of achieving this is for the ionised gas (dragging through collisional coupling associated atomic/neutral material with it) to be centrifugally ejected along the magnetic field lines. These field lines are either anchored to the star (stellar winds), at the star-disk interface (so-called X-winds), or alternatively in the disk itself (Disk or D-winds). For a review of these various options the reader is referred to \citet{2019FrASS...6...54P} but also see \ref{Outflow_Theory}. In each case the magnetic field associated with the wind/outflow, while starting off largely poloidal, increasingly, particularly in the case of D-winds, becomes toroidal with distance as a result of the material carrying more and more angular momentum (see below). Tension in these wound-up toroidal fields then generate forces towards the axis, known as $``$hoop stresses" that focus the outflow further into a jet \citep{1997MNRAS.288..333S}.  

As explained elsewhere in this chapter (\ref{Outflow_Theory}) in order for accretion to occur in a protostellar disk, angular momentum must either be removed poloidally, for example through a magnetic tower flow (\ref{Outflow_Theory}, through a disk wind or outflow, or alternatively redistributed radially outwards within the disk (for example through gravitational torques at large disk radii, \citet{2006ApJ...650..956V,2010ApJ...719.1896V,2015ApJ...805..115V}. In any event, in recent years it has become increasingly clear that angular momentum in a disk is removed through the former mechanisms as possible sources of disk viscosity, e.g. MRI driven turbulence, seem unlikely to work due to non-ideal MHD effects (e.g. for Class 0 disks see \citet{2021MNRAS.504.5588K}). While winds/outflows from the star, star-disk interface and the disk itself are all likely to play a role in angular momentum transport \citep{2019FrASS...6...54P}, it is clear that a significant fraction of the angular momentum must be removed through a disk wind/outflow in any event for material to reach the inner disk radius and be accreted onto the star along magnetic field lines as suggested by observations \citep{2020Natur.584..547G}. Thus depending on where the wind/outflow is launched, it is expected to carry varying amounts of angular momentum per unit mass. Thus rotation is expected to be a key signature of disk generated outflows \citep{2019FrASS...6...54P}. The search for rotation in ionized/neutral atomic jets has been challenging from an observational perspective as it is difficult to spatially \amrev{resolve} such jets transversely, the maximum rotation (toroidal) velocity is only a small fraction of the poloidal velocity and the signal to noise ratio is often not high \citep[e.g.][]{2014prpl.conf..451F}. All of these issues are less of a problem with mm-interferometers such as ALMA, and even radio interferometers such as the Jansky VLA, that not only have high spatial resolution, to resolve the outflow transversely, but also the necessary high spectral resolution. Thus, for example, at the start of the ALMA era, \citet{2017NatAs...1E.146H} observed rotation in the outflow from Orion Source I. In turn, using basic magneto-centrifugal wind theory, including angular momentum conservation (for details, see \citealt{2003ApJ...590L.107A}),
they could estimate from where the outflow was launched in the disk and discovered it to be far ($\gtrsim 10$\,au) from the protostar  \citep[see also][]{2016Natur.540..406B}. That the magnetic field has the configuration and strength expected to collimate this outflow has also recently been found \citep{2020ApJ...896..157H}. Rotation has now been detected in many molecular outflows including TMC1A \citep{2016Natur.540..406B}, HH\,211 \citep{2018ApJ...856...14L}, HH\,212 \citep{2017NatAs...1E.152L}, for several outflow in NGC1333 \citep{2011ApJ...728L..34C,2016ApJ...824...72C,2018ApJ...864...76Z}, HH\,30 \citep{2018A&A...618A.120L} and OMC2/FIR6b \citep{2021ApJ...916...23M}. In the latter it was found that the high-velocity jet axis is significantly inclined from the low-velocity outflow axis, indicating that the launching radii of the high- and low-velocity components differ in an inclined disk.  In all cases a disk origin for the jet/outflow is indicated and, where observed, the direction of disk rotation matches that of the outflow (see Fig.\ \ref{HH212}) as would be expected. It would also appear that all of these observations support the direct disk wind expected from core collapse simulations.

\subsection{The role of magnetic fields in shaping protostellar disks and outflows: observational keys}
\label{Bfield_obs}

\subsubsection{Observations of magnetic fields and disks}

 Here we discuss the main clues that observations provide regarding the role played by magnetic fields in setting disk sizes and masses during the embedded phase.

In a simple scenario for the collapse of a rotating core, even a small difference in initial specific angular momentum on large scales is expected to result in a large change in the radius of the resultant rotationally-supported disk. Observations of the angular momentum on scales where the gas should contribute to the building of the disk \citep{2022ApJ...925...12S} suggest most protostars should give rise to disks that are larger, by a factor of a few, than those observed \citep{2020A&A...637A..92G}. Here it is assumed that the angular momentum contained in rotation is entirely transmitted to the disk scale. Naively such observations potentially support models with efficient magnetic braking. Note however, that several theoretical studies have shown that the observations of angular momentum in protostellar envelopes can be satisfactorily reproduced without large scale core rotation, questioning whether observed angular momentum values really stem from rotational motion \citep{2020A&A...635A.130V, 2021MNRAS.502.4911X, 2021A&A...648A.101L}. 

From an observational perspective, the relationship between the outflow axis and the mean magnetic field direction has received a lot of recent attention. This is because it is thought that the efficiency of magnetic braking in redistributing angular momentum (and hence preventing the growth of disks to large radii) when a core collapses depends on the configuration of the core's magnetic field with respect to its rotation axis \citep{2012A&A...543A.128J,2020ApJ...898..118H}. 
Starlight polarization observations show large-scale magnetic fields that are misaligned with outflows powered by evolved (Class II) young stellar objects \citep[e.\,g.,][]{2004A&A...425..973M}, but the fields are better aligned with outflows from younger protostars \citep[e.\,g.][]{1986AJ.....92..633V,2011ApJ...743...54T,2015A&A...573A..34S}. Single-dish observations examining the relative orientation of the B-field on core-scales with the outflow axis are inconclusive:  some studies suggest cores exhibit a mean B-field that is not randomly distributed with respect to the large-scale outflow axis \citep{2015A&A...573A..34S,2021ApJ...907...33Y} while others suggest no correlation  \citep{2020ApJ...899...28D}. Finally, some sub-mm polarization observations of dense cores have found the direction of the core's minor axis correlates with their magnetic field orientation \citep{2013ApJ...770..151C}. A flattening in this preferential direction could be explained by the development of large magnetic pseudo-disks \am{\citep[equatorial flattened gas structures which are not supported by rotation ][]{1993ApJ...417..220G}}. 

Likewise, observations using interferometers have not reached any firm conclusions regarding correlations between outflow, disk and magnetic field axes. On very small envelope scales, the mean B-field direction is observed to be randomly aligned with respect to the outflow axis \citep{2013ApJ...768..159H}. However, as reported by \citet{2014ApJS..213...13H}, there is evidence that cores with lower fractional polarization tend to have their outflows perpendicular to the mean B-field. This suggests the existence of a non-negligible toroidal field (caused by the core/disk rotation) in addition to the poloidal one coming from envelope-disk accretion. 
An SMA survey of 20 low-mass protostars \citep{2018A&A...616A.139G, 2020A&A...644A..47G} has found that protostellar envelopes tend to have a higher angular momentum on 1000\,au scales if the mean envelope magnetic field measured on similar scales is misaligned with respect to the rotational axis of the core (assumed to coincide with the outflow axis). In contrast, observations analyzed in \citet{2021ApJ...916...97Y} have shown no correlation of dust continuum protostellar disk radii with misalignment of the magnetic fields and outflow axes in Orion A cores. 
Both results may be consistent with the magnetic field being efficient at reducing the amount of angular momentum transmitted to the inner envelope scales ($< 1000$ au), inhibiting the formation of large hydro-like disks. \am{However, they may also support the hypothesis that} the angular momentum responsible for the formation and growth in size of protostellar disks has a more local origin, on a few hundreds of au scales, as discussed previously. 

\begin{figure*}[ht]
\begin{center}
\includegraphics[width=130mm]{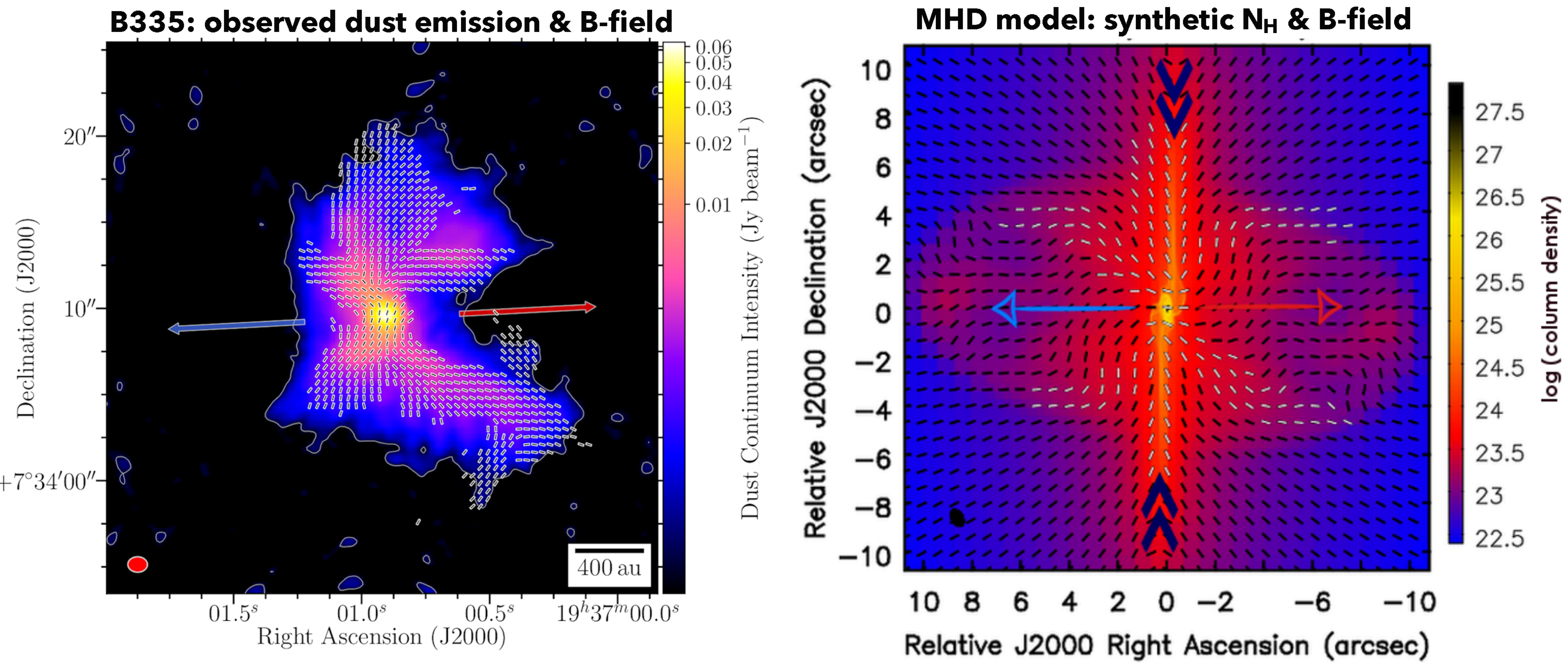}
\end{center}
\vspace{-8truemm}
\caption{Observations and best model to match the properties of the B335 protostar, from \citet{2018MNRAS.477.2760M}. These authors have demonstrated that the observed B field geometry shown in the left panel, the small disk size and the kinematics of the B335 inner envelope can only be explained by a family of MHD models where the initially poloidal field is pulled in the dominant direction of the collapse, but yet is strong enough ($\mu \sim 6$) to partially counteract the transfer of angular momentum inwards and set the disk size in this Class 0 protostar: the best model is shown in the right panel.} 
\label{fig:B335}
\end{figure*}

Comparison of observed properties with models of disk formation in both specific objects and statistical samples may also help our understanding. \citet{2018MNRAS.477.2760M} has shown that the geometry of the B field, small disk size and the kinematics of the B335 inner envelope can only be explained by a family of MHD models where the initially poloidal field is pulled in the direction of collapse, while at the same time being strong enough ($\mu \sim 6$) to partially counteract the transfer of angular momentum inwards. In B335, they show the magnetic field likely regulates the formation of the disk, constraining its size to $< 20$ au (see Figure \ref{fig:B335}). The role of the magnetic field in determining the disk size in B335 has been also proposed in follow on studies \citep{2019ApJ...871..243Y,2019A&A...631A..64B,2019ApJ...873L..21I}. Detailed analyses of the gas kinematics, disk sizes, and B field configuration in other objects are needed to ascertain whether the magnetically-regulated disk growth in B335 is common, or a special case. 

Since \textit{Protostars and Planets VI}, and despite their complexity, observations have successfully explored the properties of Class 0 disks in large part thanks to the increased spatial resolution and sensitivity of new millimeter surveys. As shown in Figure \ref{fig:Rdisks}, the vast majority of observations now point towards a population of young protostellar disks that are primarily compact. From a statistical point of view this, and the confirmed presence of magnetic fields on the scales where disks form, may naively point to magnetic fields having a key dynamical role in regulating the \am{early} growth of stellar embryos and their surrounding disks. Indeed, such disk sizes are difficult to reconcile with purely hydrodynamical models of disk formation and, if the B-field is well coupled, magnetic braking naturally produces such small disks \citep[see, e.g.][for a statistical comparison of observed disk sizes and protostellar disk sizes in magnetized models]{2021ApJ...917L..10L}. 
However, observations are still largely unable to characterize how effective magnetic fields are at coupling on the smallest scales where protostellar disks form.  This coupling is mostly due to small grains and the ionized species in the gas (see Section \ref{sec:theory}). \amrev{Measuring the amount of these two key ingredients in and around embedded disks is thus fundamental to setting constraints to the role of the magnetic field, and the importance of diffusive processes, such as ambipolar diffusion, in counteracting the outward transport of angular momentum from magnetic braking.} In embedded sources, the ionization rate from cosmic rays is inferred using various chemical signatures, often HCO$\rm{^{+}}$, and N$\rm{_{2}}$H$\rm{^{+}}$ \citep{2021A&A...645A..28B}. Measurements suggest typical cosmic ray ionization rates $\zeta \sim 10^{-17}-10^{-15} \, \rm{s}^{-1}$ with large uncertainties \citep{2014ApJ...790L...1C,2014A&A...565A..64P,2017A&A...608A..82F,2018ApJ...859..136F}. Such measurements however remain rare, not only because they are difficult to make from an observational perspective but also because of chemical model degeneracy. In any event such high values seem inconsistent with ionization from external radiation only: galactic cosmic rays, mostly relativistic protons, are the dominant source of ionization in dense molecular gas where ultraviolet radiation cannot penetrate but their flux should be severely attenuated inside dense cores \citep{2015ARA&A..53..199G,2018A&A...614A.111P}. Several recent studies \citep{2015A&A...582L..13P,2018ApJ...863..188S,2021A&A...649A.149P,2021ApJ...915...43F,2018ApJ...861...87G} have investigated the role of shocks at the protostellar surface and \am{magnetic acceleration} within jets as efficient forges to locally accelerate low-energy cosmic rays, and their role in increasing the ionization rate of the shielded protostellar material. \am{While the ionization can be increased locally, for example at the accretion shock, the typical gas densities in these regions should lead to a significant decrease in the ionization fraction. Observational studies of the chemical composition of the dense gas, and other properties could shed light on the small scale conditions that determine the coupling of the magnetic field to the disk gas.} Note that, in a recent study of the TMC1-A Class I protostar, \citet{2021A&A...646A..72H} estimate an ionization \am{from cosmic rays} $\zeta \sim 10^{-17} s^{-1}$ at the disk surface. This low ionization suggests the B-field is weakly coupled and accretion through the disk may not be driven by the magneto-rotational instability. 

Future radio observations will undoubtedly bring more constraints on the dynamical role of magnetic fields in regulating protostellar collapse and in setting pristine disk properties. Future large surveys will allow us to statistically investigate the dynamical role magnetic fields play by careful examination of their relationship with the angular momentum and mass transport processes from envelopes to disks \citep{2020A&A...644A..47G, 2021ApJ...916...97Y}. The true spatial distribution of magnetic fields on disk-forming scales may become routinely observed thanks to the development of polarimetric capabilities and modeling tools, e.g. in connection with polarized dust emission. Finally, observations with, for example, the next generation of mid-infrared facilities will help characterize the small scale structures of these young disks, still largely unresolved spatially. These studies may allow us to put observational constraints on the models of magnetic field properties and coupling efficiency inside protostellar disks \citep{2021ApJ...922...36L}, and test how mass accretion onto a protostar might be mediated through its disk. 

 \subsubsection {Observations of Magnetic Fields in Outflows Versus Prediction}
 \label{Obs_v_Pred}
 
 Perhaps the most challenging physical quantity to determine in an outflow from a young star is the strength and direction of its magnetic field. Measuring the Zeeman effect for example using optical/near-infrared atomic lines is not practical as any splitting is so weak. Interestingly, however, in the near future it may be possible to use molecular lines, e.g., those of sulphur monoxide SO \citep{2017A&A...605A..20C}.  This of course might be particularly pertinent in the case of molecular outflows from less evolved sources. In general, however, indirect methods have to be employed to assess field strengths.  For example, as pointed out many years ago, magnetic fields can cushion an outflow shock, altering the degree of compression in the post-shock radiative cooling zone and changing observed emission line ratios \citep{1994ApJ...436..125H}. Just measuring line ratios however is not sufficient as the shock solutions are degenerate. Instead additional parameters, such as the spatial extent of the cooling zone are needed, to break the degeneracy and this requires very high spatial resolution, for example as afforded by HST \citep{2015ApJ...811...12H}. 
 
 An alternative approach, that at least can be applied in some outflows, is to measure their non-thermal radio emission. Of course most of the radio flux from outflows is thermal in origin and is simply the free-free emission from the collimated ionized jet. That said a number of outflows show pockets of non-thermal emission, usually well away from the parent YSO \citep{2018A&ARv..26....3A}. A classic example is HH\,80/81 which has been shown through VLA observations to display polarized synchrotron emission \citep{2010Sci...330.1209C}. This indicates, perhaps somewhat surprisingly, the presence of relativistic particles which presumably have been energised through diffusion shock acceleration even though the shock velocities are low \citep{2016ApJ...818...27R}. In any event minimum energy requirements including a equipartition field, in conjunction with the know non-thermal flux, can lead us to estimates of the magnetic field (for example following \citet{2011hea..book.....L}). These values, typically 20--200\,$\mu$G, appear consistent with what we might expect in the regions observed far away from the collimation zone.  More recently it has emerged that low frequency radio observations of non-thermal emission can offer us a novel method of deriving at least the field strength, if not its direction. This technique relies on the fact that in an outflow there can be a mixture of relativistic particles and thermal electrons. The latter, depending on the electron density, de-collimates the forward-beamed radiation from the non-thermal electrons, giving rise to a decrease in flux, or effectively a low frequency turnover in the synchrotron spectrum. If the thermal electron density is known independently, for example from optical emission line diagnostics, then the magnetic field can be derived. This, so-called Razin Effect, may have been seen in an outflow for the first time \citep{2019ApJ...885L...7F}. The derived magnetic field strengths are in agreement with what might be expected based on the expansion of the outflow while allowing for some amplification due to shock compression \citep{2019ApJ...885L...7F}. 
 
 Finally, it is worth noting that the GK effect (see \ref{magnetic_measuring_techniques}) has been used to probe the magnetic field in protostellar outflows. In particular \citet{2018NatCo...9.4636L} have detected this effect in SiO observations of the HH\,211 outflow using ALMA. Despite the inherent uncertainty of 90$^{0}$ in the magnetic field orientation, the implied magnetic field geometry may well be largely toroidal as expected for a collimating field.



\section{\textbf{A unified scenario for the formation of protostars, protoplanetary disks, jets, and outflows}} 
\label{sec:theory}

As described in \S \ref{sec:observations}, the rapid progress of observational research since PP VI
has provided strong constraints on theoretical models.
In \S \ref{sec:theory}, we describe the theoretical progress and present a comprehensive scenario
for the formation and early evolution of protostars, disks, and outflows that is consistent with observational constraints.
The key mechanism of this scenario is the coupling, decoupling, and recoupling of the gas and magnetic field at appropriate scales.

\subsection{From cloud cores to protostars: formation and evolution of outflows, jets, and protoplanetary disks}
\bigskip

\newcommand{\gcm}{~{\rm g~cm}^{-3} }
\newcommand{\cmn}{~{\rm cm}^{-3} }

\newcommand{\mum}{~{\rm \mu m} }
\newcommand{\mm}{~{\rm m m} }
\newcommand{\cm}{~{\rm c m} } 
\newcommand{\dv}{\Delta \vel}
\newcommand{\vel}{\mathbf{v}}
\newcommand{\magB}{\mathbf{B}}
\newcommand{\cul}{\mathbf{J}}
\newcommand{\etacm}{~{\rm cm}^{2} ~{\rm s}^{-1} } 
\newcommand{\msun}{\thinspace M_\odot}
\newcommand{\msunyear}{\thinspace M_\odot \thinspace{\rm yr^{-1}}} 
\newcommand{\cms}{~{\rm cm~ s^{-1}} }
\newcommand{\ms}{~{\rm m~ s^{-1}} }
\newcommand{\kms}{~{\rm km~ s^{-1}} } 
\newcommand{\tstop}{t_{\rm stop}}
\newcommand{\tgrowth}{t_{\rm growth}}

\newcommand{\mic}{$\mu$m}
\newcommand{\scale}[3]{\left(\frac{#1}{#2}\right)^{#3}}
\renewcommand{\vec}[1]{\mbox{\boldmath $#1$}}

\subsubsection{Essential microscopic physics for the formation of protostars, disks, and outflows}


In this section, as a basis for the ones that follow, we will introduce the physical processes
that are crucial for the formation and evolution of protostars and protoplanetary disks.

Molecular cloud cores, from which protostars, disks, and outflows are formed, are generally weakly ionized
\citep[with typical electron abundances on the order of 10$^{-7}$;][]{BerginTafalla2007}.
Applying the Ohm's law, the electric field in the neutral frame $(\mathbf{E'})$ is tied to the current density ($\mathbf{j}$) as,
\begin{equation}
\label{Eq:ohmslaw}
\mathbf{j} = \sigma_{\parallel}  \mathbf{E'_\parallel}+\sigma_{\rm H} \frac{\mathbf{B}}{B}\times\mathbf{E'_\perp}+\sigma_{\rm P} \mathbf{E'_\perp}~,
\end{equation}
where $\sigma_{\parallel}$, $\sigma_{\rm P}$,  and $\sigma_{\rm H}$ are parallel, Pedersen, and Hall conductivities, respectively, and $\mathbf{E}'_\parallel$ and$\mathbf{E}'_\perp$ are the components of $\mathbf{E'}$ parallel and perpendicular to the magnetic field $\magB$.

\yt{The three components of the conductivity tensor can be expressed 
as \citep{NormanHeyvaerts1985, 1999MNRAS.303..239W,2007Ap&SS.311...35W},
\begin{eqnarray}
\sigma_{\parallel} & = & {{e c } \over B} \sum_i Z_i n_i \beta_{i,\rm H_2}~, \label{Eq:conduct1}\\
\sigma_{\rm P} & = & {{e c } \over B} \sum_i {{Z_i n_i \beta_{i, \rm H_2}} \over {1+\beta_{i,\rm H_2}^2}}~,\label{Eq:conduct2}\\
\sigma_{\rm H} & = & -{{e c } \over B} \sum_i {{Z_i n_i} \beta_{i,\rm H_2}^2 \over {1+\beta_{i,\rm H_2}^2}}~; \label{Eq:conduct3}
\end{eqnarray}
}
\yt{
here, $e$ and  $c$ are elementary charge and speed of light,  $Z_i, n_i, \beta_{i,\rm{H_2}}$ are the charge number, number density, and Hall parameter of the charged species $i$, respectively.
The Hall parameter $\beta_{i,\rm H_2}$, which is the ratio of the gyro-frequency and neutral collision frequency, determines the relative importance of the Lorentz and drag forces in balancing the electric force.
}
By solving equation (\ref{Eq:ohmslaw}) for the electric field and substituting it into Faraday's law.
The induction equation can be formulated as \citep[e.g.,][]{KunzMouschovias2009},
\begin{equation}
\label{Eq:induct}
\begin{split}
{\partial \mathbf{B} \over \partial t} & = \nabla \times (\mathbf{v} \times \mathbf{B}) - \nabla \times \biggl\{\eta_{\rm Ohm}\nabla \times \mathbf{B} \\
&+~\eta_{\rm Hall}(\nabla \times \mathbf{B}) \times {\mathbf{B} \over B} +~\eta_{\rm AD}{\mathbf{B} \over B} \times \left[(\nabla \times \mathbf{B}) \times {\mathbf{B} \over B}\right]\biggl\}~,
\end{split}
\end{equation}
which departs from the ideal MHD limit by the introduction of the three non-ideal MHD terms,
characterized by the Ohmic ($\eta_{\rm Ohm}$), Hall ($\eta_{\rm Hall}$), and ambipolar ($\eta_{\rm AD}$) resistivities.
\yt{The resistivities are related to the conductivities as
\begin{eqnarray}
\eta_{\rm AD} & = & {c^2 \over 4\pi} ({\sigma_{\rm P} \over \sigma_{\rm P}^2 + \sigma_{\rm H}^2} - {1\over \sigma_{\parallel}})~, \label{Eq:MHDcoef1}\\
\eta_{\rm Ohm} & = & {c^2 \over {4\pi \sigma_{\parallel}}}~, \label{Eq:MHDcoef2}\\
\eta_{\rm Hall} & = & {c^2 \over 4\pi} ({\sigma_{\rm H} \over \sigma_{\rm P}^2 + \sigma_{\rm H}^2})~. \label{Eq:MHDcoef3}
\end{eqnarray}
}
The microscopic origin of these three non-ideal MHD effects is related to how the particles interact in each scenario. Ohmic dissipation is the collision between neutral and charged particles whereas ambipolar diffusion is charged-particles slipping through neutral particles. The Hall effect is the velocity difference between positively and negatively charged particles.

\begin{figure}[ht]
    \includegraphics[width=0.50\textwidth]{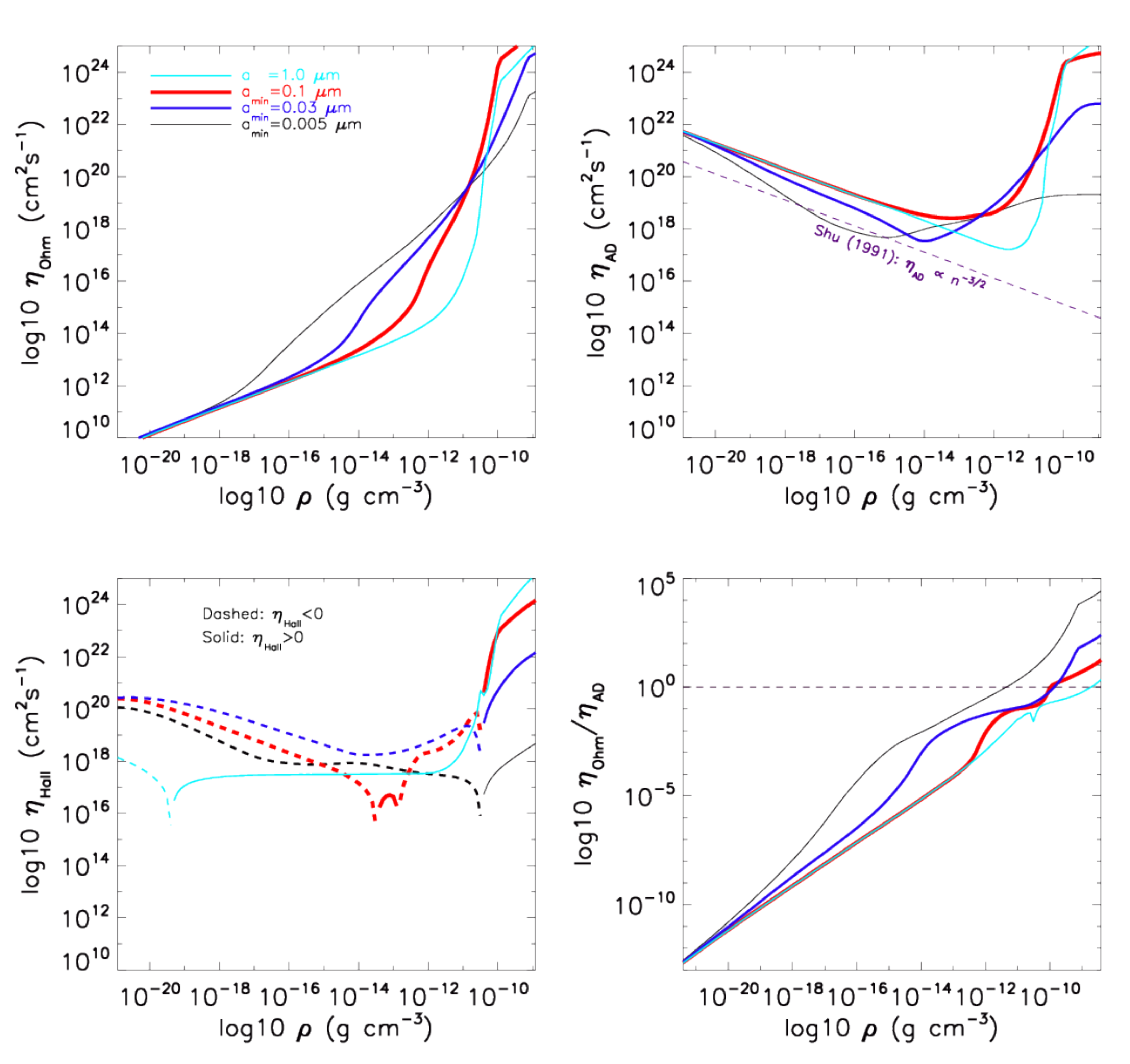}
    \caption{The magnetic resistivities computed for different dust-grain size distributions (top left: $\eta_{\rm Ohm}$; top right: $\eta_{\rm AD}$; bottom left: $\eta_{\rm Hall}$; bottom right: $\eta_{\rm Ohm}$/$\eta_{\rm AD}$)
with $a_{\rm min}$=0.005~$\mu$m and  $a_{\rm max}$=0.25~$\mu$m (MRN: black), $a_{\rm min}$=0.03~$\mu$m and  $a_{\rm max}$=0.25~$\mu$m (blue), $a_{\rm min}$=0.1~$\mu$m
and $a_{\rm max}$=0.25~$\mu$m (red), $a$=1.0~$\mu$m (cyan) from \citep{2018MNRAS.478.2723Z}.
    The reference level of ambipolar resistivity of Shu (1991) is shown in purple dashed line (top right).}
    \label{fig:etas_amin}
\end{figure}

\yt{
In the cloud core and disk, $\eta_{\rm Hall}$ (or equivalently $\sigma_{\rm H}$) is negative
over almost the entire density range (see, Figure \ref{fig:etas_amin}). 
This may be counter-intuitive.
Therefore, the following will discuss the origin of the negative $\sigma_{\rm H}$.
The key is the total dust charge. 
The total dust charge is negative in low density regions due to the large thermal velocity of the electrons.
In this case, due to charge neutrality,
total charge of ion is slightly larger than the charge of electron.
Consider the regime in which
the Hall parameter of ions ($\beta_{\rm i}$), electrons ($\beta_{\rm e}$), and dust ($\beta_{\rm d}$) obeys, $\beta_{\rm i} \gg 1$, $\beta_{\rm e} \gg 1$, $\beta_{\rm d} \ll 1$ i.e., ions and electrons are tied with magnetic field but dust is not.
In this case, $\sigma_{\rm H} \to ec /B (n_{\rm e}-n_{\rm i})$ (we can neglect the contribution of dust because  $\beta_{\rm d} \ll 1$).
Therefore, $\sigma_{\rm H}$ is negative because $n_{\rm e}<n_{\rm i}$.}
\yt{
Intuitively, it can be understood as follows.
 In a situation where negatively charged dust is coupled with neutral gas and the ions and electrons are coupled to the magnetic field,
 a relative motion of the ion electron mixture
 (which is slightly positively charged)
 and the negatively charged dust creates the electric current.
 In this case, the role of positive and negative charges is reversed compared to ordinary Hall effect, and it causes the negative $\sigma_H$.
}


In general, an ionization chemistry network \citep{OppenheimerDalgarno1974,1990MNRAS.243..103U,2002ApJ...573..199N,2016A&A...592A..18M,2016PASA...33...41W,2018MNRAS.478.2723Z,2020A&A...643A..17G}
should be properly constructed in order to obtain
the resistivities in the induction equation~(\ref{Eq:induct}). Cosmic-rays (CRs) are the main source of ionization that triggers the chain of ionization chemistry
as most of the interstellar UV-radiation is attenuated in dense molecular cores \citep[visual extinction Av$>$4~mag;][]{McKee1989}.
On the other hand, dusts act as a charge absorber and can also be a major contributor to conductivity.
As an example, Figure \ref{fig:etas_amin} shows the resistivities values for various dust distributions. \yt{Note that the Ohmic resistivity is many orders of magnitude smaller than the ambipolar resistivity in $\rho_{\rm g} \lesssim 10^{-10}\gcm$ (or in $n_{\rm g} \lesssim 10^{13} \cmn$).} 

In this section, we use the following approximated formulae for Ohmic and ambipolar resisitivities for a quantitative discussion. 
Ohmic dissipation, $\eta_{\rm Ohm}$, is approximated as \citep[][see also Figure \ref{fig:etas_amin}]{2002ApJ...573..199N,2007ApJ...670.1198M}
\begin{eqnarray}
\label{eta_Ohm_eq}
\eta_{\rm Ohm} \sim 1.6 \times 10^{13} \rho_{\rm g,10^{-16} \gcm} \sqrt{T_{10 \rm K}} ~{\rm cm^2~s^{-1}},
\end{eqnarray}
where $\rho_{\rm g}$ and $T$ denote the gas density and temperature, and $f_{X}$ means $f_{X}=(\frac{f}{X})$.
For ambipolar diffusion, the diffusion rate, $\eta_{\rm AD}$, is approximated as
\citep[][see also Figure \ref{fig:etas_amin}]{1983ApJ...273..202S,2018MNRAS.478.2723Z}
\begin{equation}
\label{eta_A_eq}
\begin{split}
\eta_{\rm AD} &\sim 2 \times 10^{18} ~{\rm cm^2~s^{-1}} \\ \times 
&\begin{cases}
 \rho^{-1/2}_{\rm g,10^{-16} \gcm} 
 & ( \rho_{g,\gcm}<10^{-16} ) \\
1 & (10^{-16} < \rho_{\rm g,\gcm}< 10^{-13}) \\
\rho_{\rm g,10^{-13} \gcm} & (10^{-13} <\rho_{\rm g,\gcm}<10^{-9}).
\end{cases}
\end{split}   
\end{equation}

Here, we assume $B=0.2 n_{\rm H}^{1/2} \mu G$ \citep{2002ApJ...573..199N} \yt{ and flux freezing during the evolution.
This is valid for a supercritical core and its isothermal collapse phase ($\rho_{\rm g} \lesssim 10^{-13} \gcm $ or $n_{\rm g} \lesssim 10^{10} \cmn $) as we will discuss in subsequent section}.


Another potentially important magnetic diffusion process is turbulent (or reconnection) diffusion.
On small scales, reconnection from turbulent vortices causes effective magnetic field diffusion.
The effective resistivity of this process is given as  \citep{2005AIPC..784...42L, 2021MNRAS.503.1290S},
\begin{eqnarray}
\eta_{\rm turb}=L_{\rm turb} \delta v_{\rm turb} {\rm min}(1, M_A^3).
\end{eqnarray}
where $L_{\rm turb}$ and $\delta v_{\rm turb}$ are the injection scale and turbulent velocity at the scale.
$M_A$ is the Alfven Mach number $M_A=\delta v_{\rm turb}/v_A$ and is typically $M_A \lesssim 1$ in cloud cores.

\bigskip


\subsubsection{Isothermal collapse phase: magnetic diffusion timescale vs free-fall timescale}
\label{isothermal_collapse_phase}

The gravitational collapse of molecular cloud core initially proceeds isothermally.
The compressional heating by gravitational contraction is immediately radiated away
via dust thermal emission and the gas keeps its temperature approximately constant at $T=10 ~$K.
In this phase, the effective polytropic index ($\Gamma_{\rm eff}$) is $\Gamma_{\rm eff} = 1$
where $P_{\rm gas}=K_{\rm s} \rho^{\Gamma{\rm eff}}$ 
 \yt{ ($K_{\rm s}$ is a constant)} is smaller than the critical value of $4/3$ for spherical gravitational collapse, and the pressure gradient force never overtakes the gravity.
\yt{As observations have shown \citep{2012ARA&A..50...29C}, most clouds
cores are supercritical.
Therefore, magnetic pressure is also not enough to support the core against gravity.}
As a result, the gas contracts with free-fall timescale.

In general, the evolution of the magnetic field is
determined by the balance between the gas dynamical timescale
and the magnetic diffusion timescale $t_{\rm diff}=L^2/\eta$ where
$L$ is the length scale of the system and $\eta$ is the magnetic resistivity.
In the case of isothermal collapse phase, the gas dynamical timescale is free-fall timescale, because 3D simulations have shown that even in the presence of a magnetic field, once gravitational collapse begins, evolution proceeds in free-fall timescale \citep[e.g.,][]{2014MNRAS.437...77B}, and observationally, the lifetime of a cores with  $n_g >10^6 \cmn$ is about free-fall time \citep{2015A&A...584A..91K}.
Therefore, by comparing the diffusion timescales of ambipolar diffusion, Ohmic dissipation, and turbulent diffusion with the free-fall time, we can understand the behavior of the magnetic field in the isothermal collapse phase.

As shown in the equation (\ref{eta_Ohm_eq}) and (\ref{eta_A_eq}), ambipolar diffusion dominates the Ohmic dissipation during the isothermal collapse phase ($\rho_{\rm g} \lesssim 10^{-13} \gcm$ or $n_{\rm g} \lesssim 10^{10} \cmn $).
Thus, let us compare the free-fall timescale and the ambipolar diffusion timescale using the equation (\ref{eta_A_eq}),
\begin{equation}
\label{tff_tAD}
\begin{split}
&\frac{t_{\rm ff}}{t_{\rm AD}}=\frac{\eta_{\rm AD} t_{\rm ff}}{\lambda_{\rm J}^2}= \\
&\begin{cases}
 2.2 \times 10^{-3} & (\rho_{\rm g,\gcm}<10^{-16}) \\
 7.1 \times 10^{-2} \rho^{1/2}_{\rm g,10^{-13} \gcm} & (10^{-16}<\rho_{\rm g,\gcm}<10^{-13})
\end{cases}
\end{split}
\end{equation}
where $t_{\rm ff}= \sqrt{3 \pi /(32 G \rho_{\rm g})}$ and 
we assume the length scale to be $L=\lambda_{\rm J}=\sqrt{\pi c_{\rm s}^2/(G \rho_{\rm g})} $, where  $\lambda_{\rm J}$ is the Jeans length.
This estimate shows $t_{\rm ff}<0.1 t_{\rm AD}~$ even at the density where ambipolar diffusion is most efficient in isothermal collapse phase,
and that the free-fall timescale is much shorter than the ambipolar diffusion timescale.
\yt{More quantitatively, to diffuse the magnetic field over a spatial scale of Jeans length $\lambda_{\rm J}=8.7 \times 10^{2} ~\rho_{\rm g, 10^{-16} \gcm}$ au, it takes about $3.4 \times 10^{11} ~\rho_{\rm g, 10^{-16}\gcm}^{-2}$ yr with Ohmic dissipation and $2.7 \times 10^6 ~\rho_{\rm g, 10^{-16}\gcm}^{-1}$
yr with ambipolar diffusion (we assume $\rho_{\rm g}>10^{-16}\gcm$ here).}
This estimate is consistent with previous studies with more realistic $\eta_O$ and $\eta_A$
\citep{1986MNRAS.221..319N,1990MNRAS.243..103U}.
Hence, this estimate shows that  ambipolar diffusion (as well as Ohmic dissipation) may not play a role during the isothermal collapse phase.

The estimate is also consistent with three-dimensional simulations.
\citet{2015MNRAS.452..278T} and \cite{2016A&A...587A..32M} compared the magnetic field evolution between  ideal and non-ideal MHD simulations and 
showed that, in the isothermal collapse phase, the magnetic field evolution is identical between
simulations confirmed that, even with Ohmic dissipation and ambipolar
diffusion, the gas and magnetic field are completely coupled during the isothermal collapse phase.

\yt{Note that equation (\ref{eta_A_eq}) and hence (\ref{tff_tAD}) assumes the flux freezing. This assumption is justified and our discussion is self-consistent because we conclude that ambipolar diffusion does not work during isothermal collapse phase.}

Then, can turbulent (or reconnection) diffusion play the role during the isothermal collapse phase?
The following inequality holds for the turbulent resistivity,
\begin{eqnarray}
\eta_{\rm turb}=L_{\rm turb} \delta v_{\rm turb} {\rm min}(1, M_A^3) \leq L_{\rm turb}  \delta v_{\rm turb}.
\end{eqnarray}
Here, we assume that the injection length is of the order of Jeans length $L_{\rm turb}\sim \lambda_{\rm J}$.

As shown in \citet{2007prpl.conf...33W},
the turbulent velocity is subsonic in the cloud core. 
Thus, we have the inequality $\delta v_{\rm turb}<c_{\rm s}$ and 
we can show that the ratio of the free-fall timescale to the magnetic diffusion timescale due to the turbulent reconnection diffusion as 
\begin{eqnarray}
  \frac{t_{\rm ff}}{t_{\rm turb}}=\frac{\eta_{\rm turb} t_{\rm ff}}{\lambda_{\rm J}^2} \leq \frac{ \delta v_{\rm turb} t_{\rm ff}}{\lambda_{\rm J}} \leq \frac{ c_{\rm s} t_{\rm ff}}{\lambda_{\rm J}}
    \sim \frac{t_{\rm ff}}{t_{\rm sound}}<1,
\end{eqnarray}
where $t_{\rm sound}=\lambda_{\rm J}/c_{\rm s}$ is the sound-crossing timescale.
The last inequality follows from the fact that the sound-crossing time is longer than the free-fall time due to the graviationally collapsing core. 
Hence, the turbulent reconnection diffusion also may not play a role in the isothermal collapse phase.
Note, however, that since supersonic turbulence exists in massive cores,
turbulent diffusion could be an important effect there. 



In summary, any magnetic diffusion may not play a role during the isothermal collapse phase,
which is consistent with previous studies
\citep{1986MNRAS.221..319N,1990MNRAS.243..103U,2015ApJ...801..117T, 2016A&A...587A..32M}.
The estimates made in this subsection show
that the magnetic field structure in the cloud core and envelope reflects the dynamics of the gas.
Therefore, it should be possible to study the dynamics of the gas from the structure of the magnetic field, and vice versa.

\subsubsection{Formation of hydrostatic first-core and magnetic diffusion}
As gravitational collapse proceeds and the central density increases, drastic changes in gas thermal evolution and magnetic diffusion occur.
Once the  density reaches $\rho_{\rm g} \sim 10^{-13} \gcm$ (or $n_{\rm g} \sim 10^{10} \cmn$ ), the gas becomes optically thick to the dust thermal emission, and the compressional heating overtakes the radiative cooling.
The effective polytropic index increases to $\Gamma_{\rm eff}=5/3$, which is larger than the critical value of $4/3$, and the pressure gradient force begins to balance the gravitational force.
As a result, the pressure supported first hydrostatic core (or first core) forms \citep{1969MNRAS.145..271L,1998ApJ...495..346M,2011A&A...530A..13C,2012A&A...545A..98C, 2013ApJ...763....6T,
2017A&A...598A.116V,2018A&A...618A..95B}, whose size and mass are determined by \yt{the Jeans length $\sim 1$ to $10$ au and} Jeans mass
$\sim 10^{-2}$ to $10^{-1} \msun$ \citep{1969MNRAS.145..271L,1998ApJ...495..346M}.

In the first core, a dramatic increase in resistivity happens, which is crucial for magnetic field evolution.
As shown in Figure \ref{fig:etas_amin} (and Equation \ref{eta_A_eq}),
the ambipolar resistivity begins to increase around the density of $\rho_{\rm g} \gtrsim10^{-13} \gcm$ (or $n_{\rm g} \gtrsim 10^{10} \cmn $),
as density increases due to adsorption of ion/electron by dusts.
Hence, in the first core,
\begin{eqnarray}
\frac{t_{\rm ff}}{t_{\rm AD}} \sim 2.2 \rho_{\rm g,10^{-12} \gcm}^{3/2}.
\end{eqnarray}
and the diffusion timescale becomes shorter than the free-fall timescale at $\rho_{\rm g} \gtrsim 10^{-12} \gcm$ (or $n_{\rm g} \gtrsim 10^{11} \cmn $).
Furthermore, gas is supported by the pressure gradient force, and the first core has a much longer lifetime than the free-fall time.
Thus, the magnetic diffusion timescale becomes significantly shorter than the lifetime of the first core,
\yt{ its magnetic field weakens and its mass-to-flux ratio increases}
mainly due to ambipolar diffusion
\citep[$\eta_{\rm AD}$ is still larger than $\eta_{\rm Ohm}$   in $\rho_{\rm g} \lesssim 10^{-10} \gcm$ or $n_{\rm g} \lesssim 10^{13} \cmn $][]{2015ApJ...801..117T,2015MNRAS.452..278T,2016A&A...587A..32M, 2017PASJ...69...95T}.
Simultaneously, decoupling between the gas and magnetic field in the first-core suppresses
angular momentum removal by magnetic tension and allows the first-core to retain its rotation,
which is essential for the formation of a circumstellar disk at the protostar formation epoch \citep{2011MNRAS.413.2767M,2012A&A...541A..35D,2015MNRAS.452..278T,2018MNRAS.475.1859W,2018A&A...615A...5V}.

\begin{figure}[ht]
\includegraphics[width=80mm]{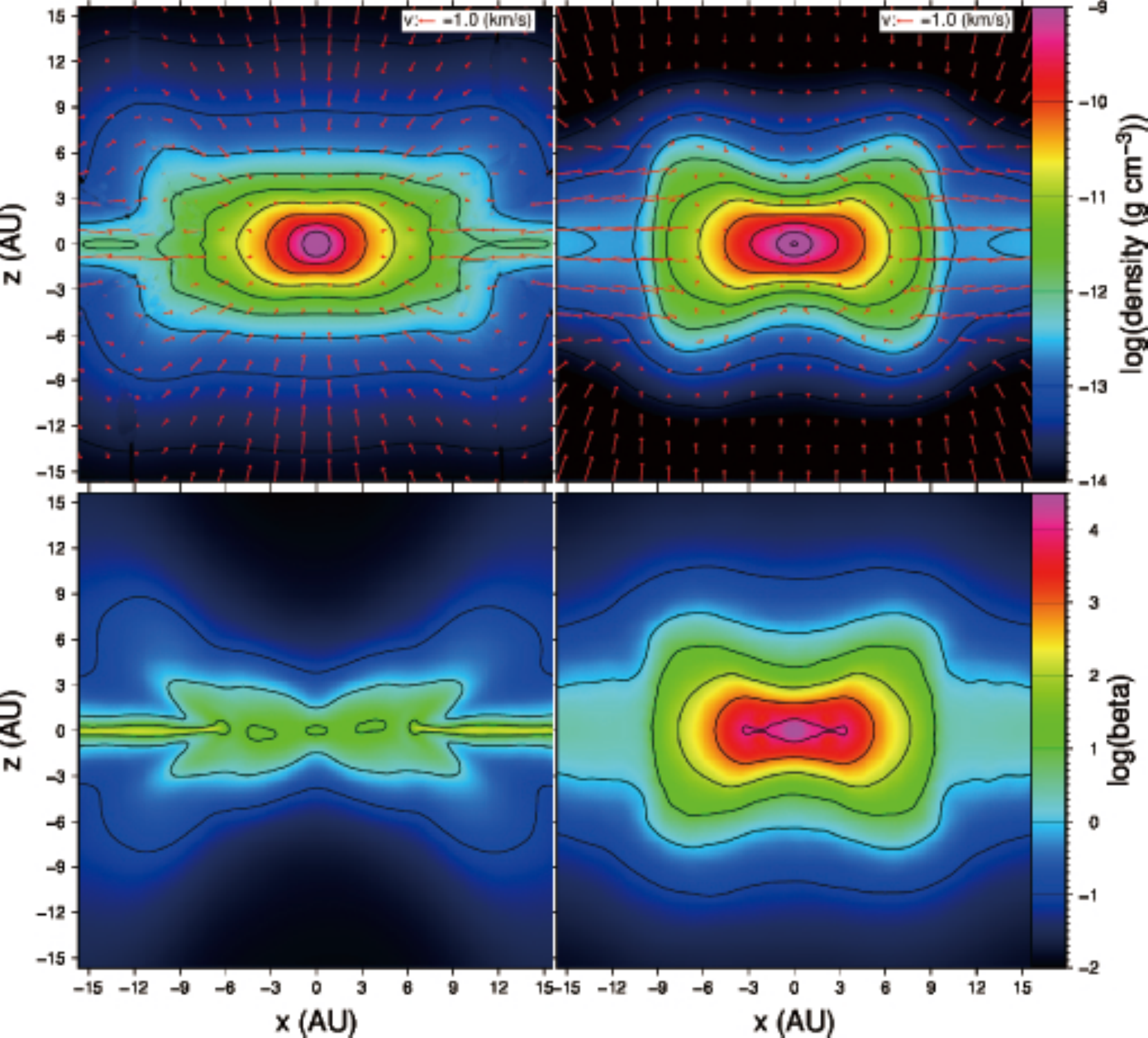}
\caption{
Density (top panels) and plasma $\beta$ (bottom panels) map of the first core in the ideal MHD (left panels) and non-ideal MHD (right panels)
simulations \citep[][]{2015MNRAS.452..278T} from edge-on view.
Red arrows show the velocity field.
}
\label{fig_first_core}
\end{figure}

Figure \ref{fig_first_core} shows the density and plasma $\beta (=P_{\rm gas}/P_{\rm mag})$ (where $P_{\rm gas}$ and $P_{\rm mag}$ gas and magnetic pressure)
maps of the first core formed in ideal MHD simulation and non-ideal MHD simulation (with Ohmic dissipation and ambipolar diffusion).
We can see the striking difference of plasma $\beta$ in the first core between the two simulations.
Without magnetic diffusion, the plasma $\beta$ inside the first
core is $\beta\sim 10$, and a strong magnetic field retains in the first core.
On the other hand, the plasma $\beta$ at the center is $\beta \sim 10^4$ in non-ideal MHD simulations.
This significant difference induced by magnetic diffusion significantly
alters the fate of the first core, and the formation process of the protostars and circumstellar disk.

\begin{figure}[ht]
\includegraphics[clip,width=90mm]{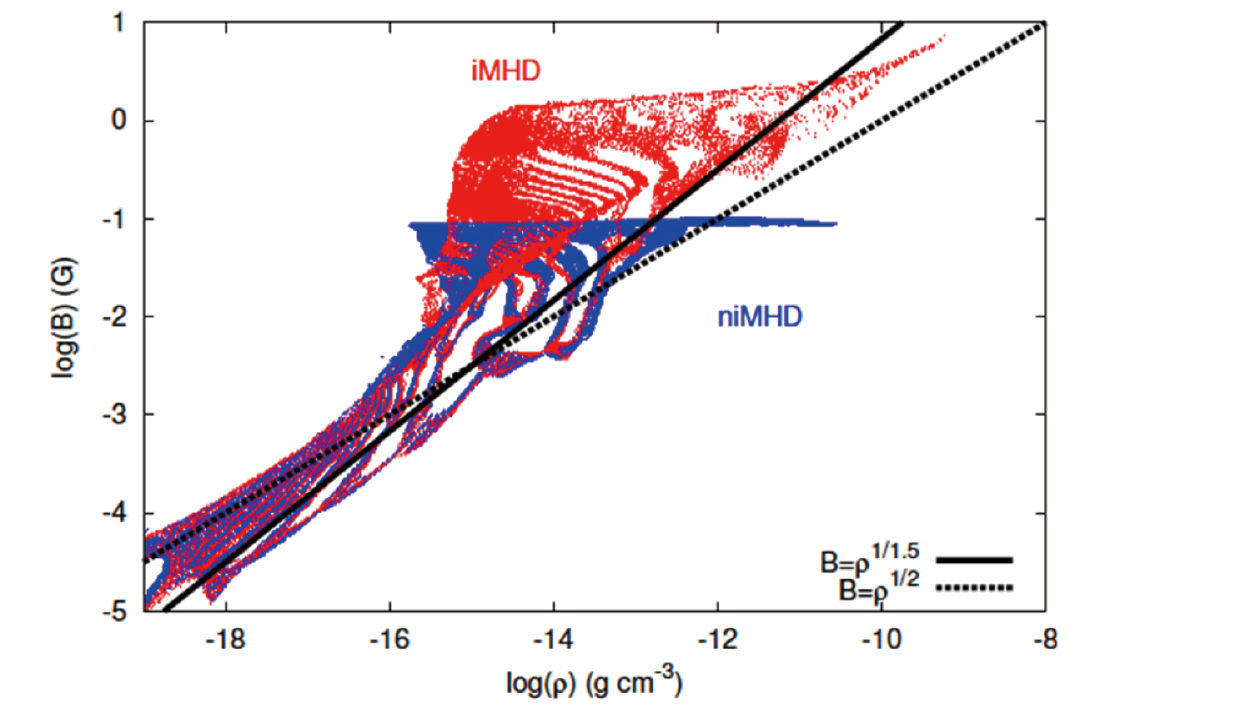}
\caption{
Magnetic field strength as a function of the density at the first-core formation epoch
in ideal MHD (red) and non-ideal MHD (blue) simulations
at 2000 years after first core formation. The solid and dashed black lines show $B \propto  \rho^{2/3}$ and $B \propto  \rho^{1/2}$, respectively \citep[from][]{2016A&A...587A..32M}.
}
\label{fig_Bfield_firstcore}
\end{figure}

Recent simulations also show that the magnetic field saturates in the first core and has a universal value of $B\sim 0.1 ~\rm G$
\citep{2016A&A...587A..32M,2017PASJ...69...95T,2020A&A...635A..67H,2021MNRAS.502.4911X}.
Figure \ref{fig_Bfield_firstcore} shows the comparison of the magnetic field strength of the first-core in ideal MHD and non-ideal MHD simulation (with ambipolar diffusion).
Ambipolar diffusion becomes stronger as the magnetic flux is brought into the first core.
This causes negative feedback on magnetic field amplification. As a result, the magnetic field saturates at the specific value of $B \sim 0.1 ~\rm G$.

\begin{figure*}[ht]
\includegraphics[width=140mm]{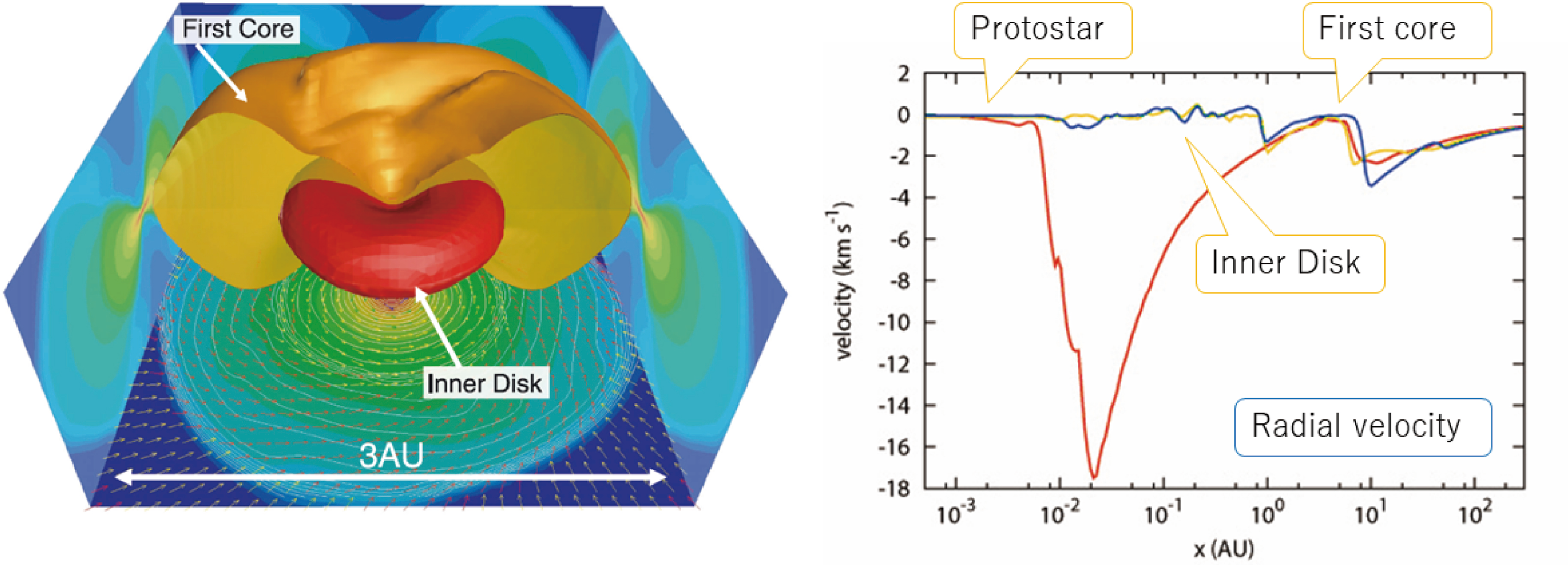}
\hspace{-50truemm}
\vspace{-5truemm}

\caption{
Left panel shows the first core (orange iso-density surface) and inner disc (red iso-density surface) at $t = 1.7$ yr after the formation of the protostar  \citep[from][]{2011MNRAS.413.2767M}.
The density distribution on the $x = 0$, $y = 0$ and $z = 0$ plane is projected on to each wall surface.
The velocity vectors on the $z = 0$ plane are also projected on to the bottom wall surface.
Right panel shows the radial velocity just after the formation of the protostar \citep[from][]{2015MNRAS.452..278T}.
The red, orange, and blue lines show the radial velocity in ideal MHD, non-ideal MHD with Ohmic dissipation, and non-ideal MHD with Ohmic dissipation and ambipolar diffusion, respectively.
}

\label{fig_disk_formation}
\end{figure*}

\begin{figure*}[ht]
\centering
\includegraphics[width=0.8\textwidth]{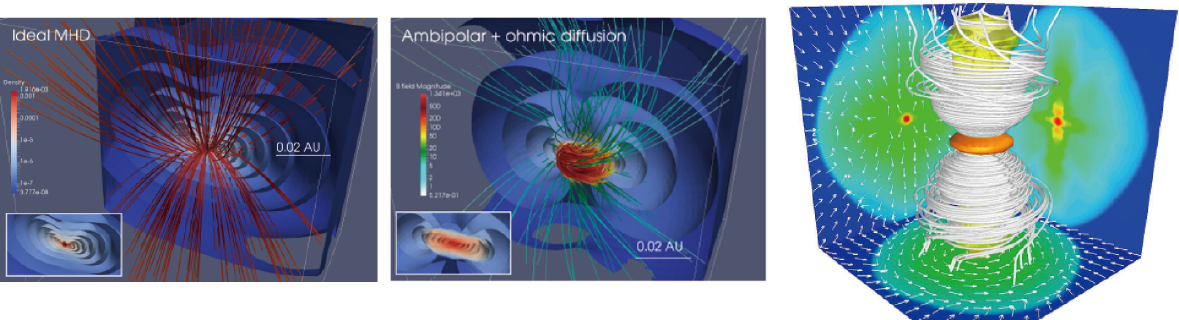}
\vspace{-4truemm}
\caption{
Left and middle panels show density isosurfaces
around the newly born protostar  in the ideal MHD (left) and
non-ideal MHD (middle) simulations. The magnetic field morphology is shown as lines and are 
colored according to the magnitude of the magnetic field \citep[from][]{2018A&A...615A...5V}.
The right panel shows the magnetic field profile and jet launching
after the protostar formation \citep[from][]{2013ApJ...763....6T}.
The edge of the right panel is $\sim 0.27$ au. The high-density region ($\rho > 10^{-5} \gcm$ ) is visualized with
the orange surface. White arrows denote the direction of the fluid motion and
white lines show the magnetic field. Fast outflowing gas ($v_z > 3 \kms$ ) is
volume rendered in pale yellow.
}
\label{fig_protostar}
\end{figure*}

\subsubsection{Second collapse and birth of protostar and circumstellar disk}
\label{second collapse}
The seeds of protostar and jet begin to sprout in the inner part of the first core as it evolves.
The first core adiabatically evolves and its central temperature increases
as $T\propto \rho^{\Gamma_{\rm eff}-1}$ \citep{1969MNRAS.145..271L,1998ApJ...495..346M,2012A&A...543A..60V,2015ApJ...801..117T}.
When the central temperature reaches $T \sim 10^3$ K at $\rho_{\rm g} \sim 10^{-8} \gcm$ (or $n_{\rm g} \sim 10^{15} \cmn $),
the magnetic field and gas recouple due to thermal ionization of potassium and go back to the ideal MHD regime
but with a weak magnetic field ($\beta \gg 1$) due to magnetic diffusion occurring at the first core scale.
Subsequently, the molecular hydrogen begins to dissociate, an endothermic reaction, at $\sim 2\times 10^3$ K and decreases the effective polytropic index $\Gamma_{\rm eff} \sim 1.1$,
which is smaller than the critical value. As a result, the gravitational collapse in the first core resumes, and is called as the "second collapse".
The second collapse lasts until the density reaches $\rho_{\rm g} \sim 10^{-2} \gcm$ (or $n_{\rm g} \sim 10^{21} \cmn $), and the molecular hydrogen has wholly dissociated.
Then the effective polytropic index becomes $\Gamma_{\rm eff} \sim 5/3$, and the pressure supported (second) core or protostar forms, whose mass is determined by the Jeans mass at $\sim 10^{-3}$ to $10^{-2} \msun$  \citep{1969MNRAS.145..271L,2000ApJ...531..350M, 2017A&A...598A.116V}. This is the birth of a protostar.

The magnetic field amplification during the second collapse determines the magnetic field strength of the newly born protostar.
Due to the conservation of the magnetic flux, the magnetic field in the ideal MHD regime evolves as $B \propto R^{-2}$.
Thus, during the second collapse the central part of the first core ($\sim 1$ au) collapses to the protostellar size of $0.01$ au, and the magnetic field strength increases from the critical value of the magnetic field in the first core, $|\magB| \sim 10^{-1} ~\rm  G$ to $ \sim 10^{3} ~\rm G$, which is consistent with protostar observations \citep{2009ApJ...700.1440J}.
\yt{ On the other hand, when we conduct the ideal MHD simulations of the collapsing cloud core,
the magnetic field strength of the central protostar is more than ten times larger than that in the non-ideal MHD simulations}
\citep[][]{2015MNRAS.452..278T,2018A&A...615A...5V,2018MNRAS.481.2450W,2019ApJ...876..149M}.
Therefore, the Ohmic dissipation and ambipolar diffusion are the key  to solve the classical ``magnetic flux problem" \citep{1985ApJ...291..772M} in star formation.

During the second collapse, most of the gas in the first core is distributed around the protostar with \yt{angular momentum
that is sufficient to allow the centrifugal force to balance gravity.}
Therefore, most of gas in the first core quickly transforms into a rotationally supported disk.
This is the birth of the circumstellar disk.
Figure \ref{fig_disk_formation} shows this transformation.
The left panel shows the bird's eye view of the forming protoplanetary disk (red region)
inside the first core (\yt{orange} region) shortly after the formation of the protostar.
The right panel shows the infall velocity inside the first-core shortly after the protostar formation.
In the ideal MHD simulation (red line), gas is directly accreted to the protostar ($\sim 0.01$ au),
whereas in the non-ideal MHD simulation, gas accretion stops at $1$ au--
indicating that the centrifugal force balances with gravity at $\sim 1$ au, and the protoplanetary disk is forming.
The size of the newly born circumstellar disk is approximately 1 to 10 au, which corresponds to the size of the first core.

\subsubsection{Outflow/jet launching from the newly born disk and protostar}
\label{Outflow_Theory}
The formation process of protostars and circumstellar disks described in the previous section is also closely related to the formation of outflows and jets. Protostellar outflows and jets have been ubiquitously observed in star-forming regions. 
Wide-angle and slow-velocity flows are called `molecular outflow' or `low-velocity outflow' in molecular line emissions.
On the other hand, well-collimated high-velocity flows are called `collimated jet' or `optical jet' and are observed by molecular and atomic line emissions from radio to optical wavelengths. 
Hereafter, we refer to the former as `the outflow' and the latter as `the jet.'


Magnetically driven wind scenarios have been proposed as the driving mechanism of these flows \citep[e.g.,][]{1986ApJ...301..571P}. These winds have been investigated in past theoretical and simulation studies, in which the outflow and jet are magnetocentrifugally driven \citep{1982MNRAS.199..883B}
or driven by magnetic pressure gradient force \citep{1985PASJ...37..515U}. 
However, in these past studies, the settings for driving the outflow and jet were highly idealized.
Considerably idealized (or an unrealistic) circumstellar disk and magnetic field configurations
were setup for the emergence of winds. 
For example, only poloidal fields (or dipole magnetic fields) were assumed within a thin disk as the initial state in past studies.   
The problem is that such an idealized condition (artificial distribution of density or magnetic field) is not actually realized in the star formation process.

\citet{1998ApJ...502L.163T,2000ApJ...528L..41T,2002ApJ...575..306T} successfully reproduced
the outflow in the star formation simulation with a two dimensional ideal MHD code, for the first time. 
These studies simulated the gravitational collapse process of star forming cores, spatially resolving both prestellar core and protostar.
In this simulation, the outflow and jet naturally appear in gravitationally collapsing clouds. 
Furthermore, driving the outflow and jet has been comprehensively investigated in many star-formation simulations \citep{2004MNRAS.348L...1M,2006ApJ...647L.151M,2006ApJ...641..949B,2008ApJ...676.1088M,2008A&A...477...25H,2010MNRAS.409L..39C,2011MNRAS.417.2036B,2012MNRAS.423L..45P,2012A&A...543A.128J,2012MNRAS.422..347S,2013ApJ...763....6T,2014MNRAS.437...77B,2017MNRAS.467.3324L,kolligan:18,2018MNRAS.475.1859W,2021MNRAS.502.4911X}.
With these studies, the driving mechanism of outflow and jet in the star formation process can be established. 

In a first core, the gravitational collapse slows and the collapse timescale is longer than the rotation timescale.
Thus, magnetic field lines are twisted around the first core, and the low-velocity outflow emerges due to the magnetocetrifugal force before the formation of the protostar. 
The outflow driven by the first core continues to be produced by the rotationally supported disk after the protostar has formed. 

During the main accretion phase, 
the disk gas intermittently falls onto the protostar due to the episodic accretion, inducing an episodic mass ejection or a time-variable outflow (and jet) \citep{2019ApJ...876..149M}. 
The outflow physical parameters such as mass, momentum, momentum flux, kinetic energy and kinetic luminosity, are in good agreement with observations, suggesting that the molecular outflow is directly driven by the disk \citep[e.g.,][]{2013MNRAS.431.1719M}. 
A wire-like (relatively strong), radially-distributed magnetic field guides the outflow,  which is mainly driven by the magnetocentrifugal mechanism and creates a wide opening angle flow.

On the other hand, a weak magnetic field around the protostar, due to Ohmic dissipation and ambipolar diffusion,
is passively twisted by the rotation of protostar and disk inner edge and generates a strong magnetic pressure gradient force to drive the jet.
Figure \ref{fig_protostar} illustrates this evolution.
The left panel shows the magnetic field configuration around the protostar
just after its formation in the ideal MHD simulation, and shows that an hourglass-shaped magnetic field has been realized. This indicates that the strong magnetic field suppresses
the gas rotation and that the gas motion can not twist the magnetic field.
Conversely, the weak magnetic field due to magnetic diffusion
is easily twisted during the second collapse.
The middle panel shows the magnetic field configuration in the non-ideal MHD simulation,
and shows that the magnetic field is passively twisted by the rotation.
This passive twisting of the magnetic field drives the jets from the vicinity of the protostar, as shown in the right panel.
Therefore, magnetic diffusion also plays an essential role in driving the protostellar jet.
The mass-loss rate of the jet obtained from the 3D simulation is $\sim 10^{-5}~\msunyear$ \citep{2019ApJ...876..149M}, and 
is in agreement with the observed value in a protostar, $\sim 3 \times 10^{-6}~\msunyear$  \citep{2020A&ARv..28....1L}.

Lastly, we comment on the classical scenario for driving the outflow. 
It had been considered that the low-velocity outflow is entrained by the high-velocity (primary) jet (so-called the entrainment scenario). 
In this scenario, the jet appeared near the protostar and provided linear momentum to the ambient gas through some mechanism \citep{2007prpl.conf..245A}. 
The ambient gas entrained by the jet is then observed as the low-velocity outflow. 
Recently, observations of outflow rotation could resolve the long-standing debate on what is driving the outflow. 
It had been considered that we would unveil the outflow driving mechanism if this rotation could be measured. 
In the entrainment scenario, the primary jet, which provides both linear and angular momentum to the outflow (or the entrained gas), has only a small amount of angular momentum, because the jet driving region is located near the protostar.
Thus, it is expected that the outflowing gas receives a very small amount of angular momentum from the primary jet. 
On the other hand, in the direct disk driven scenario proposed from the core collapse simulations,
the outflow should have a large amount of angular momentum because the outflow driving region
is much far from the protostar where the rotation velocity and angular momentum are large \citep{2021ApJ...907L..41L}. 


\subsection{Evolution of protoplanetary disks from Class 0 to the end of the Class I phase in YSOs}
\label{disk_evolution}

At the time of PP VI, many researchers were seriously considering the possibility
that highly efficient magnetic braking completely suppresses the disk formation in the early phase of protostar evolution
(so-called magnetic braking catastrophe (MBC) \citep[see the review of PP VI][]{2014prpl.conf..173L}.
MBC is, in the strict sense, resolved by the scenario described in \S \ref{second collapse} because the circumstellar disk with a size of $\sim 1$ au simultaneously forms at the formation epoch of protostar
\yt{\citep[Note that a disk with a size of $\sim 1$ au is out of the scope of the studies of MBC by][because the inner boundary is at $r \sim 6$ au in their simulations]{2008ApJ...681.1356M,2009ApJ...698..922M,2011ApJ...738..180L}}.



However, theoretically, magnetic braking is estimated to be very efficient, and findings of large disks with size of several $10$ to $100$ au in the embedded phases by 
\am{
recent observational surveys (see section\ref{ssec:obs_radii})} force researchers to revisit the disk size evolution more precisely.

In this subsection, we review the physical mechanisms which weaken the magnetic braking and \am{may play key roles in determining} the disk size evolution during \am{the early phases of star formation}.

\subsubsection{Misalignment between magnetic field and rotation vector}
In many theoretical studies, it is assumed for simplicity that the
angular momentum of the parent cloud core is aligned with its magnetic field.
However, in real molecular cloud cores, the magnetic field and rotation vector are often observed to be misaligned
\citep{2013ApJ...768..159H, 2014ApJS..213...13H,2021ApJ...907...33Y}.

\citet{2009A&A...506L..29H} first reported that the rotationally supported disk
more easily forms in cores with misalignment between the magnetic field and the angular momentum vector.
More quantitatively, \citet{2012A&A...543A.128J} showed that the mean specific angular momentum of the central region
in a perpendicular core (\yt{in which the initial magnetic field and angular momentum are perpendicular}) is about two times larger than that in a parallel core
(\yt{in which the initial magnetic field and angular momentum are parallel}).
The influence of misalignment is also investigated by \citet{2013ApJ...774...82L}.
The authors obtained results consistent with \citet{2009A&A...506L..29H}, and concluded that disk formation becomes
possible when $\mu\gtrsim 4$, where $\mu$ is the mass-to-flux ratio of the core normalized by its critical value.

By examining the angular momentum evolution of fluid elements during the cloud core collapse,
\citet{2018ApJ...868...22T} elucidated three mechanisms by which misalignment affects the angular momentum evolution (and ultimately the size) of the disk, 
\begin{enumerate}
\item the selective accretion of fluid elements with large (small) angular momentum to the central perpendicular (parallel) core;
\item magnetic braking in the isothermal-collapse phase;
\item magnetic braking in the disk;
\end{enumerate}

\yt{
The selective accretion is a process in which the magnetic tension suppresses
mass accretion from the direction perpendicular to the magnetic field and selectively
enhances mass accretion parallel to the field.
As a result, the fluid elements with a larger (small) angular momentum accrete onto the central region in a perpendicular (parallel) core.
This effect may play a role until the envelope disappears.
}
\yt{
Thus, this effect can lead to the misinterpretation of the impact of misalignment on magnetic braking efficiency on time scales shorter than the time scale of envelope dissipation,
and it makes the magnetic braking appear 
stronger (weaker) in parallel (perpendicular) cores.
(In reality, it is just that the angular momentum of the accreted gas is smaller (larger))}

\citet{2018ApJ...868...22T} showed that the magnetic braking
during the isothermal collapse phase is stronger in perpendicular cores than in parallel cores.
This result is opposite to that expected from the previous studies such as \citet{2009A&A...506L..29H} and \citet{ 2012A&A...543A.128J}, but is consistent with the classical discussion of magnetic braking by  \citet{1985A&A...142...41M} and the simulation results of prestellar collapse phase \citep{2004ApJ...616..266M}.
The inconsistency between the trend of magnetic braking in the isothermal collapse phase
and the results of \citet{2009A&A...506L..29H} and \citet{ 2012A&A...543A.128J}, \yt{ which investigate the disk evolution after protostar formation, may suggest that magnetic braking during the prestellar collapse phase (or in the envelope)} is not the dominant mechanism determining the evolution of the disk size, \yt{otherwise the size of the disks should be opposite in the parallel and perpendicular cores}.

Of the three mechanisms, magnetic braking in the disk seems to play the dominant role for disk size evolution in the main accretion phase
because the disk is supported by the centrifugal force and the magnetic field
has a much longer time than the local free-fall time to extract the gas angular momentum.
Furthermore, because the magnetic field fans out around the central region and has an hour-glass-like geometry,
the magnetic braking in the disk with the parallel magnetic field can be enhanced
\citep{1985A&A...142...41M,2009A&A...506L..29H,2018ApJ...868...22T}.
Note also that \citet{2013ApJ...774...82L} \citep[see also ][]{2020ApJ...898..118H} point out that the angular momentum
removal by disk outflows also plays an important role in the parallel core which further favors the growth of the disk in the misaligned core in which outflow formation is suppressed \citep[e.g.,][]{2012A&A...543A.128J,2013ApJ...774...82L,2016MNRAS.457.1037W,2017PASJ...69...95T}. Note, however, that \citet{2020ApJ...900..180M} demonstrated that magnetic braking is primary and outflows are secondary in angular momentum transport.

In summary, because mechanisms 1 and 3 work effectively during the (early) main accretion phase,
misalignment seems to promote disk growth.
On the other hand, in the prestellar collapse phase,
the presence of misalignment seems to enhance the angular momentum removal from the central region due to mechanism 2.

Note, however, that it has been pointed out that the impact of misalignment is not so prominent once ambipolar diffusion is considered \citep{2016A&A...587A..32M}, probably because of the suppressed magnetic braking inside the disk.
To observationally confirm that misalignment plays a significant role in disk evolution, it is important to investigate a correlation between outflow activity
(rather than direction), magnetic field orientation, and disk size.
Several studies suggest that misalignment suppresses outflow formation, and that there is a correlation among the disk size,
outflow activity, and misalignment \citep{2013ApJ...774...82L,2020ApJ...898..118H}.
If such correlation is observed, it provides evidence that misalignment affects disk evolution.

\subsubsection{Turbulence\label{sec:turb}}
\citet{2012ApJ...747...21S} suggested that turbulence in the cloud core weakens magnetic braking.
They compared the simulation results for a coherently rotating core and a turbulent core
and found that a rotationally supported disk is formed only in the turbulent cloud core.
Similar results were obtained by \citet{2012MNRAS.423L..40S,2013MNRAS.432.3320S}.
\citet{2012ApJ...747...21S} suggested that random motions due to turbulence cause small-scale magnetic reconnection events and provide an effective magnetic
diffusion that efficiently removes magnetic flux, and disks with a size of $r\sim 100$ au are formed even in the ideal MHD limit.
However, their results are obtained with relatively coarse numerical resolution of $r\gtrsim 1$ au.
With such a coarse resolutions in ideal MHD, the numerical magnetic diffusion
can promote the magnetic flux loss and can artificially suppress the magnetic braking.

\citet{2013A&A...554A..17J} investigated the impact of turbulence with a higher numerical resolution of $\sim 0.4$ au.
Their results suggests that the impact of turbulence is limited.
For example, they compared the mass-to-flux ratio of the central region (e.g., $r\sim 100$ au)
between cores with subsonic turbulence and without turbulence, with a realistic magnetic field strength of $\mu \le 5$.
Their results show that the difference is about the factor of two and is not so significant.
Furthermore, they showed that the mass of the disk decreases as the numerical resolution increases.
In other words, they point out that the problem of numerical convergence cannot be resolved
even with resolutions below 1 au in the case of ideal MHD simulations with turbulence.

Furthermore, it is pointed out that the impact of turbulence becomes less important
once the non-ideal MHD effect is considered \citep{2019MNRAS.489.5326L,2020MNRAS.495.3795W}.
These results, as well as the analytic estimate of turbulent (reconnection) diffusion rate in \S \ref{isothermal_collapse_phase},
suggest that the turbulent diffusion may not play a significant role for the disk size evolution.

\subsubsection{Ohmic dissipation and ambipolar diffusion}
\label{ambi_sec}
So far, we have examined the mechanisms that weaken magnetic braking in the ideal MHD limit.
However, the ideal MHD approximation is not valid in real YSOs because of their low ionization degree.
Non-ideal MHD effects can decouple the magnetic field and gas, and can decrease the magnetic braking efficiency in the disk.
Thus, non-ideal effects may affect the size evolution of circumstellar disks.
In this section, we review the influence of Ohmic dissipation and ambipolar diffusion on early disk evolution.

In the context of disk size evolution,
Ohmic dissipation becomes effective in the dense inner region of the disk of $\rho_{\rm g}>10^{-10} \gcm $ (or $n_{\rm g}>10^{13} \cmn$).
\citet{2011PASJ...63..555M} used 3D simulations with Ohmic dissipation and found that the disk size
is $r\lesssim 10$ au in $t \lesssim 10^4$ yr after the formation of the protostar (or Class 0 phase where the envelope is very massive).
This result suggests that it is still difficult to form disks with sizes of several tens of au in the Class 0 phase by Ohmic dissipation alone.

Ambipolar diffusion is a more efficient magnetic diffusion process in the disk and envelope.
It has two important properties.
One is that ambipolar resistivity $\eta_{\rm AD}$ is much larger than the Ohmic resistivity $\eta_{\rm Ohm}$ in almost the entire region of the disk.
The ambipolar resistivity becomes an increasing function of density at $\rho_{\rm g} \gtrsim 10^{-12} \gcm$ (or $n_{\rm g} \gtrsim 10^{11} \cmn$) and dominates
the Ohmic resistivity in $\rho_{\rm g} \lesssim 10^{-9} \gcm$ (or $n_{\rm g} \lesssim 10^{14} \cmn$) \citep[e.g.,][see also Figure \ref{fig:etas_amin} and equations (\ref{eta_Ohm_eq}) and (\ref{eta_A_eq})]{2015ApJ...801..117T,2015MNRAS.452..278T,2016A&A...592A..18M,2018MNRAS.478.2723Z}.
3D simulations have shown that ambipolar diffusion allows the formation of a relatively extended
disk with a size of $r \gtrsim 10$ au even in the early disk evolution phase \citep[$t \lesssim 10^4$ yr after protostar formation,][]{2016A&A...587A..32M,2018MNRAS.473.4868Z,2020ApJ...896..158T}.
They also showed that the disk is massive and the spiral arms induced by gravitational instability form despite a relatively strong magnetic field.

\yt{Since ambipolar diffusion is also effective in the envelope, the magnetic flux brought into the central region by accretion is distributed around the disk.
It has been proposed that the accretion towards the magnetic flux tube results in the formation of a shock \citep[ambipolar diffusion shock][]{1996ApJ...464..373L,1998ApJ...504..257C}.}




The resistivities ($\eta_{\rm Ohm}, \eta_{\rm AD},$ and $\eta_{\rm Hall}$) strongly depend
on the dust properties as well as the cosmic-ray (CR) ionization rate, and 
any change of these parameters is expected to affect the disk size evolution
\citep{2016MNRAS.460.2050Z,2018MNRAS.476.2063W,2017A&A...603A.105D,2019MNRAS.484.2119K,2020A&A...643A..17G} 

The CR ionization rate ($\zeta_{\rm CR}$) generally gauges the overall abundances 
of charge species.
The typical level of 
$\zeta_{\rm CR}$ in dense cores (inferred from 
chemical analysis) ranges from a few times 10$^{-18}$~s$^{-1}$ to a few times 10$^{-16}$~s$^{-1}$ 
\citep{Caselli+1998,Padovani+2009,Ivlev+2019}, with $\zeta_{\rm CR}\approx10^{-17}$~s$^{-1}$ as the 
``standard" rate. However, the current methods for constraining $\zeta_{\rm CR}$ rely on 
the measured abundance of molecular ions such as HCO$^+$, DCO$^+$ and N$_2$H$^+$  \citep{Caselli+1998,vanderTak+vanDishoeck2000,Doty+2002}, 
which can be largely uncertain due to the freeze-out of molecules in cloud cores. 
Nevertheless, recent non-ideal MHD simulations show that disk formation is possible for such a range of $\zeta_{\rm CR}$.
If only ambipolar diffusion and Ohmic dissipation are considered,
a slightly higher CR ionization rate ($\zeta_{\rm CR}$ $\gtrsim$a few times $10^{-17}$~s$^{-1}$) than the standard rate at core scale can suppress the formation of disks with sizes of several tens of au \citep{2016MNRAS.460.2050Z,2020A&A...639A..86K}.


\yt{Dusts plays an important role in both determining the abundance of ions and electrons and in determining the conductivity.
Dust adsorbs charged particles and determines the ionization degree of the gas phase.
On the other hand, a large population 
of charged small grains ($\lesssim$100 \AA)
with Hall parameter close to unity can determine the conductivities
\citep{Li1999,Padovani+2014,2016MNRAS.460.2050Z,2017A&A...603A.105D},}
and decrease the ambipolar and Hall resistivities at envelope and disk densities \citep{2018MNRAS.478.2723Z,Koga+2019,2020ApJ...900..180M}. 
The main reason for the suppression of disk formation in early non-ideal MHD work \citep{2009ApJ...698..922M,2011ApJ...738..180L}
is largely attributed to the use of ionization chemistry assuming the presence of a large number of small charged grains. 
As shown in analytical work, the smallest grains are rapidly depleted in cold dense environments \citep{Ossenkopf1993,Hirashita2012,Kohler+2012,2020A&A...643A..17G,Silsbee+2020}, 
which is supported by the non-detection of spinning dust grain emission (produced by small dust grains of $\lesssim$100~\AA) in recent Galactic cold core surveys \citep{Tibbs+2016}.
Therefore, given the observational and chemical constraints in cloud cores on the microphysical parameters $\zeta_{\rm CR}$ and the minimum grain sizes, 
the formation of circumstellar disks in low-mass protostellar cores should be relatively universal.

Although the microscopic physics which determines the magnetic diffusion rate, and hence disk evolution, looks complicated,
there is a simple essence that determines disk size evolution
under a strong magnetic field with magnetic diffusion,
which is the balance of coupling and decoupling between the magnetic field and the gas in the disk.
\citet{2016ApJ...830L...8H} showed that ambipolar diffusion limits the disk size to a few tens of au during Class 0 phase  by assuming that the generation timescale of toroidal magnetic field equals to the ambipolar diffusion timescale and the magnetic braking timescale equals to the rotation timescale.
This analytic estimate reproduces well in many simulations of disk evolution during the Class 0 phase \citep{2020A&A...635A..67H,mignon:21a}.

The microscopic origin of ambipolar diffusion is ion-neutral drift.
In other words, if ambipolar diffusion works effectively, a velocity difference between ions \citep[e.g., H$^{3+}$, HCO$^+$,][]{2012ApJ...754....6T,2018MNRAS.478.2723Z} and neutral gases (e.g., CO) is expected to occur.
Thus, observations of ion-neutral drift are essential to test the importance of ambipolar diffusion.
Theoretical studies have suggested that the velocity difference is possibly observable \citep{2018MNRAS.473.4868Z,2020ApJ...896..158T, 2020ApJ...900..180M} \yt{in the disk and inner envelope (i.e., $100$ to $1000$ au scale)} of late Class 0 to Class I YSOs.
Although recent attempts to observe ion-neutral drift have resulted
in non-detection for B335 \citep{2018A&A...615A..58Y},
possibly because it is too young \citep[as shown in][ion-neutral drift is more likely observable in more evolved objects]{2020ApJ...896..158T}, or because it is highly ionized. Future observational attempts to detect ion-neutral drift are extremely important.

\subsubsection{Hall effect}
Unlike other non-ideal effects, which are just magnetic diffusion,
the Hall effect has a unique feature.
The Hall effect can actively induce a clockwise rotation  
(when $\eta_{\rm Hall}<0$) around a magnetic field
by generating a toroidal magnetic field from the poloidal magnetic field
\citep{1999MNRAS.303..239W,2012MNRAS.427.3188B,2012MNRAS.422..261B}.
\yt{
During gravitational collapse, the magnetic field 
is dragged toward the center and
an hourglass-shaped magnetic field structure is formed. 
At the ``neck" of the hourglass-shaped magnetic field,
a toroidal current exists.
Then, the Hall effect drags the
magnetic field toward the azimuthal direction as if the gas rotates with the velocity $\vel_{\rm H}=-\eta_{\rm Hall} (\nabla \times \magB)/B$
(see equation (\ref{Eq:induct})).
The generated toroidal magnetic field exerts a 
toroidal magnetic tension on the gas and induces the rotation.
}
The combination of the rotation induced by the Hall effect
and the inherent rotation of the cloud core sometimes
assists the disk size growth, and causes various
interesting velocity structures in YSOs.

\citet{2011ApJ...733...54K} used two-dimensional simulations to investigate the impact of the Hall effect in the context of disk formation for the first time.
They focused on the dynamical behavior induced by the Hall effect and showed that a circumstellar disk with a size of $r\gtrsim 10$ au can be formed 
with sufficiently high $\eta_{\rm Hall}$. Another finding of \citet{2011ApJ...733...54K} is the formation of a counter-rotating envelope opposite to the disk rotation.
\yt{
  As a back reaction to the Hall effect induced rotation in the midplane,
  angular momentum conservation causes the upper layer
  to rotate in the opposite direction of the induced rotation.
  As a result, a counter-rotating (upper) envelope emerges.
  }
\citet{2011ApJ...738..180L} investigated the effect of the Hall effect in two-dimensional simulations
that included all the non-ideal MHD effects using realistic magnetic diffusion coefficients.
They confirmed that Hall-induced rotation occurs even when other non-ideal effects are considered.
\yt{They also showed the formation of the counter-rotating region opposite
  to the initial rotation in the upper region of the envelope at $\sim 100$ au.}

\citet{2015ApJ...810L..26T} conducted three-dimensional simulations of disk formation which included all the non-ideal effects as well as a radiative transfer for the first time.
They followed the evolution until shortly after the formation of the protostar.
They showed that a disk $\sim 20$ au in size is formed at the protostellar formation epoch when the magnetic field and initial rotation vector of the core are anti-parallel (in this case, the Hall induced rotation and initial rotation has the same direction).
On the other hand, a disk $1$ au in size formed with the parallel configuration
(in this case, the Hall induced rotation and initial rotation has the opposite direction).
\yt{Because the Hall effect strengthens or weakens magnetic braking depending on whether the angle between the magnetic field and the angular momentum of the cloud core is acute or obtuse,} they suggest that disk size could be bimodal in the Class 0 phase with the Hall effect.

More recently, \citet{2016MNRAS.457.1037W} investigated the impact of the Hall effect several thousand years after the formation of the protostar
with three-dimensional simulations.
They also confirmed that disk size could be bimodal. 
\citet{2018A&A...619A..37M,2019A&A...631A..66M} also reproduced the evolution induced by the Hall effect described above.
On the other hand, they pointed out that the Hall effect is difficult to handle numerically,
and that it is necessary to pay attention to whether the conservation of angular momentum is satisfied in the implementation.
\citet{2017PASJ...69...95T} investigated the impact of the Hall effect in the misaligned core.
They showed that, in a core with an acute angle misalignment between the magnetic field and the angular momentum,
the disk size growth is suppressed, while in a core with an obtuse angle, the disk size growth is promoted.

\citet{2020MNRAS.492.3375Z,2021MNRAS.505.5142Z} investigated the impact of dust models on the Hall effect and the resulting evolution of the disk.
When Hall effect is included, the upper limit of $\zeta_{\rm CR}$ for disk formation increases to a few times $10^{-16}$~s$^{-1}$, matching the upper range of the estimated $\zeta_{\rm CR}$ in cloud cores. 
They also showed that the Hall effect  could flip the rotation direction at $\sim 100$ au  from initial forward rotation to the counter rotation, making a counter-rotating disk.
This kind of counter-rotating disk has recently been observed \citet{2018ApJ...865...51T}.
If future observations reveal that counter-rotating structures are relatively common, it will be evidence of the importance of the Hall effect.

\subsubsection{Disappearance of the envelope}
As discussed in \S \ref{ambi_sec}, the balance between magnetic braking and magnetic diffusion essentially limits the size of the disk to a few tens of au \am{when a relatively massive protostellar envelope is present. However, this envelope will progressively be accreted and dissipated while the star is forming}.

\begin{figure}[ht]
\includegraphics[width=75mm]{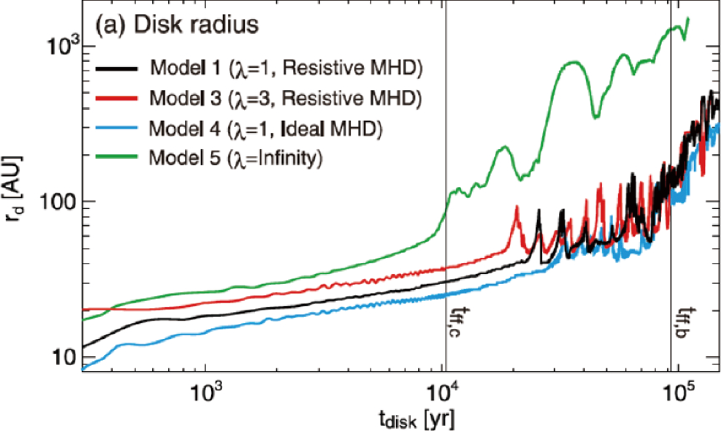}
\caption{
  Time evolution of disk radius of  non-ideal MHD simulations
  with $\mu=1$ (black), $\mu=3$ (red), ideal MHD simulation with $\mu=1$ (blue),
  and hydrodynamics simulation (green) against the elapsed time after the circumstellar disk formation where $\mu$ is the mass-to-flux ratio normalized by its critical value.
  The free-fall timescale at the center of the cloud, $t_{\rm ff, c}$, and at the cloud boundary, $t_{\rm ff, b}$ ($t_{\rm ff, b}$ roughly corresponds to the epoch at which the disk mass
  becomes larger than the envelope mass), are plotted in each panel (from \citet{2011PASJ...63..555M}).
}
\label{fig_disk_size}
\end{figure}

Magnetic braking is a process that transfers the angular momentum from the disk to the envelope.
Thus, once the envelope loses its ability to receive angular momentum, the magnetic braking \am{becomes less efficient} \citep{2011PASJ...63..555M, 2020A&A...635A..67H}.
The long-term simulations of \citet{2011PASJ...63..555M}, which cover from the prestellar core to the end of Class I phase
($10^5$ yr after protostar formation) \am{have provided some predictions on how the disk sizes evolve during the late embedded stages, as a consequence of the envelope losing an order of magnitude of mass.} 
Figure \ref{fig_disk_size} shows the long-term evolution of disk size
($t\gtrsim 10^5$ yr after protostar formation) in ideal and non-ideal MHD simulations.
In the early stage of protostar formation ($t<10^4$ yr), in which the envelope is massive,
the disk size is only about 10 au.  
On the other hand, as the system evolves to $t>10^4$ yr, which corresponds
to the epoch when the envelope mass becomes smaller than that of protostar and disk,
the disk size begins to increase and reaches  $\sim 100$ au.
This disk size evolution process may explain the $100$ au scale disks in some Class 0/I YSOs even with a relatively strong magnetic field in the natal cloud cores. 

\subsection{Similarities between low- and high-mass star formation}

Young massive protostars differ from their low-mass counterparts essentially because of their intense luminosity throughout their evolution. By chronological order, the accretion luminosity \citep[up to stellar masses of $\simeq 10 \msun$, e.g.,][]{hosokawa:09}, the internal luminosity and eventually the ionising radiation \citep[from mass $>30  \msun$][]{kuiper:18} shape the evolution, the accretion and the ejection of young massive protostars. It is has been shown that magnetic fields  regulate disk and outflow formation within massive protostars as well \citep[e.g.,][]{banerjee:06,2013MNRAS.432.3320S}. It is thus of prime importance to estimate the relative balance of the radiative and magnetic fields effects throughout the early evolution of young massive protostars in order to characterise the accretion and ejections processes. Since last PP VI, a lot of progress has been made in order to study the combined effects of magnetic fields and radiative feedback on the formation of disks and outflows in young massive protostars. In the following, we review state-of-the-art simulations, focusing on the small scales of collapsing massive dense core (typically  $<1000$ au) leading to the formation of isolated star-disk-outflow systems. \bc{We do not review the recent work focusing on the large scale fragmentation of collapsing massive dense cores as this goes beyond the scope of this chapter.}

\subsubsection{Disk formation and properties}

Various processes that shape disk formation in young protostars have been investigated in the recent years. In the first class of models, magnetic fields have been neglected. Rather, these models focus on the accurate treatment of the radiative protostellar feedback, using hybrid irradiation schemes \citep{kuiper:10,klassen:14,rosen:17,mignon:20}. \cite{rosen:16} and \cite{rosen:19} report 3D radiation-hydrodynamics models of the collapse of $150~\msun$  massive dense cores. They show that accretion disks form regardless of the core's initial virial state and that the disk supplies material to the star, especially at late times ($M_\star>25 \msun$). The typical disk they report has a radius $<1000$~AU and  eventually becomes gravitationally unstable, leading to disk fragmentation at late times. Similar results on  disk properties have been reported in \cite{klassen:16} and \cite{mignon:20} for clouds without initial turbulence, except that the later studies do not report disk fragmentation. This discrepancy might originate from the choice of initial rotation level as well as of the numerical resolution \citep{meyer:18}. Lastly, \cite{kuiper:18} also report disk formation and late evolution in 2D axisymetric  models including photoionization. In this case, the disk gets destroyed in the late evolutionary stages as the outflow broadens when photoionization starts operating. 

The picture drawn above changes dramatically once magnetic fields are taken into account. In the first series of studies, ideal MHD has been introduced. As with low-mass protostars, magnetic braking is operating and disk formation is suppressed in high magnetization \citep[$\mu<10$, ][]{seifried:11}. With initial turbulence, \cite{2013MNRAS.432.3320S,myers:13} report the formation of 50-150 AU Keplerian disks. Similar to low-mass star formation, magnetic fields greatly affect the disk size. However, these models suffer from  convergence issues; in particular regarding  the disk properties, since diffusion is mainly controlled by numerical resolution (see \S \ref{sec:turb}). In the past five years, the non-ideal MHD framework developed for low-mass star formation has then been applied to massive protostars. \cite{matsushita:17} present 3D simulations of collapsing massive dense cores in the aligned rotator configuration  \bc{(the angular momentum of the parent cloud core is aligned with its initial magnetic field)} which include the effects of Ohmic resistivity and mimic the gas thermal behaviour with a barotropic EOS. Disks form in all cases, with sizes 50-300 AU and masses $<10 \msun$-- their exact properties depend on the initial conditions (mass, magnetisation) as well as on the resistivity (larger disks form with larger resistivity). Only the most massive case, with an accretion rate $\dot{M}>10^{-2}~\msun~\rm yr^{-1}$ exhibits fragmentation. Similarly, \cite{kolligan:18} report the formation of 100-1000 AU disks in 2D axisymmetric  simulations of the collapse of $100~\msun$ dense cores using Ohmic diffusion and an isothermal equation of state. In these models, the disks do not fragment. Given the differences between the numerical methods and setups of the two latter studies, it is difficult to draw a firm conclusion on the effect of Ohmic diffusion on disk fragmentation around massive protostars.

More recently, \cite{mignon:21a} and \citet{commercon:21} present $100~\msun$ dense core collapse simulations including the effect of radiative transfer (continuum and protostellar feedback) and magnetic fields with ambipolar diffusion. \cite{commercon:21} perform a comparison between hydrodynamical, ideal MHD, and ambipolar diffusion cases, and report similar properties regarding disk formation and early evolution as for low-mass star formation in the aligned rotator case. Disk fragmentation occurs only in the hydrodynamical case.  The disk  formed in the non-ideal MHD case is characterized by plasma $\beta>1$, Keplerian rotation,  and has essentially a vertical magnetic field in the inner regions ($R < 200$ AU),  while the toroidal magnetic pressure dominates in ideal MHD with $\beta<1$. Regarding magnetic field evolution, the central accumulation of magnetic flux gets limited when ambipolar diffusion operates, with a plateau similar to the one found in the low-mass regime \citep{2016A&A...587A..32M}, but with a larger amplitude around 1~G \citep{commercon:21}. Finally, the disk radius evolution is consistent with the ambipolar diffusion regulated scenario \citep{2016ApJ...830L...8H}.  \cite{mignon:21a} further confirm the disk properties regulated by ambipolar diffusion in similar models but include initial turbulence and a more accurate radiative transfer scheme for protostellar feedback \citep[hybrid irradiation instead of grey FLD,][]{mignon:20}. In the case of initial super-Alfv\'enic turbulence, \cite{mignon:21a} report disk fragmentation events that lead to the formation of stable binary systems with separation $300 - 700$~AU. Interestingly, individual disks form around secondary fragments, whose properties are consistent with magnetic regulation  as well.

\subsubsection{Outflow mechanisms}

Outflows are ubiquitous in star forming regions. Contrary to the case of low-mass star formation, various mechanisms can be at play to accelerate outflowing gas in young massive protostars: magnetic fields \citep[e.g.,][]{banerjee:07}, radiative force \citep[e.g.,][]{yorke:02} and ionization \citep[e.g.,][]{dale:05}. While the radiative force and the ionization are regulated by the luminosity emerging from the young massive protostar that has already entered the Pre-Main Sequence evolution ($M_\star \gtrapprox 20 \msun$), the magnetic outflows are expected to be launched earlier. \cite{banerjee:06} have shown in  ideal MHD models that magnetic outflows are launched in massive protostars similarly to low-mass case. Interestingly, these outflows produce cavities out of which radiation pressure can be released \citep{krumholz:05}. \cite{2012MNRAS.422..347S} studied in more detail the mechanisms at the origin of magnetic outflows and jets (magneto-centrifugal acceleration versus magnetic pressure gradients, see section \ref{Outflow_Theory}). They report that both mechanisms contribute to the acceleration of the outflowing material:  the  outflows are launched from the disks via centrifugal acceleration and then the toroidal magnetic pressure progressively  contributes to the further acceleration  away from the disks. These ideal MHD models have been revised in recent works which include resistive effects. \cite{matsushita:17} report that the outflow is driven by the outer disk region, i.e., where the ionization is sufficient  for the magnetic field and neutral gas to be well coupled. They also measure a ratio between the mass outflow rate and the mass accretion rate that is nearly constant, from 10 to 50$\%$, throughout the stellar mass spectrum they consider ($30-1500~\msun$). \cite{commercon:21} and \cite{mignon:21b} report the first 3D models of outflow formation which account for both radiative feedback and ambipolar diffusion.  They show that magnetic processes dominate the outflow launching at early stages up to masses $\simeq 20~\msun$. \cite{mignon:21b} report similar results to \cite{2012MNRAS.422..347S} on the origin of the magnetic outflows: both the centrifugal acceleration and the magnetic pressure gradients contribute to the outflow. Interestingly, \cite{mignon:21b} find that initial turbulence greatly affects the outflow launching and report monopolar outflows in their most (supersonic) turbulent case. Lastly, outflows are launched perpendicularly to the disk, but with no correlation to the parent cloud magnetic field orientation.

Nonetheless, further work is required in order to remark on the origin of magnetic outflows and their impact on massive star formation since both numerical resolution and time integration remain insufficient. From their 2D resistive models, \cite{kolligan:18} conclude that sub-au resolution is required to obtain converged results on the magneto-centrifugal jets.  Additionally, further time integration is needed in order to remark about the fate of the magnetic acceleration once the the pressure and later ionization begin to dominate the acceleration ($M_\star > 20 \msun$).

Altogether, these results strongly suggest that the same mechanisms are at play in the prostostellar disk and outflow formation process for both low- and high-mass star formation. The scenario we present in this chapter might thus also apply to the early evolution of young massive protostars. 


\section{\textbf{
\yt{Dust evolution: a key feature to interpret observed disk properties and MHD models}}}
\label{sec:dust}


\subsection{The ``Missing mass problem in Class 0/I YSOs"}
\label{missing_mass_problem}
As we have described in previous sections, \am{the recent theoretical studies predict disk sizes in relatively} good agreement with the \am{observed sizes of embedded disks.}
However, there is still a discrepancy between the observations and \am{predictions from the} theoretical studies. \am{The most salient issue is the one of disk masses. Observational estimates of embedded disk masses find typical median gas masses $0.001-0.01 \msun$ (see the detailed discussion in \S \ref{ssec:obs_masses}). Taken at face value, Class 0/I disks are thus on average an order of magnitude lighter than that suggested from non-ideal MHD simulations.}
\yt{
In this section, we discuss the recent studies regarding disk masses and the discrepancy between observations and simulations. We explore one possible solution to this discrepancy: dust evolution and its impact on both disk evolution models and their comparison to observations.}

\begin{figure}[ht]
\includegraphics[width=70mm]{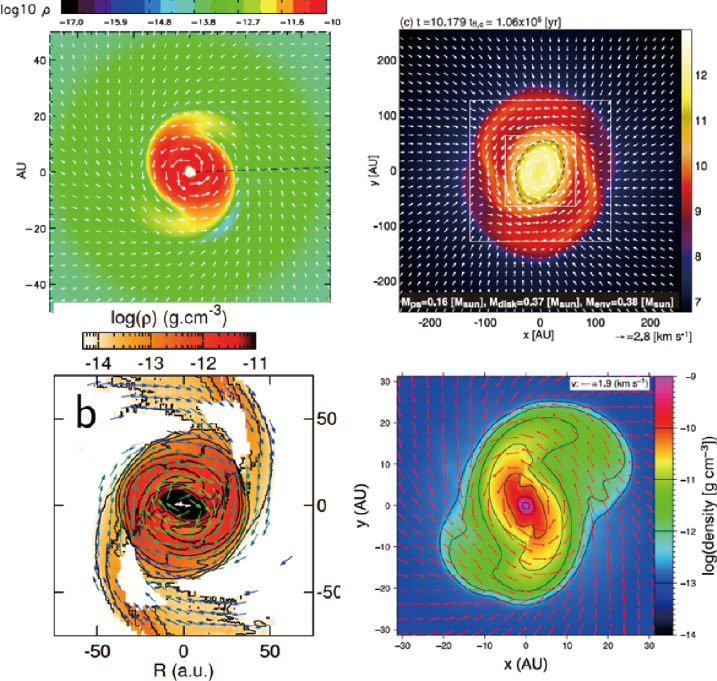}
\caption{
  Examples of the disks formed in non-ideal MHD simulations
  \citep[from][]{2011PASJ...63..555M,2015ApJ...810L..26T,2016A&A...587A..32M,2018MNRAS.473.4868Z}.
  The spiral arms are caused by gravitational instability.
  }
\label{fig_GI_disk}
\end{figure}

Non-ideal MHD simulations have shown that, even with a relatively strong magnetic field,
disks tend to become massive and gravitationally unstable (or Toomre's $Q$ value is $Q\sim 1$)
\citep{2011PASJ...63..555M,2015ApJ...810L..26T,2015MNRAS.452..278T, 2015ApJ...801..117T, 2016A&A...587A..32M,2016MNRAS.457.1037W, 2015ApJ...810L..26T,2015MNRAS.452..278T, 2017ApJ...846....7K, 2017ApJ...835L..11T,2018MNRAS.473.4868Z,2019ApJ...876..149M,2021MNRAS.502.4911X}.
Figure \ref{fig_GI_disk} shows examples of simulation results in which the gravitationally unstable disk is formed.
As shown in this figure, the disk becomes massive and spiral arms induced by gravitational instability form.
The simulations employed different numerical methods/codes, inner boundaries, and initial conditions, yet obtained consistent results. Note that the spiral arms are transient and gravitationally unstable disks are not always non-axisymmetric \citep[][]{2017ApJ...835L..11T,2017ApJ...838..151T}.
Thus, whether a disk is massive or not cannot be determined strictly from the fact that its morphology is axisymmetric \citep[although non-axisymmetric structure, which can be interpreted as the result of gravitational instability, have been observationally found in e.g., Elias 2-27][]{2016Sci...353.1519P,2017ApJ...835L..11T, 2021ApJ...914...88P}, and obtaining a mass estimate from the radiative flux is necessary.

The formation of massive disks is also supported by the formation scenario
that disks are formed from first cores. As mentioned above, most of gas in the first core
does not directly accrete to the protostar, but transforms to a disk around the protostar.
Since the mass of the first core ($\sim 10^{-2}$ to $10^{-1} \msun$) is about 10 times
larger than the mass of the protostar at its formation
epoch $\sim 10^{-3}$ to $10^{-2} \msun$, the disk is expected to
be very massive immediately after the formation of the protostar
\citep{2010ApJ...718L..58I,2012PTEP.2012aA307I}.

\begin{figure}[ht]
  \includegraphics[width=70mm]{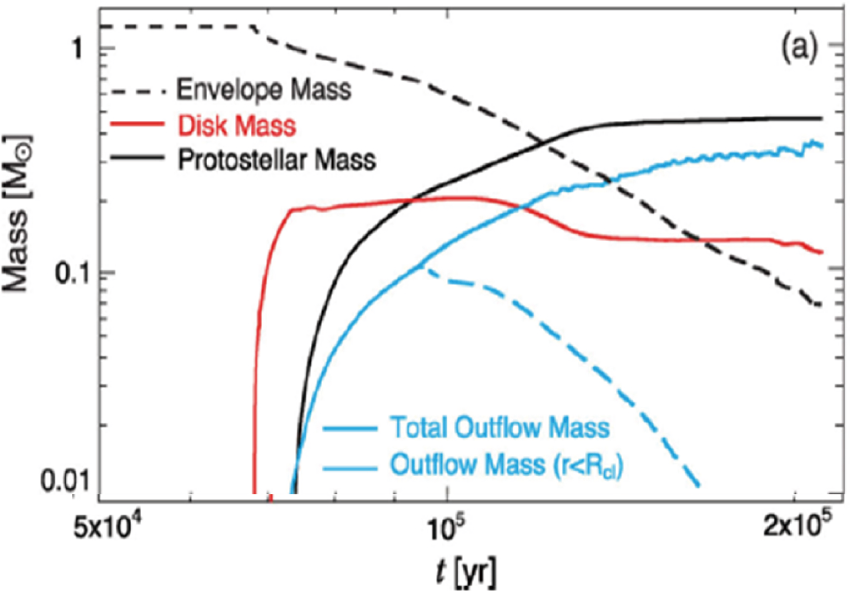}
  \caption{
    Long term evolution ($\gtrsim 10^5$ yr after the formation of the protostar) of mass in disks, protostars, and envelopes \citep[from][]{2013MNRAS.431.1719M}.
    The red, solid black, dashed black, and blue lines represent the mass of the disk, protostar, envelope, and outflow, respectively.
}
\label{fig_disk_mass_evoluiton}
\end{figure}

The massive disk seems to not be short-lived (e.g., $\lesssim 10^4$ yr).
\citet{2011PASJ...63..555M} and \citet{2013MNRAS.431.1719M} investigated very long-term evolution of circumstellar disks
(until $10^5$ years after the formation of the protostar, corresponding to the end of Class 0 to in Class I phase)
and showed that gravitationally unstable disks may frequently form even in strongly magnetized
cloud cores once it grows to several $10$ au.
Figure \ref{fig_disk_mass_evoluiton} shows the long-term ($t\sim 10^5$ yr) mass evolution of the protostar, disk, and envelope.
The figure shows that the disk has relatively large mass ($\gtrsim 0.1 M_\odot$) even $\sim 10^5$ yr after the formation of the protostar.
Therefore, from a theoretical point of view,  it is expected that gravitationally unstable disks \am{should be frequently observed} in the Class 0/I phase.
Thus, there is a discrepancy between disk masses obtained from simulations and those from observations, 
dubbed the ``Missing mass problem in the Class 0/I phase".



A simple theoretical estimate is useful to understand why the disk mass in the simulations tends to be so large. 
Considering the current estimates of protostellar lifetimes ($\lesssim 10^6$ yr), large mass accretion rates of $\dot{M}_{\rm disk} \sim 10^{-6} \msunyear$ in the disk, which is also similar value to the accretion rates estimated from observations \citep[e.g., ][]{2014prpl.conf..195D,2020A&ARv..28....1L}, on average are required to grow the protostar to $\sim 1 \msun$ in the required timescale.
\yt{Using this fact and considering a viscous accretion disk model \citep{1973A&A....24..337S},
one can estimate the time and vertically averaged viscous $\alpha$ value of disk \citep[see for more details,][]{2017ApJ...838..151T}}, 
\begin{eqnarray}
\label{alpha_Q_relation2}
 \alpha = 1.2 \dot{M}_{\rm disk,3 \times 10^{-6} \msunyear}Q_{10} T_{30 \rm K}^{-3/2},
\end{eqnarray}
where we assume $c_{\rm s}=\sqrt{k_{\rm B}T/m_{\rm g}}=1.9 \times10^4~T_{10 \rm K}^{1/2} \cms$.
The equation (\ref{alpha_Q_relation2}) shows that there is a trade-off between $\alpha$  and $Q$ of the disk.
For a given mass accretion rate, a large $\alpha$ is required to achieve a large $Q$ value (or small disk mass).

This estimate shows that an extremely efficient angular momentum transport of $\alpha \sim 1$ \yt{on average} is necessary \yt{over the entire period of evolution } to achieve  $\dot{M}_{\rm disk} \sim 10^{-6} \msunyear$ in a disk with $Q \sim 10$.
Note that mass supplied by the envelope accretion to disk
is on the order of  $\dot{M}_{\rm envelope} \sim 10^{-6} \msunyear$ in Class 0/I YSOs, and that
$\dot{M}_{\rm disk} \sim 10^{-6} \msunyear$ is also necessary to keep the disk mass small, otherwise mass accumulates in the disk and its mass inevitably increases.
However, such a highly efficient mechanism of angular momentum transport is not realized in the above-mentioned simulations because of non-ideal MHD effects.
As a result, $Q$ tends to become small ($Q\sim 1$) in the simulations to reduce $\alpha$ to the realistic level ($\alpha \sim 0.1$).
\yt{
  Note that this estimate neglects the temporal variation of the mass accretion rate (episodic accretion),
  which may lead to realistic $\alpha$ even for low-mass disks during quiescent periods, but requires ultra-efficient angular momentum transport during burst periods.
  }

A possible mechanism that has been overlooked in previous theoretical studies 
and that can significantly change the mass evolution of the disk is the effect of dust evolution on the magnetic resistivity.
\am{Most numerical simulations of disk evolution consider sub-micron-sized dust to compute the resistivities: this results in highly efficient Ohmic dissipation and ambipolar diffusion in the disk. However, if the dust grains are significantly larger than the typical ISM dust size, the magnetic resistivity can become many orders of magnitude smaller, causing the recoupling between the magnetic field and the gas inside the disk.} 

\yt{Note also that disk mass in the simulations is sensitive to the treatment of inner boundary or sink \citet{2014MNRAS.438.2278M,2016MNRAS.463.4246M,2020A&A...635A..67H,2021MNRAS.502.4911X}, and 
there is a study which reported formation of a long-lived lightweight disk with the sink radius of $\sim 4$ au \citep{2021A&A...648A.101L}. 
Thus, further research on appropriate inner boundary conditions is needed.}

\yt{On the other hand, disk mass estimates from observations of dust thermal emission also include a number of assumptions on dust properties subjecting them to large errors (such as the dust opacity and the dust-to-gas mass ratio, see \S \ref{ssec:obs_disks} for more detail).
We stress that an issue with dust properties may hamper our understanding of Class II disk masses, for which observations suggest the dust mass is smaller than the Minimum Mass for Solar Nebula and is insufficient for planets to form (this is so-called ``missing mass problem", see the PPVII chapter of Miotello {\it et al.}, and discussion in \S \ref{ssec:obs_masses})}. 

Note also that the disk mass estimate from CO, which is another widely used tracer of disk mass, also suffers from possible CO depletion and uncertainty in the CO-to-H$_2$ ratio \citep[e.g., ][]{1996ApJ...467..684A, 2001ApJ...561.1074T,2016ApJ...833..105Y,2018ApJ...868L..37D,2019ApJ...878..116P}, which is highlighted by the discrepancy between the mass estimate using HD 
and using CO \citep{2013Natur.493..644B,2013ApJ...776L..38F, 2016A&A...592A..83K, 2016ApJ...831..167M}.

In the following \S \ref{sec:dust_obs}, \am{we describe the tantalizing thread of observational evidence that may suggest that the dust properties in protostellar envelopes and disks differ significantly from the assumed dust properties implemented in models, possibly because of dust evolution.} 
We then describe more quantitatively the important consequences caused by dust growth, in \S \ref{dust_theory}.

\subsection{Observations of dust grain properties in embedded protostars}
\label{sec:dust_obs}

In the last decade, observations have attempted to put constraints on dust properties in star-forming structures by measuring the spectral indices of the thermal dust emission at (sub-)millimeter wavelengths  \citep{2012ApJ...751...28M,2016ApJ...826...95C,2016A&A...588A..30S,2017A&A...606A.142C}. \am{In the regime of optically thin dust emission, and in the Rayleigh-Jeans regime (high dust temperatures compared to $h\nu/k$), the dust emissivity scales as a power law with frequency and thus the flux density of the thermal dust emission $F(\nu)$ depends on the frequency as $\nu^{(2+\beta)}$, with $\beta$ the dust emissivity. Dual-frequency observations in the submm and mm regimes are thus routinely used to measure the dust emissivity $\beta$}. Observational studies show that star-forming clouds and cores present significant variations in 
$\beta$ at $\sim 0.1$ pc scales. Moreover, a fair fraction of the dense material involved in the star formation process exhibit $\beta$ values much lower than the typical values of $\sim 1.7$ observed in the diffuse ISM, although the flattening of the spectral energy distribution may depend on the choice of wavelengths used to measure the spectral index underlying the determination of $\beta$. Studies in the infrared also suggest that \am{dust grains are significantly larger than the typical $0.1\mu$m sizes  in the ISM, with indications of} micrometer-size grains in the outer layers of molecular cores \citep{2015A&A...582A..70S,2021A&A...647A.109S}.
At the other end of the star formation process, observations of T-Tauri circumstellar disks reveal $\beta$ from 2 to 0, compatible with the presence of millimeter grains (see e.\,g., the review by \citealt[][]{2014prpl.conf..339T} and the recent survey by \citealt[][]{2021MNRAS.506.5117T}). ALMA observations suggest that some 1-2 Myr old disks may already be hosting planets \citep{2016ApJ...818...76J, 2018A&A...618L...3M, 2018ApJ...866L...6C}, which would mean that the formation of planetesimal seeds already occurs in the first million years after the onset of collapse, while the disk and the star are being built concomitantly. 
Observations have also suggested that partial grain growth up to sizes 10-100 \mic\ could have already occurred in a few objects during the Class I phase \citep[see e.g.][]{2014A&A...567A..32M,2018NatAs...2..646H,2020ApJ...889...20L,2021A&A...646A..18Z}. 

\begin{figure*}[!h]
\begin{center}
\includegraphics[width=0.35\linewidth]{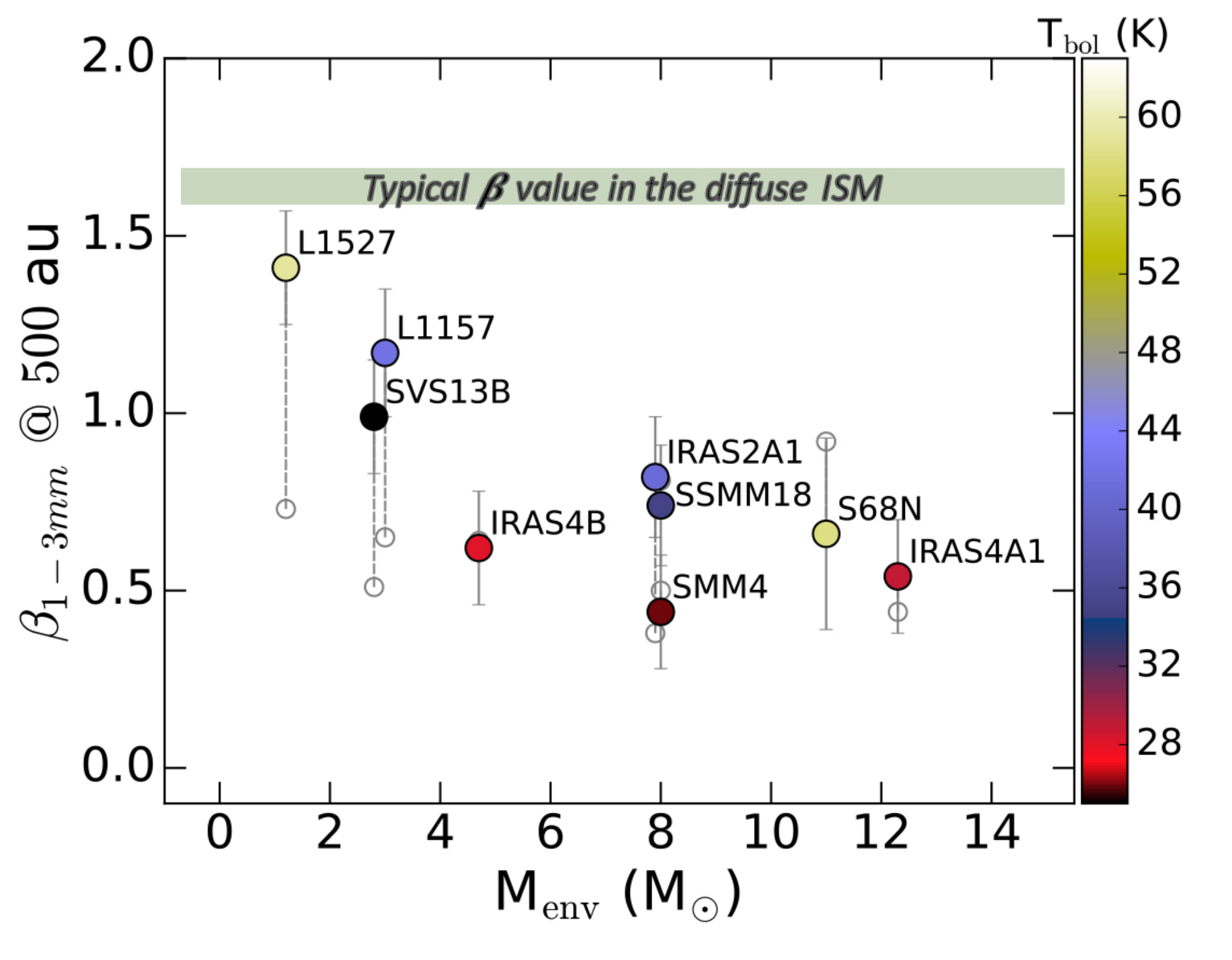}
\hspace{0.1cm}
\includegraphics*[width=0.45\linewidth]{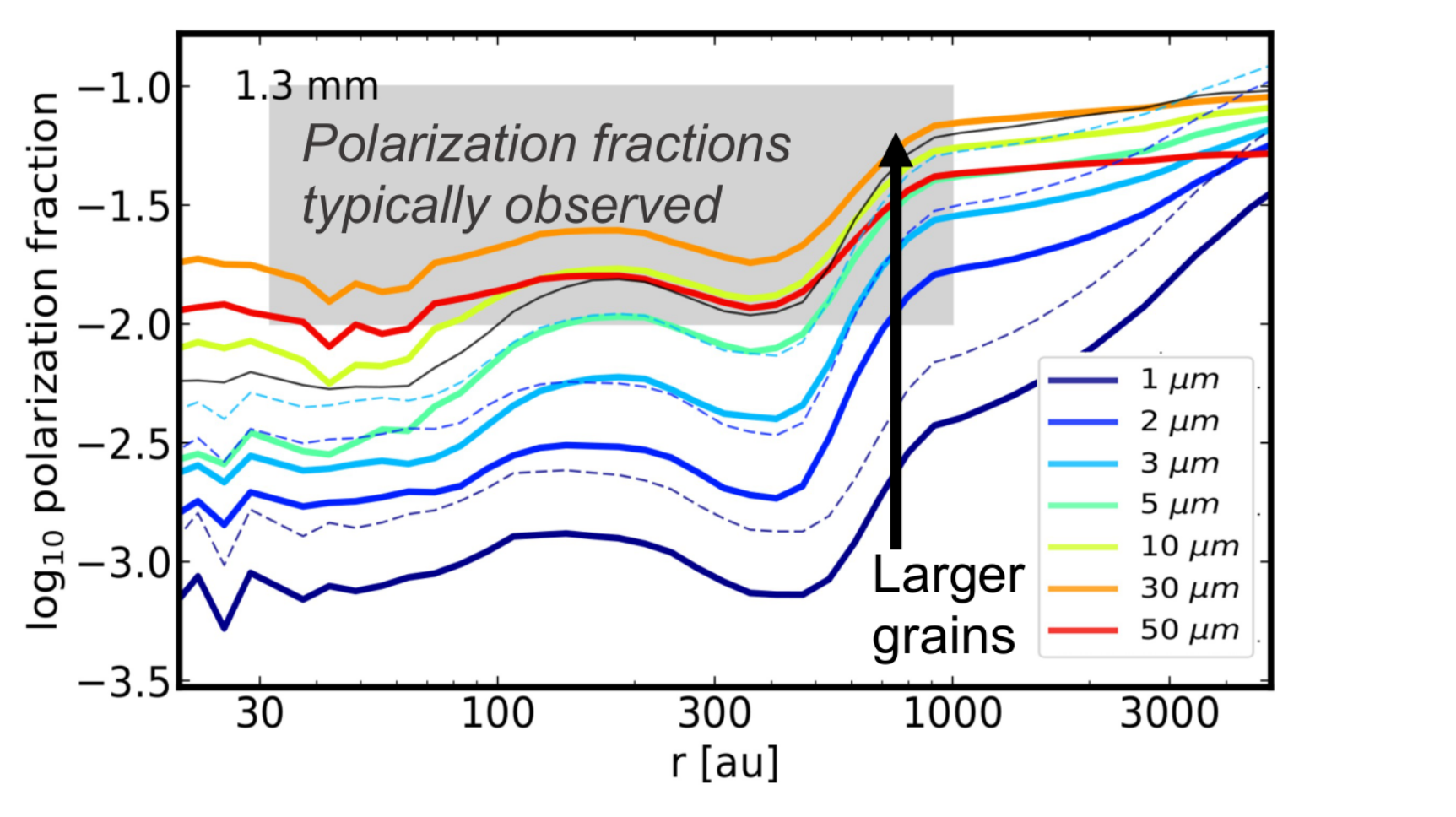}
\end{center}
\vspace{-1.0cm}
\caption{Left: Observed millimeter dust emissivity index in a sample of young protostars \citep{2019A&A...632A...5G}. All protostellar envelopes probed at radii $\sim 500$ au show lower $\beta$ values than the progenitor diffuse ISM. Right: Radial profile of the polarization fraction from the $1.3~\mathrm{mm}$ dust emission predicted from \am{magnetically aligned grains (RATs)} in a typical low-mass protostellar envelope \citep{2019MNRAS.488.4897V}. Large ($>10 \,\mu$m) dust grains are required to match the observations of the polarized emission (grey area).}
\label{fig:galametz_valdivia_dust}
\end{figure*}

We have shown, earlier in this review chapter, that the progenitors of planet-forming disks form at the Class 0 stage, but start very compact in size. Thus, Class 0 protostars are ideal candidates to study the pristine properties of the dust grains building these disks. Recent studies analyzing the dust properties in Class 0 protostars at different scales have suggested that the grain properties may already be significantly different from typical ISM dust, in these objects less than a million year old.
At the small radii probed by interferometric observations where the dust emission is a combination of envelope and disk emission, \citet{2020A&A...640A..19T} find a mean $\beta \sim 0.4$ in the compact dusty components of Perseus protostars, believed to trace mostly the dust enclosed in young protostellar disks. Interferometric observations have also found low spectral indices at millimeter wavelengths in the dusty $\sim 60$ au disk surrounding L1527, a Class 0/I protostar and in the Class 0 protostar, L1157 \citep{2020ApJ...895L...2N, 2012ApJ...756..168C}. In the Orion region, recent ALMA observations confirmed the low spectral indices of dust emission in 16 protostars at scales $<100$ au \citep{2021arXiv210710743B}.  
Analyzing the dust properties of 12 nearby Class 0 protostars observed at 1.3 and 3.2 mm with the PdBI (CALYPSO survey), \citet{2019A&A...632A...5G} found that most Class 0 protostars, sampled from different star-forming regions, show significantly shallow dust emissivities at envelope radii $\sim 100-500$ au, with $\beta < 1$ (see \am{left panel of} Figure \ref{fig:galametz_valdivia_dust}). 
These observations have also uncovered a radial evolution of the dust emissivity, with many objects showing a decrease of $\beta$ toward small envelope radii. 
Their analysis of dust emissivity in the u-v domain limits the biases due to interferometric mapping reconstruction (i.e. filtering effects depending on wavelengths) and also rules out the contamination from small-scale protostellar disks in decreasing the dust emissivity index in inner envelopes-- these results hence suggest the dust contained in young protostellar envelopes has different emissivity than the typical dust observed in the diffuse ISM. 

A mixing of dust temperatures along the line of sight could artificially produce decrease in $\beta$, but the typical temperature range in embedded protostars at scales probed by these interferometric observations (envelope radii 50-1000 au) is large enough to produce such a significant flattening of the spectral index at millimeter wavelengths. An origin of these low spectral indices due to different grain composition than typical ISM grains is another possibility, although it is very challenging for models to significantly change the grain composition during the $\sim$1 Myr that the prestellar phase lasts. 
However, such low emissivities are expected from models of fluffy aggregates and grain coagulation in dense clouds:
recent studies on various interstellar dust analogues suggest that mm $\beta$ values $<$1, like those observed in these young protostars, can only be produced by grains larger than 100 \mic\ \citep{2019A&A...631A..88Y}. 
Significant grain growth in Class 0 envelopes are also consistent with the theoretical predictions of the polarized radiative transfer models \citep[][see left panel of Fig.\ref{fig:galametz_valdivia_dust}]{2019MNRAS.488.4897V}, showing large $>100$ \mic\ grains are  required to produce millimeter polarized dust emission at \am{polarization} fractions similar to those observed routinely in protostellar envelopes. 

If these low spectral indices measured in the millimeter wavelengths do indeed trace a population of large dust grains in the inner layers of protostellar envelopes, \am{such presence of large grains has many implications. Indeed, these embedded protostars are less than 0.1 Myr old, thus} the timescales to grow grains up to sizes $>1$ \mic\ in star-forming envelopes at typical gas densities $n_{\rm H_2} \sim 10^6 ~\rm cm^{-3}$ may need to be revised. \am{As an illustration, current models predict that at such gas densities, it takes a bit less than a Myr to grow grains up to a few microns  \citep{2009A&A...502..845O}. While considering porous dust grains may help in speeding the process, taken at face values the observations of millimeter size grains are not consistent with the dust evolution models.} If small grains are not only coupled to the gas but also to the magnetic field, a magnetized collapse may also help to segregate the grains with different charges, and ultimately shorten the grain coagulation timescale \citep{2020A&A...643A..17G}. A more likely explanation involves the high density conditions that reign in  \am{protostellar} disks, making them efficient forges to build up big dust grains that are re-injected in envelopes due to protostellar winds and outflows \citep{2016PASJ...68...67W,2017MNRAS.465.1089B, 2020A&A...641A.112L,2021ApJ...907...80O,2021ApJ...920L..35T}. \am{Such a scenario may also explain the apparent lack of millimeter grains in the L1544 prestellar core \citep{2017A&A...606A.142C}, since a disk would be required to build big grains in short timescales.}

Dust properties also 
heavily influence magnetic field coupling as the small grains are the main charge carriers, and their disappearance to form bigger grains significantly changes the efficiency of magnetic braking \citep{2020A&A...643A..17G,2021MNRAS.505.5142Z}. Moreover, the size of the dust grains have an effect on both the physical structure of the accretion shock and on the associated chemistry \citep{2007A&A...476..263G,2017ApJ...839...47M}-- the increase in grain size leads to weaker coupling between gas and dust, lower dust temperatures, and dust being less efficient at cooling the shock material.
Therefore, characterizing dust properties and early evolution in embedded protostars is crucial to understand the physics at work to form the end product, i.e. the stars, disks, and planets: it will undoubtedly be one of the main focuses of research in the field in the coming years. The following section describes some key consequences of the early dust evolution in protostellar environments, as expected from numerical models of protostellar formation.

\subsection{Impact of dust growth on magnetic field evolution and dust opacity:possible solutions to the missing mass problem}
\label{dust_theory}

As discussed in previous section, observations of the dust thermal emission in the Class 0/I YSOs  have shown the possible dust growth.
Theoretically, the \am{dust is also expected to grow in the disk of Class 0/I, with the growth timescale of several $10^3$ yr, which is about 100 times shorter than the age of Class 0/I YSOs} \citep[see, e.g.,][]{2012ApJ...752..106O,2017ApJ...838..151T}. 

\subsubsection{Impact of dust growth on charge state and resistivity}
Dust growth can change magnetic resistivities which are the fundamental parameters for the formation and evolution of disks.
\yt{
As we discussed above, the dust has two different roles in determining resistivities.
The first is that the small dust is
responsible for the conductivity.
The second is that it is the absorber of charged particles.}
\citet{2018MNRAS.473.4868Z} showed that the omission of tiny dust
increases the resistivity by decreasing dust 
conductivity, hence promoting disk growth (\S \ref{disk_evolution}).
\yt{However,
  when the dust grows to $a_{\rm d}>1 \mum$ and when maximum dust size determines the total surface area, the (absence of ) second role becomes important in determining the resistivities.
  }
In this case, dust growth reduces the total surface area
and the grains lose their ability to absorb ions and electrons. As a result, the dust growth is expected 
\yt{to change the resistivity to what it would be in the absence of dust. }
Here, we quantitatively investigate the effect of dust growth on the resistivities.

\begin{figure}[ht]
  \includegraphics[clip,trim=0mm 0mm 0mm 0mm,width=55mm,,angle=-90]{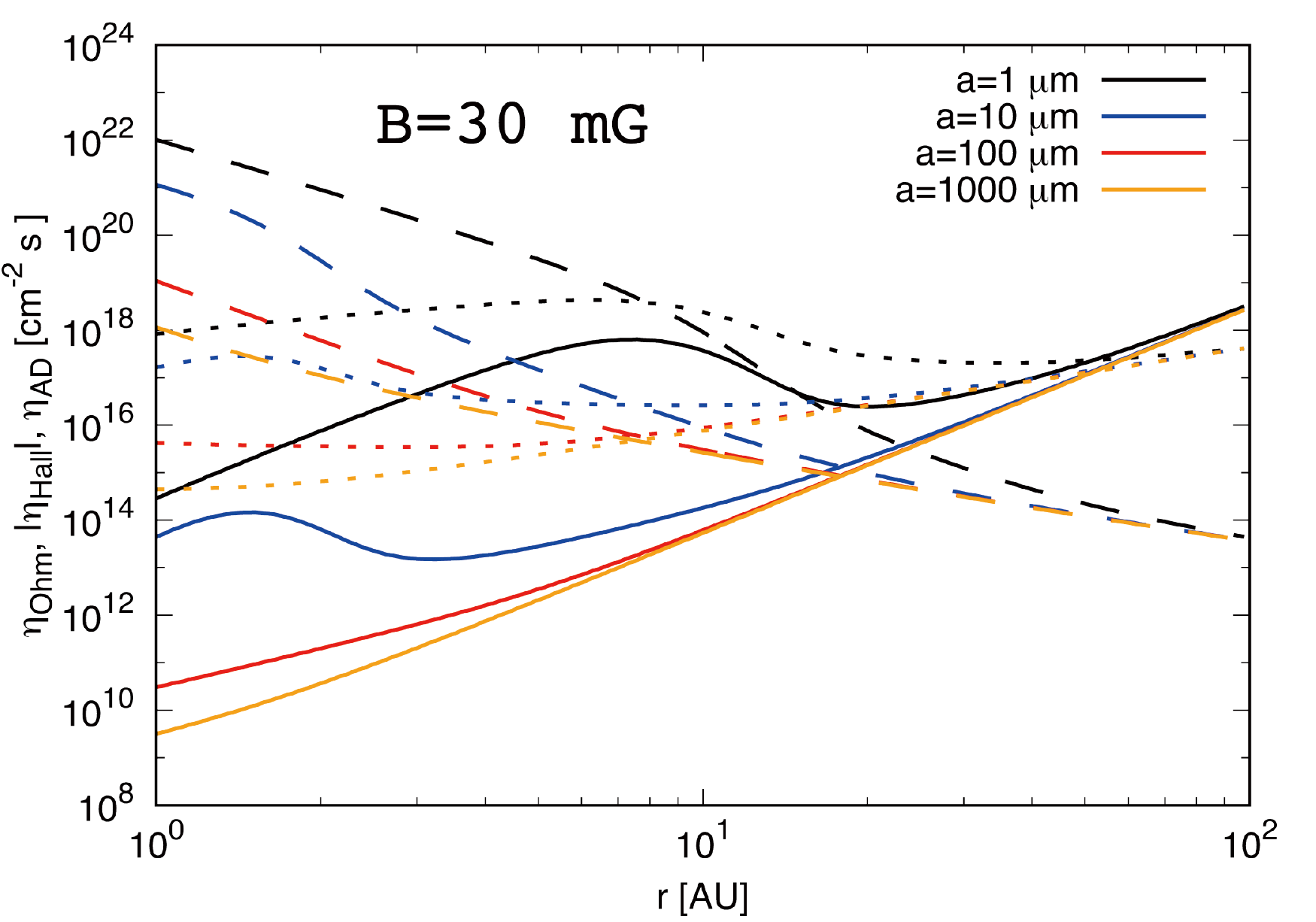}
  \caption{
    $\eta_{\rm AD}$ (solid), $\eta_{\rm Ohm}$ (dashed), and $\eta_{\rm Hall}$ (dotted) at the midplane of a disk.
    The black, blue, red, and orange lines show
    the resistivties of $a_d=1 \mum,~ 10 \mum,~ 100 \mum,$ and $ 1000 \mum$, respectively. 
    The disk surface density is assumed to be $\Sigma_g=45.6 ~r_{50 au}^{-12/7} {\rm g cm^{-2}}$,
    which corresponds to the gravitationally unstable disk with $M_*=0.3 \msun$.
    The disk temperature and midplane density are given as $T=150~  r^{-3/7}_{AU}~{\rm K}$ and $\rho_{g}=\Sigma(r)/H_{\rm scale}$, respectively,  where $H_{\rm scale}=c_s/\Omega$ is the disk scale height.
    The magnetic field strength is assumed to be $B=30~ {\rm m G}$. We consider the cosmic-ray ionization rate of $\xi_{\rm CR}=10^{-17}~ {\rm s^{-1}}$ as the only ionization source. The dust-to-gas mass ratio is fixed to be $10^{-2}$.
}
\label{rho_eta_large_dust}
\end{figure}

In Figure \ref{rho_eta_large_dust}, we calculate the resistivities with large (single sized) dust
at the midplane of a disk  \citep[using the method of][outlined in their Appendix A]{2021ApJ...913..148T}.
The figure shows that the resistivity in the disk
decreases as the dust size increases (especially in the inner dense region) and converges to a single power law
which solely determined by the equilibrium of cosmic-ray ionization and gas-phase recombination.
The interpretation of this result is straightforward; large dust of $a_{\rm d}\gtrsim 100 \mum$ does not contribute to the determination of the resistivity
(even though the dust is significantly charged for $a_{\rm d}>10 \mum$, which is properly calculated in our plot, for example, $\langle Z \rangle\sim -10^4$ for $1 \mm$ dusts at $T=100$ K).

Although the result is rather simple, its impact on disk evolution is possibly significant.
For example, at $r\sim 10$ au,  $\eta_{\rm Ohm}$ and $\eta_{\rm AD}$ are reduced
by a factor of $10^{2}$ to $10^{4}$.
With such a small resistivity, it is expected that the system essentially behaves as ideal MHD.
Therefore, the disk evolution may be significantly affected by dust growth.
Such a ``co-evolution of disk and dust" has not been investigated with multi-dimensional simulations and is a promising future research area.

\yt{In particular, this could be a possible solution for the missing mass problem in Class 0/I YSOs by promoting mass accretion in the disk.
Once the disk goes back to the ideal MHD regime, the strong magnetic braking in the disk and the development of magneto-rotational instability (MRI) over a large area of the disk are expected.
This will significantly enhance the mass accretion from the disk to the protostar.
As a result, the mass of the disk is likely to be reduced. Therefore, the dust growth and subsequent resistivity decrease may solve the missing mass problem in Class 0/I YSOs.}

Note that, in Figure \ref{rho_eta_large_dust}, the effect of dust growth is possibly exaggerated compared to the realistic case in which dust grains have a size distribution.
When the dust grains have a size distribution and there is a certain amount of small dust, we expect them to increase the resistivity. This would be especially true if such small dusts determine the total surface area.


\subsubsection{Impact of dust growth on the dust thermal emission}

The mass of most of the disk in Class 0/I YSOs is estimated from its dust thermal emission
by assuming a dust-to-gas mass ratio, dust opacity, and that disk is optically thin.
However, recent studies have suggested that these assumptions
may break down in at least some disks \citep{2018ApJ...868...39G,2018A&A...612A..54L}.
For convenience, we  estimate the optical depth of the circumstellar disk.
Assuming the dust absorption opacity at (sub-)mm as
$\kappa_{\rm {mm}} \sim 2 \cm^2 \rm g^{-1}$ \citep[e.g.,][]{1990AJ.....99..924B,1994A&A...291..943O,2018ApJ...869L..45B}
and a dust-to-gas mass ratio of 0.01, the optical depth of disk with constant $Q$ value is estimated as:
\begin{eqnarray}
\tau_{\mm}=0.9  Q_1^{-1} M_{*, 0.3 \msun}^{1/2} r_{50 \rm au}^{-12/7}.
\end{eqnarray}
where $M_*$ is the prototar mass and $r$ is the disk radius. We assume $T=150 ~r_{1 \rm au}^{-3/7} K$ \citep{1997ApJ...490..368C}.

This indicates that a massive disk with $Q\sim 1$ and size of $\lesssim 50$ au is optically thick in almost the entire region with (sub-)mm wavelengths,
and that the assumption of optically thin is not valid.
Note that this estimate is for a face-on disk, and a disk with inclination has a larger optical depth.

This estimate is reminiscent of the inconsistency of disk mass in Orion between
from ALMA 0.87 mm and from VLA 9 mm \citep{2020ApJ...890..130T}.
As shown in Figure 71 of \citet{2020ApJ...890..130T}, the estimate of the Class 0/I
disk mass in the Orion region using ALMA 0.87 mm is typically about 10 times smaller
than that using VLA 9 mm, possibly suggesting that the disk tends to be optically thick at sub-mm wavelengths.

A natural curiosity may be that, since the dust (absorption)
opacity at 9 mm is highly uncertain,
extrapolating the opacity from mm to 9 mm with dust spectral index $\beta=0$ (due to dust growth) should yield a consistent mass estimate.
However, this idea does not work. This is because the decrease in $\beta$ due to dust growth
(e.g., to $a_{\rm d}>1 $ cm) is not caused by an increase in dust opacity in the longer wavelength,
but by a decrease in dust opacity in the shorter wavelength (see \citet[][]{1993Icar..106...20M} or Figure 4 of \citet[][]{2018ApJ...869L..45B} for example),
or because (mono-sized) dust opacity obeys $\kappa_\lambda={\rm const}$ for $a_{\rm d}/\lambda \ll 1$
and $\kappa_\lambda \propto a_{\rm d}^{-1}$ for $a_{\rm d}/\lambda \gg 1$ \citep{1998asls.book.....B,2007ipid.book.....K}.
\yt{Therefore, as the dust grows from micron sized, the grains first reach mm sizes and the dust opacity at mm wavelengths begins to decrease.
Then, when the dust grains reach cm sizes, the dust opacity at cm wavelength begins to decrease.}
In other words, dust growth primarily causes an underestimation of disk mass at (sub-)mm, and not an overestimation at cm.


Furthermore, it has been shown that dust scattering,
in an optically thick disk
causes the underestimate of optical depth, and the disk appears optically thin and lightweight.
The dust scattering at (sub-)$\mm$ wavelengths has recently been widely recognized as new evidence of dust growth \citep{2015ApJ...809...78K,2016MNRAS.456.2794Y}.
For example, the change of the polarization pattern in Class I YSOs HL Tau shows that dust scattering is effective and that the dust may grow to $\sim 100 \mum$ \citep{2016ApJ...820...54K,2017ApJ...844L...5K} \yt{(Note, however, that recent laboratory experiments
have shown that detailed size estimates should be made with caution because of the strong influence of the dust shape \citep{2021ApJS..256...17M}).}
Dust scattering decreases the dust thermal emission from
an optically thick region of the disk to be $I_\nu \sim \sqrt{1-\omega_\nu}B_\nu(T)$, where $\omega_\nu$ is the single scattering albedo \citep{1993Icar..106...20M,2019ApJ...877L..18Z},
because scattering reduces the depth where photons can escape.
\citet{2019ApJ...877L..18Z} show that the estimated disk mass with $Q= 1$ and a radius of $r=50$ au is a tenth smaller than the actual mass (see their Figure 7) in (sub-)mm dust thermal emission. Thus, an optically thick, heavy disk with scattering can be
misidentified as an optically thin, lightweight disk if the dust scattering is ignored.
Hence, it is possible that a significant amount of disk mass is hidden from (sub-)mm observations if dust grows to $\sim 100 \mum$.

If dust grows further to $a_{\rm d} \gtrsim 1 \cm$, which is expected from the short growth timescale in the disk,
it inevitably causes radial drift, which also leads an underestimate of disk mass due to dust depletion.
It has been recognized that dust disks evolve on much shorter time scales than gas disks and that dust
can be depleted at a relatively early stage of disk evolution,
when dust growth and dust radial migration are taken into account \citep{2005ApJ...627..286T,2012A&A...539A.148B}.
\citet{2017ApJ...838..151T} investigated the evolution of dust-to-gas mass ratio in a disk undergoing envelope mass accretion.
They showed that the dust-to-gas mass ratio of the entire disk could be 3-10 times smaller than the ISM dust-to-gas mass ratio as the dust grows.
Furthermore, once the dust size exceeds the observation wavelength, the opacity begins to decreases. Therefore, the dust thermal emission further decreases.
This dust depletion and decrease of the opacity may also cause the underestimate of disk mass.


We believe that these recent studies regarding a change of resistivity and dust thermal emission
due to the dust growth have the potential to ease the tension between observations and simulations, and reconcile the discrepancy.

\section{Conclusions}
\yt{
In this chapter, we have examined the latest observational results and theoretical scenarios on the formation and evolution processes of protostars, protoplanetary disks, and outflows. Many observed properties of protostar/disk systems are in agreement with magnetized models, and could be a consequence of the coupling, decoupling, and recoupling processes of the magnetic field and gas at different scales and density regimes. 

For example, the interplay of magnetic braking and magnetic diffusion naturally explains the radii of observed Class 0/I disks.
The scenario that outflows are ejected directly from the disk via magnetic torque, rather than by the entrainment of the protostar jet,
is also consistent with recent observations of outflow rotation,
and evidence of outflow ejection around the edge of the disk from high-resolution observations of launching regions.}
\yt{ Furthermore, the relatively weak magnetic field of the protostar ($\sim 1$ kG) could be due non-ideal MHD effects, causing a decoupling in the vicinity ($<1$ au) of the protostar. According to the models, the weak magnetic field in and nearby the protostar could also be key for jet formation.
}

\yt{
However, some salient issues remain to be addressed in the future. The relatively massive protostellar disks predicted from non-ideal MHD simulations are not observed, although these findings could be reconciled if dust properties need to be revised as suggested by recent observational works. It also remains uncertain how protostellar disk masses evolve with time, and what are the best observational tracers to characterize disks at different stages, as they evolve. Finally, the importance of dust evolution on both the magnetic processes and the mechanisms of planet formation, which may have already begun in these early protostellar evolutionary stages, are challenging questions yet to be adressed by future theoretical and observational investigations.}




\textbf{Acknowledgments.} TR acknowledges support from the European Research Council (ERC) under advanced grant number 743029. AM acknowledges support from the European Research Council (ERC) under Starting grant MagneticYSOs (number 679937).

\bigskip
\parskip=0pt
\baselineskip=11pt
\bibliographystyle{pp7.bst}

\bibliography{bibreview.bib,article.bib,biblio_commercon.bib,biblio_zhao.bib}

\end{document}